\documentclass[12pt]{article}
\pdfoutput=1
\usepackage{amsfonts,amsthm,amsmath,amssymb,graphicx,hyperref,yfonts,cite}
\usepackage{xcolor}
\usepackage[vcentermath]{youngtab}
\newcommand{\bea}{\begin{eqnarray}\displaystyle}
\newcommand{\eea}{\end{eqnarray}}
\newcommand{\Tr}{{\rm Tr}}

\begin{document}
\makeatletter
\@addtoreset{equation}{section}
\makeatother
\renewcommand{\theequation}{\thesection.\arabic{equation}}
\vspace{1.8truecm}

{\LARGE{ \centerline{\bf Generating Functions for Giant Graviton Bound States}}}

\vskip.5cm 

\thispagestyle{empty} 
\centerline{{\large\bf Warren Carlson$^b$\footnote{{\tt Warren.Carlson@wits.ac.za}}, Robert de Mello Koch$^{a,b}$\footnote{{\tt robert@zjhu.edu.cn}} and Minkyoo Kim$^{b,c,}$\footnote{{\tt mimkim80@gmail.com}}}}

\vspace{.8cm}
\centerline{{\it $^a$ School of Science, Huzhou University, Huzhou 313000, China}}

\vspace{.8cm}
\centerline{{\it $^b$ School of Physics and Mandelstam Institute for Theoretical Physics}}
\centerline{{\it University of the Witwatersrand, Wits, 2050, }}
\centerline{{\it South Africa }}

\vspace{.8cm}
\centerline{{\it $^{c}$ Center for Quantum Spacetime (CQUeST),}}
\centerline{{\it Sogang University Seoul, 121-742,}}
\centerline{{\it South Korea}}

\vspace{1truecm}

\thispagestyle{empty}

\centerline{\bf ABSTRACT}

\vskip.2cm 
We construct generating functions for operators dual to systems of giant gravitons with open strings attached. These operators have a bare dimension of order $N$ so that the usual methods used to solve the planar limit are not applicable. The generating functions are given as integrals over auxiliary variables, which implement symmetrization and antisymmetrization of the indices of the fields from which the operator is composed. Operators of a good scaling dimension (eigenstates of the dilatation operator) are known as Gauss graph operators. Our generating functions give a simple construction of the Gauss graph operators which were previously obtained using a Fourier transform on a double coset. The new description provides a natural starting point for a systematic ${1\over N}$ expansion for these operators as well as the action of the dilatation operator on them, in terms of a saddle point evaluation of their integral representation. 

\setcounter{page}{0}
\setcounter{tocdepth}{2}
\newpage
\tableofcontents
\setcounter{footnote}{0}
\linespread{1.1}
\parskip 4pt

{}~
{}~

\section{Introduction}\label{introduction}

The AdS/CFT correspondence has generated significant progress in understanding both quantum gravity in negatively curved spacetimes, as well as strongly coupled Yang-Mills theory \cite{Maldacena:1997re,Gubser:1998bc,Witten:1998qj}. For observables with a dimension that is fixed as we take $N\to\infty$ integrability was discovered \cite{Minahan:2002ve,Beisert:2010jr} giving unprecedented control, at both weak and strong coupling, over the planar limit of the theory. For these observables, a complete matching with the dual string is evident (see, for example, \cite{Beisert:2003ea}) and we now know an exquisite matching between string states in the gravity Hilbert space and single trace gauge invariant operators in super Yang-Mills theory.

There are many observables whose bare dimensions scale with a power of $N$ as we take $N\to\infty$, that are of interest in holography. These operators are very heavy states and they probe non-perturbative aspects of the dual string theory. Nice examples include giant gravitons \cite{McGreevy:2000cw,Hashimoto:2000zp,Grisaru:2000zn} which are dual to operators constructed with order $N$ fields \cite{Balasubramanian:2001nh,Corley:2001zk,Berenstein:2004kk}, and new spacetime geometries which are dual to operators constructed using order $N^2$ fields \cite{Lin:2004nb}. The construction of the large $N$ expansion for these operators is not straight forward. We don't have a nice analogue of the statement that Feynman diagrams become ribbon graphs and that the genus of the graph determines the dependence on $N$. The basic problem is that the number of ribbon graphs of a given genus is now dependent on $N$ so that the $N$ dependence is no longer determined by the genus of the ribbon graph being summed. The combinatorics is much more involved and there does not seem to be a nice subset of diagrams that, if summed, produce the leading effect at large $N$. 

Without a better understanding of how to perform a systematic approximation, one could try to sum everything, producing results that are exact to all orders in ${1\over N}$. Remarkably, the paper \cite{Corley:2001zk} showed that this can actually be done in practice, for the free theory. A key insight of \cite{Corley:2001zk} was to show that the combinatorics of ribbon graphs can be reformulated using symmetric groups and their characters. This initiated an active area of research, opening new directions that are still being developed. The original paper \cite{Corley:2001zk} showed that Schur polynomials provide a basis for the local gauge invariant operators, that they diagonalize the free field two point function and that the free field two point function can be explicitly computed, to all orders in ${1\over N}$. In addition, all extremal correlators were also explicitly computed, again to all orders in ${1\over N}$. Schur polynomials are labelled by Young diagrams, which may have columns with a maximum length of $N$ as well as rows that are arbitrarily long. In the same way that there is a beautiful matching between the Hilbert space of strings and single trace operators in the planar limit, \cite{Corley:2001zk} argued for a matching between long columns and sphere giant gravitons, as well as between long rows and AdS giant gravitons, generalizing and completing the observations of \cite{Balasubramanian:2001nh} which related single sphere giant gravitons to subdeterminant operators. See also \cite{Aharony:2002nd,Berenstein:2003ah}. These results were quickly generalized to multi-matrix models in a series of papers \cite{Kimura:2007wy,Brown:2007xh,Brown:2008ij,Bhattacharyya:2008rb,Bhattacharyya:2008xy}, as well as to other gauge groups \cite{Caputa:2013hr,Caputa:2013vla,Kemp:2014apa,Kemp:2014eca,Pasukonis:2013ts,Mattioli:2016gyl,Ramgoolam:2018xty,Barnes:2021ehp,Barnes:2022qli}.

Giant gravitons branes can be excited by attaching open strings to them. The gauge theory operators dual to these states are described using decorated Young diagrams \cite{Balasubramanian:2004nb,deMelloKoch:2007rqf}. The spin chain language that worked so well for closed strings in the planar limit has to be modified when describing the open strings attached to giant gravitons. The new effect that needs to be included is that single fields can hop between the giant graviton and onto the string, so that we are forced to consider the dynamics of a spin chain defined on a dynamical lattice. A creative solution to this problem was put in place in \cite{Berenstein:2005fa,Berenstein:2006qk} for strings attached to single sphere giant gravitons, and then extended to more general brane bound states in \cite{deMelloKoch:2007nbd,Bekker:2007ea}. By now, a compelling physical picture is evident \cite{Berenstein:2014isa,Lin:2014yaa,Koch:2015pga,Berenstein:2020grg,Holguin:2021qes,Berenstein:2022cju}.

As we have just discussed, combinatorial approaches based on permutation groups and their characters, compute correlation functions exactly to all order in ${1\over N}$. This is a valuable source of information and has stimulated a lot of progress, but it hasn't given any guidance into how to develop a systematic ${1\over N}$ expansion to the physics of excited giant graviton systems. Since we don't expect any except the most symmetric problems to be exactly solvable, this is an obstacle to progress. One attempt at a simplification has been to use a displaced corners approximation \cite{Carlson:2011hy,deMelloKoch:2011wah}. The idea is that the physics of the open strings should simplify in the limit in which the giant gravitons are well separated in spacetime. This corresponds to a limit in which the lengths of distinct rows (or columns) of the Young diagram differ by an amount $\sim N$. In this limit one can simplify the description of the operators considered, at large $N$ \cite{Carlson:2011hy,deMelloKoch:2011wah}, and some mixing problems can be solved exactly \cite{deMelloKoch:2011ci,deMelloKoch:2012ck}. There is a nice harmony between these results and another complimentary approach, based a coherent state description of the giant gravitons \cite{Berenstein:2013md,Berenstein:2013eya,Berenstein:2014pma,Berenstein:2014zxa}. However, these small steps do not really improve the situation and a systematic ${1\over N}$ expansion of the giant plus open string physics, in which, for example, we see the splitting and joining of open strings, remains to be developed.

An important recent development for the case of maximal giants was reported in the series of papers \cite{Jiang:2019xdz,Jiang:2019zig,Yang:2021kot}. The maximal giant graviton is described by a determinant. The idea of \cite{Jiang:2019xdz,Jiang:2019zig,Yang:2021kot} is to write the determinant as a fermionic integral. The rewriting replaces the symmetric group projector onto the representation given by a column of length $N$, written in terms of symmetric group characters, with an integral over some new auxiliary Grassmann fields. By writing the symmetric group projector as an integral one opens the possibility of performing a saddle point analysis which can provide the starting point for a systematic ${1\over N}$ expansion. Similar integrals have been constructed for the projectors onto representations labelled by a single column or row, with an arbitrary length \cite{Chen:2019gsb,Chen:2019kgc}. Another development \cite{Berenstein:2022srd,Holguin:2022drf,Lin:2022wdr,Holguin:2022zii}, which is closely related, has shown that many of the results from the combinatorial approach can be reproduced by writing suitable generating functions of coherent states as a matrix integral. In some cases these matrix integrals are related to the well known Harish-Chandra-Itzykson-Zuber integral and can be computed using localization. In this approach, non-trivial combinatorial information is repackaged in the manipulations of the coherent state generating function. This is a promising development, relevant to systematic ${1\over N}$ expansions for heavy operators.

The central question motivating us in this paper, is the systematic ${1\over N}$ expansion for giant graviton systems. Physically this expansion is naturally described in terms of splitting and joining interactions of open strings. Towards this end, we develop integral representations for generating functions of Schur polynomials labelled by Young diagrams that have two rows (corresponding to a boundstate of two AdS giants) or two columns (corresponding to a boundstate of two sphere giants). Our approach to the ${1\over N}$ expansion entails developing a saddle point expansion of these integral representations \cite{ongoing}. Apart from this application, we will see that the new representations provide a dramatically simpler description of excited giant graviton systems. To demonstrate this, we explain how to recover known results in what follows. The class of problems we focus on is rather rich because a bound state of two giant gravitons is described by interesting dynamics. The world volume theory of $p$ branes is a U$(p)$ gauge theory \cite{Witten:1995im}, so in this situation with $p=2$ the effective Hamiltonian coming from the dilatation operator should reproduce a U$(2)$ non-Abelian gauge theory \cite{Balasubramanian:2004nb,deMelloKoch:2020agz}. To verify this expectation, we must necessarily correctly describe the splitting and joining of open strings, suggesting that this two giant bound state system is a good testing ground for the methods we are trying to develop.

Constructing the operators dual to a bound state of two giant gravitons requires symmetric group projectors labelled by Young diagrams that have two columns or two rows. The key new progress reported in this paper starts with the construction of an integral representation for these projectors. To motivate the construction, in Section \ref{youngProjectors} we describe how Schur polynomials are constructed using Young projectors. It is the Young projectors that can be given an integral representation. Since there is a direct connection between Schur polynomials and characters of the symmetric group, the same integral representations also provide new formulas for character generating functions. A simple generalization beyond the case of a single matrix produces integral representations for operators that are dual to excited giant graviton bound states. A detailed comparison of the operators constructed show that they correspond to the Gauss graph operators discovered in \cite{deMelloKoch:2011ci,deMelloKoch:2012ck}. We further develop this connection in Section \ref{actionofD}, where we evaluate the action of the dilatation operator on the excited giant graviton bound states. We discuss our results and suggest how they might be extended in Section \ref{conclusions}. The Appendices collect technical details needed to perform the computations discussed in the body of the paper.

\section{Projectors and Generating Functions}\label{youngProjectors}

Half BPS operators in ${\cal N}=4$ super Yang-Mills theory, dual to a system of giant graviton branes, are given by Schur polynomials $\chi_R(Z)$. The label $R$ is a Young diagram with at most $N$ rows. Each long column\footnote{By a ``long column'' of a Young diagram $R$, we mean a column that has order $N$ boxes in it. The number of boxes in each column is bounded by $N$. Similarly, a ``long row'' refers to a row with order $N$ boxes. Row lengths are unbounded.} in the Young diagram corresponds to a sphere giant graviton and each long row to an AdS giant graviton\footnote{Recall that a giant graviton is a spherical S$^3$ brane that carries a D3-brane dipole moment and the same quantum numbers as a point graviton. A sphere giant graviton is one for which the world volume S$^3\subset$S$^5$, while an AdS giant graviton is one for which S$^3\subset$AdS$_5$.}. An innovative framework for thinking about the construction of these Schur polynomials and their correlation functions, was developed in \cite{Corley:2002mj} where the role of projection operators was stressed and an elegant diagrammatic formalism was established. The relevant projection operators used to construct an operator composed of $n$ $Z$ fields can be written in terms characters of the symmetric group S$_n$, as well as permutations acting on $V^{\otimes\, n}$ where $V$ is the vector space that the matrix $Z$ acts on. These characters and permutations are determined by the Young diagram $R$ of the Schur polynomial as well as the number $n$ of fields in the operator. The parameter $N$ does not enter transparently in the construction\footnote{$N$ does appear as a cut off on the number of rows of $R$.}. Consequently, simplifications as $N\to\infty$ are not immediate and it is not clear how one would pursue a systematic ${1\over N}$ expansion. To go beyond the half BPS operators, we need to go beyond the single matrix $Z$ sector. The simplest way to do this is by constructing operators which are systematically small deformations of the half BPS operators. These are nice systems to study as we expect that they will be simpler than the generic system that might be considered. In the dual gravity description, the giant graviton system is excited by attaching open strings to the giant gravitons. These generalizations also have a nice description in terms of projectors which proves useful when evaluating correlation functions and matrix elements of the dilatation operator, but again it is not obvious how a systematic large $N$ expansion can be developed. 

In this section we briefly review the projection operator construction and its relation to Young symmetrizers. The language of Young symmetrizers gives a construction of the projectors which employs simple symmetrization and antisymmetrization operations on the indices of the $Z$ fields appearing in the operator. Since these operations can be realized by introducing extra fermionic and bosonic auxiliary fields, we are able to construct an integral representation for the projection operators. More precisely, the integral representation gives a generating function for all Schur polynomials with two rows or two columns. This novel generating function provides a starting point for a systematic saddle point approximation. The generating function can also be interpreted as a generating function for characters, suggesting that one might be able to generate a systematic ${1\over N}$ expansion of the characters themselves. There are natural ways to generalize the generating functions of Schur polynomials, to generating functions describing the Schur polynomials with strings attached. Remarkably, as we will demonstrate, the operators constructed in this way have an immediate and direct connection to other constructions of excited giant graviton operators based on restricted Schur polynomials and Gauss graph operators. We will develop this connection in the Section \ref{actionofD} where we evaluate the action of the dilatation operator on the excited giant graviton states.

\subsection{Young Symmetrizers}\label{YoungSymmetrizers}

Consider the construction of a Schur polynomial $\chi_R$ which is a function of a single complex adjoint scalar $Z$. The Young diagram $R$ labelling the Schur polynomial specifies the system of D-branes as well as an irreducible representation $R$ of the symmetric group $S_n$ where $n$ is the number of boxes in the diagram $R$, which is also the number of $Z$ fields used to construct the operator. A simple formula for the Schur polynomial employs a projection operators $P_R$ as follows \cite{Corley:2002mj}
\bea
\chi_{R}(Z)={\rm Tr}_{V^{\otimes\, n}}(P_R Z^{\otimes n})
\eea
Here $Z$ acts on the vector space $V$ and both $P_R$ and $Z^{\otimes n}$ are linear operators mapping $V^{\otimes\, n}$ to itself. The image of the projector corresponds simultaneously to a representation of the symmetric group $S_n$ and a representation of the unitary group $U(N)$, both labelled by the same Young diagram\footnote{The projector projects to a subspace of dimension $d_R{\rm Dim}_R$, where $d_R$ is the dimension of symmetric group representation $R$ and ${\rm Dim}_R$ is the dimension of $U(N)$ representation $R$. States in this subspace carry both an $S_n$ state label and a $U(N)$ state label.} $R$. A concrete expression for the projection operator is given by\footnote{As written $P_R$ is proportional to a projection operator. For a correctly normalized projection operator obeying $P_RP_S=\delta_{RS}P_R$ the right hand side of (\ref{projop}) must be multiplied by $d_R$.}
\bea
P_R={1\over n!}\sum_{\sigma\in S_n}\chi_R(\sigma)\sigma\label{projop}
\eea
where the matrix elements of the permutation $\sigma$ are given by
\bea
\sigma^{i_1 i_2\cdots i_n}_{j_1j_2\cdots j_n}=\delta^{i_1}_{j_{\sigma(1)}}\delta^{i_2}_{j_{\sigma(2)}}\cdots\delta^{i_n}_{j_{\sigma(n)}}
\eea
so that
\bea
{\rm Tr}( \sigma Z^{\otimes \, n})
=Z^{i_1}_{i_{\sigma(1)}}Z^{i_2}_{i_{\sigma(2)}}\cdots Z^{i_n}_{i_{\sigma(n)}}
\label{actionofsn}
\eea
Thus, given the characters of $S_n$ a formula for $P_R$ easily follows. These formulas can be extended to study operators constructed using more than one matrix, making it possible to consider perturbations of these half BPS operators. The characters that are needed are less well known and the combinatorics quickly becomes complicated. Nevertheless, at the level of two point functions we obtain a complete generalization of the Schur polynomial results i.e. we are able to construct a complete basis for the local operators, this basis takes finite $N$ effects into account and this basis diagonalizes the free field theory two point function \cite{Bhattacharyya:2008rb}.

An alternative (but completely equivalent) approach uses the Young symmetrizer associated to $R$ to construct the projector $P_R$. A Young symmetrizer is an element of the group algebra of the symmetric group. It is an endomorphism of the vector space $V^{{\otimes n}}$ which uses the action (\ref{actionofsn}) of $S_{n}$ on $V^{{\otimes n}}$ given by permuting indices. We illustrate the Young projectors making use of a specific example. Consider a Young diagram $R$ that has two rows, with three boxes in the first row and two in the second. Each box corresponds to a field. The boxes in $R$ are labelled, with a unique integer, as follows
\bea
\young(123,45)
\label{exampleR}
\eea
There is no rule for the placement of integers in boxes, except that each box should be assigned a unique integer. Each box corresponds to the indices of a field in $Z^{\otimes 5}$ as shown below
\bea
Z^{i_1}_{j_1}  Z^{i_2}_{j_2}   Z^{i_3}_{j_3}   Z^{i_4}_{j_4}   Z^{i_5}_{j_5}
\label{5tensored}
\eea
For every labelled Young diagram $R$ we can define a ``row group'' for each row and a ``column group'' for each column. Each row group permutes the indices in the row, in all possible ways. Thus, for the example given in (\ref{exampleR}) the two row groups are
\bea
{\cal G}_{\rm row 1}=\{ 1,(12),(13),(23),(123),(132)\}\qquad\qquad {\cal G}_{\rm row 2}=\{1,(45)\}
\eea
where we have written the permutations in cycle notation. The product of these groups is denoted ${\cal G}_{\rm row}$. For the example we consider here ${\cal G}_{\rm row}$ is of order 12. Similarly, the column groups are given by
\bea
{\cal G}_{\rm col 1}=\{1,(14)\}\qquad\qquad {\cal G}_{\rm col 2}=\{1,(25)\}
\eea
and their product, denoted ${\cal G}_{\rm col}$, has order 4. The Young symmetrizer is now given by
\bea
Y_{\tiny\yng(3,2)}=\sum_{\sigma_1\in {\cal G}_{\rm row}}\sum_{\sigma_2\in {\cal G}_{\rm col}}\,\, {\rm sgn} (\sigma_2)\sigma_1 \sigma_2
\eea
where ${\rm sgn} (\sigma)$ is the parity of $\sigma$. A cycle of length 2 is called a transposition. Any permutation can be written as a product of transpositions. A particular permutation is even or odd if it can be expressed using an even or an odd number of transpositions. For an even permutation ${\rm sgn} (\sigma)=1$ and for an odd permutation ${\rm sgn} (\sigma)=-1$. When applied to the tensor product (\ref{5tensored}) we find the $\sigma_2$ factor above antisymmetrizes over the pairs $i_1,i_4$ and $i_2,i_5$ of indices (i.e. the indices appearing in the columns of (\ref{exampleR})) while the $\sigma_1$ factor symmetrizes over the $i_1,i_2,i_3$ indices and over the $i_4,i_5$ indices (i.e. the indices associated to the rows of $R$). After a simple sum we find that
\bea
{\rm Tr}(Y_{\tiny\yng(3,2)} Z^{\otimes 5})&=&{\rm Tr}(Z^5)+2\,{\rm Tr}(Z^2){\rm Tr}(Z)^3-4\,{\rm Tr}(Z)^2{\rm Tr}(Z^3)+3\,{\rm Tr}(Z){\rm Tr}(Z^2)^2\cr\cr
&&-6\,{\rm Tr}(Z){\rm Tr}(Z^4)+4\,{\rm Tr}(Z^2){\rm Tr}(Z^3)
\eea
The Schur polynomial corresponding to this Young diagram is given by
\bea
\chi_{\tiny\yng(3,2)}(Z)&=&{1\over 5!}
\left(5\,{\rm Tr}(Z^5)+10\,{\rm Tr}(Z^2){\rm Tr}(Z)^3-20\,{\rm Tr}(Z)^2{\rm Tr}(Z^3)
+15\,{\rm Tr}(Z){\rm Tr}(Z^2)^2\right.\cr\cr
&&\left.-30\,{\rm Tr}(Z){\rm Tr}(Z^4)+20\,{\rm Tr}(Z^2){\rm Tr}(Z^3)\right)
\eea
which illustrates the general formula
\bea
\chi_{R}(Z)={d_R\over n!}{\rm Tr}(Y_{\tiny\yng(3,2)} Z^{\otimes 5})
\label{schurtoyoung}
\eea
Thus, up to an unimportant overall constant, we can construct Schur polynomials by using Young projectors. This observation can be used to give an integral representation of the Young projectors needed to construct the two giant graviton bound states.

\subsection{Bound states of two AdS Giants}\label{TwoRowSchur}

A bound state of two AdS giants, one of momentum $p_1$ and a second of momentum $p_2$ with $p_2\le p_1$, corresponds to a Schur polynomial labelled by a Young diagram with two rows. There are $p_1$ boxes in the first row and $p_2$ boxes in the second. Denote this Young diagram by listing row lengths, i.e. $R=(p_1,p_2)$. To describe a system of two giant gravitons, both $p_1$ and $p_2$ are of order $N$. The Young projector must symmetrize over all indices in row 1 and all indices in row 2, as well as antisymmetrize the pair of indices belonging to each column. The symmetrization to be performed represents the bulk of the work. Fortunately, this symmetrization of indices is easily accomplished with the help of a Gaussian integral. Indeed, a simple application of Wick's theorem implies that
\bea
 {1\over\pi^N}\int d^N\bar{\phi}\int d^N\phi\,\, e^{-\bar{\phi}^a\phi_a} \bar{\phi}^{i_1}\bar{\phi}^{i_2}\cdots \bar{\phi}^{i_k}\phi_{j_1}\phi_{j_2}\cdots\phi_{j_k}=\sum_{\sigma\in S_k}\delta^{i_1}_{j_{\sigma(1)}}\delta^{i_2}_{j_{\sigma(2)}}\cdots 
\delta^{i_k}_{j_{\sigma(k)}}\label{bosonicresult}
\eea
\bea
d^N\phi \equiv \prod_{a=1}^N d\phi_a \qquad\qquad
d^N\bar{\phi} \equiv \prod_{a=1}^N d\bar{\phi}^a
\eea

To construct the generating function, we perform the anti-symmetrization in each column by hand and use $(\varphi_i,\bar{\varphi}^i)$ to symmetrize indices in the first row and $(\phi_i,\bar{\phi}^i)$ to symmetrize indices in the second row. The two giant graviton bound state generating function can then be written as follows (as usual, repeated indices are summed)
\bea
Z_{\rm AdS}(t_1,t_2)&=&\sum^\infty_{p_1=p_2}\sum_{p_2=1}^\infty t_2^{2p_2}t_1^{p_1-p_2}{(p_1+p_2)!\over d_{(p_1,p_2)}p_2!(p_1-p_2)!}\chi_{(p_1,p_2)}(Z)\cr\cr\cr
&=& {1\over \pi^{2N}}\int d^N\bar{\varphi}\int d^N\varphi\int d^N\bar{\phi}\int d^N\phi\cr\cr
&&\qquad\qquad 
e^{-\bar{\varphi}^i\varphi_i-\bar{\phi}^i\phi_i+t_1\bar{\varphi}^iZ_i^j\varphi_j
+t_2^2\bar{\varphi}^i\bar{\phi}^j (Z^k_i Z^l_j-Z^l_i Z^k_j)\varphi_k\phi_l}
\label{twoAdSgiants}
\eea
The coefficient of the Schur polynomial in the first line above combines factors arising from power expanding the exponential on the second line above (so that $t_2^{2p_2}$ comes with $1/p_2!$ and $t_1^{p_1-p_2}$ comes with $1/(p_1-p_2)!$), as well as from the coefficient in (\ref{schurtoyoung}). In Appendix \ref{2AdS} we check this result by performing the integral to second order in $t_1$ and $t_2$ and show that we reproduce the correct formulas for the relevant Schur polynomials. 

This generating function can also be interpreted as a generating function for characters of the symmetric group $S_{p_1+p_2}$ in the representation labelled by Young diagram $R=(p_1,p_2)$. Permutations with a given cycle structure all belong to the same conjugacy class and the character is a function on these classes. We follow the standard notation $(1^{n_1},2^{n_2},\cdots,(p_1+p_2)^{n_{p_1+p_2}})$ to denote the cycle structure of the permutation which has $n_k$ $k$-cycles. In what follows we use the shorthand $\{ i^{n_i}\}$ for this cycle structure. As usual we have
\bea
\sum_{k=1}^{p_1+p_2} k n_k =p_1+p_2
\eea
for permutations $\sigma\in S_{p_1+p_2}$. Denote the symmetric group $S_{p_1+p_2}$ character, of a permutation with cycle structure $\{ i^{n_i}\}$ in irreducible representation $R=(p_1,p_2)$ by $\chi_{(p_1,p_2)}(\{ i^{n_i}\})$. The number of permutations with this cycle structure is given by
\bea
{(p_1+p_2)!\over 1^{n_1}n_1!2^{n_2}n_2!3^{n_3}n_3!\cdots (p_1+p_2)^{n_{p_1+p_2}}n_{(p_1+p_2)!}}\label{CCsize}
\eea
We can then write the generating function for characters of the irreducible representations $R=(p_1,p_2)$ as follows
\bea
Z_{\rm AdS}(t_1,t_2)&=&
\sum_{p_1=p_2}^\infty \sum_{p_2=1}^\infty\sum_{\{n_i\}}\,\,
{t_2^{2p_2}t_1^{p_1-p_2}\over 1^{n_1}n_1!2^{n_2}n_2!\cdots (p_1+p_2)^{n_{p_1+p_2}}n_{(p_1+p_2)!}}
{(p_1+1)!\over (p_1-p_2+1)!}\cr\cr\cr
&&\qquad \times\chi_{(p_1,p_2)}(\{i^{n_i}\}) {\rm Tr}(Z)^{n_1}{\rm Tr}(Z^2)^{n_2}\cdots {\rm Tr}(Z^{p_1+p_2})^{n_{p_1+p_2}}
\eea
where $\sum_{\{n_i\}}$ denotes a sum over the conjugacy classes of $S_{p_1+p_2}$. As a simple illustration of this formula, expanding to linear order in $t_1$ and $t_2^2$, we find
\bea
Z_{\rm AdS}(t_1,t_2)&=&
1+t_1\chi_{\tiny\yng(1)}((\cdot))\Tr(Z)+t_2^2\chi_{\tiny\yng(1,1)}((\cdot)(\cdot))\Tr(Z)^2+t_2^2\chi_{\tiny\yng(1,1)}((\cdot\cdot))\Tr(Z^2)+\cdots\cr\cr
&=&1+t_1\Tr(Z)+t_2^2\Tr(Z)^2-t_2^2\Tr(Z^2)+\cdots
\eea
reproducing the correct characters for $S_1$ and $S_2$
\bea
\chi_{\tiny\yng(1)}((\cdot))=1\qquad \chi_{\tiny\yng(1,1)}((\cdot)(\cdot))=1\qquad 
\chi_{\tiny\yng(1,1)}((\cdot\cdot))=-1
\eea
This integral representation is a nice repackaging of the computation of characters using Young symmetrizers. Since the computation of the power series expansion of $Z_{\rm AdS}(t_1,t_2)$ involves evaluating moments of a Gaussian integral, there is no obstacle to computing characters for any reasonable value of $p_1$ and $p_2$. Evaluating $Z_{\rm AdS}(t_1,t_2)$ explicitly is frustrated by the fact that the term with coefficient $t_2^2$ is quartic in the integration variables. By using the simple identity
\bea
{1\over \pi}\int_{-\infty}^{\infty} dx_1 \int_{-\infty}^{\infty} dx_2 e^{-(x_1+ix_2)(x_1-ix_2)+A (x_1+ix_2)+B(x_1-ix_2)}&\equiv&{1\over \pi}\int dx \int d\bar{x} e^{-x\bar{x}+A x+B\bar{x}}\cr\cr
&=&e^{AB}
\eea
we find
\bea
Z_{\rm AdS}(t_1,t_2)&=& 
{1\over \pi^{2N+2}}\int d^N\bar{\varphi}\int d^N\varphi\int d^N\bar{\phi}\int d^N\phi\int dx\int d\bar{x}\int dy\int d\bar{y}\cr\cr
&&
e^{-\bar{\varphi}^i\varphi_i-\bar{\phi}^i\phi_i-x\bar{x}-y\bar{y}+t_1\bar{\varphi}^iZ_i^j\varphi_j
+t_2x\bar{\varphi}^iZ^k_i\varphi_k+t_2\bar{x}\bar{\phi}^j Z^l_j\phi_l
+t_2y\bar{\varphi}^iZ^l_i \phi_l-t_2\bar{y}\bar{\phi}^j Z^k_j\varphi_k}
\nonumber
\eea
Notice that after this rewriting the integral over both $\varphi_i,\bar\varphi^i$ and $\phi_i,\bar\phi^i$ are Gaussian so that we can do these integrals exactly. The result is
\bea
Z_{\rm AdS}(t_1,t_2)
&=& {1\over \pi^2}\int dx \int d\bar{x}\int dy \int d\bar{y}
{e^{-x\bar{x}-y\bar{y}}\over\det M}
\eea
where
\bea
M&=&{\bf 1}_2\otimes {\bf 1}_N-m\otimes Z\qquad\qquad
m\,\,\,=\,\,\,\left[\begin{array}{cc} t_1+t_2 x &t_2 y\\ -t_2\bar{y} &t_2\bar{x}\end{array}\right]
\eea
and ${\bf 1}_k$ is the $k\times k$ identity matrix. The eigenvalues $\lambda_a$ with $a=1,2$ of $m$ are given by
\bea
 \lambda_{1,2}={1\over 2}\left(t_1+t_2(x+\bar{x})\pm\sqrt{(t_1+t_2(x+\bar{x}))^2-4(\bar{x}t_1t_2+t_2^2(x\bar{x}+y\bar{y}))}\right)\label{eigens}
\eea
Denote the eigenvalues of $Z^i_j$ by $z_i$. A straight forward computation now shows that
\bea
\det M&=&\prod_{i=1}^N {1\over (1-\lambda_1 z_i)(1-\lambda_2 z_i)}\cr\cr
&=&\exp\left(\sum_{n=1}^\infty {\lambda_1^n+\lambda_2^n\over n}\Tr (Z^n)\right)\cr\cr
&=&\exp\left(\sum_{n=1}^\infty f_n \Tr (Z^n)\right)
\eea
where $f_n$ is the following polynomial
\bea
f_n&=&\sum_{r=0}^{{\rm int}\left({n\over 2}\right)}{(-\beta)^r\over r!}{\alpha^{n-2r}\over (n-2r)!}(n-r-1)!\cr\cr
\alpha&=&t_1+t_2(x+\bar{x})\qquad\qquad
\beta\,\,=\,\, t_1t_2\bar{x}+t_2^2 (x\bar{x}+y\bar{y})\label{definefn}
\eea
where ${\rm int}\left({n\over 2}\right)$ is an instruction to take the integer part of ${n\over 2}$. Thus, we have
\bea
Z_{\rm AdS}(t_1,t_2)
&=& {1\over \pi^2}\int dx \int d\bar{x}\int dy \int d\bar{y}\,\,
e^{-x\bar{x}-y\bar{y}+\sum_{n=1}^\infty f_n\Tr (Z^n)}\label{finalZads}
\eea
This implies the following formula for the character
\bea
\chi_{(p_1,p_2)}(\{i^{n_i}\})&=&[t_1^{p_1-p_2}t_2^{2p_2}]\left({1\over\pi^2}\int dx
\int d\bar{x}\int dy\int d\bar{y}\, e^{-x\bar{x}-y\bar{y}}\,\,{(p_1-p_2+1)!\over (p_1+1)!}
\prod_{i=1}^{p_1+p_2}(f_{n_i})^{n_i}\right)\cr\cr
&&
\eea
where $[x^n]p(x)$ stands for the coefficient of $x^n$ in the power series expansion of the polynomial $p(x)$. This integral formula for characters of the symmetric group is a new result. Lets test this for the case that $p_2=0$, in which we case we know that
\bea
 \chi_{(p_1,0)}(\{i^{n_i}\})&=&1\label{charres}
\eea
When $t_2=0$ we easily find $f_n=t_1^n/n$ and it is possible to compute (\ref{finalZads}) explicitly
\bea
Z_{\rm AdS}(t_1,0)&=&e^{\sum_{n=1}^\infty {t_1^n\over n}\Tr (Z^n)}\cr\cr
&=&\sum_{p_1=0}^\infty\sum_{\{ n_i\}}t_1^{p_1} {1\over n_1!}{1\over 2^{n_2}n_2!}\cdots {1\over 2^{n_{p_1}}n_{p_1}!}\Tr (Z)^{n_1}\Tr (Z^2)^{n_2}\cdots \Tr (Z^{p_1})^{n_{p_1}}\cr\cr
&=&\sum_{p_1=0}^\infty\sum_{\{ n_i\}}t_1^{p_1} {1\over n_1!}{1\over 2^{n_2}n_2!}\cdots {1\over 2^{n_{p_1}}n_{p_1}!}\chi_{(p_1,0)}(\{ i^{n_i}\})\Tr (Z)^{n_1}\Tr (Z^2)^{n_2}\cdots \Tr (Z^{p_1})^{n_{p_1}}\nonumber
\eea
which immediately implies (\ref{charres}). As a second example, consider $p_1+p_2=6$ and choose a character from the conjugacy class $(2^3)$. We easily find
\bea
{1\over\pi^2}\int dx
\int d\bar{x}\int dy\int d\bar{y}\, e^{-x\bar{x}-y\bar{y}}\,\,
(f_2)^3\,\,=\,\,t_1^6-6t_1^4 t_2^2+60t_1^2t_2^4-72t_2^6
\eea
Thus we find
\bea
\chi_{\tiny\yng(3,3)}([2^3])&=&[t_2^6]\left({1\over 24}{1\over\pi^2}\int dx
\int d\bar{x}\int dy\int d\bar{y}\, e^{-x\bar{x}-y\bar{y}}\,\,
(f_2)^3\right)\,\,=\,\, -3\cr\cr
\chi_{\tiny\yng(4,2)}([2^3])&=&[t_2^4t_1^2]\left({1\over 20}{1\over\pi^2}\int dx
\int d\bar{x}\int dy\int d\bar{y}\, e^{-x\bar{x}-y\bar{y}}\,\,
(f_2)^3\right)\,\,=\,\, 3\cr\cr
\chi_{\tiny\yng(5,1)}([2^3])&=&[t_2^2t_1^4]\left({1\over 6}{1\over\pi^2}\int dx
\int d\bar{x}\int dy\int d\bar{y}\, e^{-x\bar{x}-y\bar{y}}\,\,
(f_2)^3\right)\,\,=\,\, -1\cr\cr
\chi_{\tiny\yng(6)}([2^3])&=&[t_1^6]\left({1\over\pi^2}\int dx
\int d\bar{x}\int dy\int d\bar{y}\, e^{-x\bar{x}-y\bar{y}}\,\,
(f_2)^3\right)\,\,=\,\, 1
\eea
which are indeed the correct results. This demonstrates that the computation of characters of $S_n$ is reduced to evaluating moments of a Gaussian integral.

An interesting extension of these ideas would be to consider characters for $R=(p_1,p_2)$ with both $p_1-p_2$ and $p_2$ of order $N$. The relevant character can be extracted from $Z_{\rm AdS}(t_1,t_2)$ by performing the following contour integrals
\bea
\oint_{\cal C_1}{dt_1\over t_1^{1+p_1-p_2}}\oint_{\cal C_2}{dt_2\over t_2^{1+2p_2}}
\,\, Z_{\rm AdS}(t_1,t_2)
\eea
Employing a saddle point evaluation would give a ${1\over N}$ expansion of the characters. The leading term in this expansion should agree with the results of \cite{asymptoticchar}.

\subsection{Excited bound states of two AdS Giants}\label{excitedAdS}

\begin{figure}[h]%
\begin{center}
\includegraphics[width=0.65\columnwidth]{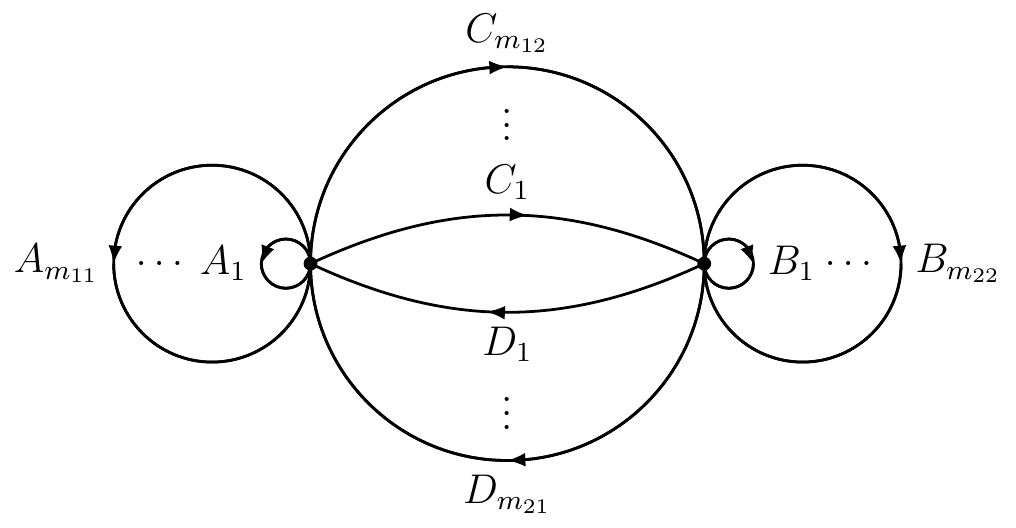}%
\caption{This Figure gives a graphical representation of the excited brane bound states described by the generating function (\ref{ZAdSW}). The node on the left represents the brane with momentum $p_1$ while the node on the right is for the brane of momentum $p_2$.}
\label{Fig:Strings}
\end{center}
\end{figure}

To extend the analysis of the previous subsection we consider excited giant graviton boundstates obtained by attaching open strings to the brane system. Open strings are described by words composed of $O(\sqrt{N})$ letters \cite{Berenstein:2002jq}. The letters can be any of the scalar or fermion fields of ${\cal N}=4$ super Yang-Mills theory, the field strength tensor or covariant derivatives of them. Attach $m_{11}$ open strings described by the words  $W_{A_1},W_{A_2},\cdots,W_{A_{m_{11}}}$ to the first brane (of momentum $p_1$). Since indices of the first brane are symmetrized by the $\bar{\varphi}^i$, $\varphi_i$ fields attaching these strings is achieved by including the factor
\bea
\prod_{r=1}^{m_{11}}\bar{\varphi}^{i_r} (W_{A_r})_{i_r}^{j_r}\varphi_{j_r}
\eea
in the integrand of the generating function. We also attach $m_{22}$ open strings, described by the words $W_{B_1},W_{B_2},\cdots,W_{B_{m_{22}}}$, to the second brane (of momentum $p_2$). Since these fields are in the second row of the Young diagram describing the brane system, we must anti-symmetrize these words with the $Z$ field that lives in the same column but in the first row. Thus, attaching these strings is achieved by including the factor
\bea
\prod_{s=1}^{m_{22}}(\bar{\varphi}^{i_s} Z_{i_s}^{j_s}\varphi_{j_s}\,\bar{\phi}^{k_s}(W_{B_s})_{k_s}^{l_s}\phi_{l_s}-\bar{\varphi}^{i_s}Z_{i_s}^{j_s}\phi_{j_s}\, \bar{\phi}^{k_s}(W_{B_s})_{k_s}^{l_s}\varphi_{l_s})
\eea
in the integrand. We can also attach strings that stretch between the two branes. To do this, we start with the product of a factor which attaches a string to the first brane and a factor which attaches a string to the second brane. From this we obtain a pair of strings stretched between the branes by permuting column indices, thereby swapping the endpoints of the two strings. There is a non-trivial fact \cite{Balasubramanian:2004nb} about the Hilbert space of states on a two giant graviton brane bound state that this construction must respect. The world volume theory of a system of $p$ branes is a U$(p)$ gauge theory. One consequence of this gauge symmetry is the Gauss Law, which states that for a giant graviton brane, which has a compact world volume, the total charge on the brane vanishes. In the open string language, this forces the number of strings leaving any particular brane to be equal to the number of strings terminating on the brane. We implement the Gauss Law constraint by adding open strings in pairs, one stretching from 1 to 2 and one from 2 to 1. Describe the $m_{12}$ open strings stretching from the first to the second brane using the words $W_{C_1},W_{C_2},\cdots,W_{C_{m_{12}}}$ and describe the $m_{21}$ ($=m_{12}$) open strings stretching from the second to the first brane using the words $W_{D_1},W_{D_2},\cdots,W_{D_{m_{21}}}$. Thus the generating function for this class of excited bound states, using an obvious dot product notation, is
\bea
Z_{\rm AdS}(t_1,t_2,\{W\})&=& {1\over \pi^{2N}}\int d^N\bar{\varphi}\int d^N\varphi 
\int d^N\bar{\phi}\int d^N\phi\,\, e^{-\bar{\varphi}\cdot\varphi-\bar{\phi}\cdot\phi}\cr\cr
&&\times
e^{t_1\bar{\varphi}\cdot Z\cdot\varphi
+t_2^2\bar{\varphi}\cdot Z\cdot\varphi \bar{\phi}\cdot Z\cdot\phi
-t_2^2\bar{\varphi}\cdot Z\cdot\phi \bar{\phi}\cdot Z\cdot\varphi}
\prod_{r=1}^{m_{11}}\bar{\varphi}\cdot W_{A_r}\cdot\varphi\cr\cr
&&\times
\prod_{s=1}^{m_{22}}(\bar{\varphi}\cdot Z\cdot \varphi\,\bar{\phi}\cdot W_{B_s}\cdot \phi-\bar{\varphi}\cdot Z\cdot\phi\, \bar{\phi}\cdot W_{B_s}\cdot\varphi)\cr\cr
&&\times
\prod_{t=1}^{m_{12}}(\bar{\varphi}\cdot Z\cdot\varphi\,\bar{\phi}\cdot W_{D_t}\cdot \varphi\,\bar{\varphi}\cdot W_{C_t}\cdot \phi
-\bar{\varphi}\cdot Z\cdot \phi\,\bar{\phi}\cdot W_{D_t}\cdot \varphi\,\bar{\varphi}W_{C_t}\,\varphi)\cr\cr
&&
\label{ZAdSW}
\eea
A graphical representation of this generating function is given in Figure \ref{Fig:Strings}. The dependence on $t_1$ and $t_2$ continues to tell us about the brane bound state with no open strings attached, i.e. the coefficient of the term $t_1^{p_1-p_2}t_2^{2p_2}$ is a two giant bound state for which the first brane has momentum $p_1$ and the second momentum $p_2$. To simplify our discussion, we focus on the su(2) sector, in which case the open string words are constructed using two complex adjoint scalars, $Z$ and $Y$. Further, the last and first letter of the word are both $Y$s. By acting on these operators with the dilatation operator, the dynamics of the open string words is described by a Hamiltonian for a spin chain, with non-trivial boundary conditions. In Section \ref{actionofD} we consider this Hamiltonian. 

Operators dual to a giant graviton system with strings attached were constructed in \cite{deMelloKoch:2007rqf}. For operators constructed using $n$ $Z$ fields and $q$ open string words, the construction of \cite{deMelloKoch:2007rqf} starts with a chain of subgroups $S_{n}\subset S_{n+1}\subset S_{n+2}\subset\cdots\subset S_{n+q}$. Using this chain of subgroups, we can define a restriction of a representation $R$ of $S_{n+q}$ to a representation $r$ of $S_n$, by giving the representation obtained as one successively restricts to each subgroup in the chain. The sequence of representations of the subgroups specify where the open strings are attached, and the representation $r$ defines the giant graviton bound state. All of this information can be summarized by decorating the Young diagram label of the operator. For example, a two AdS giant graviton bound state with four strings attached to the first giant graviton is given by the operator
\bea
\chi_{\tiny\young({\,}{\,}{\,}{\,}{\,}{\,}4321,{\,}{\,}{\,}{\,})}
\label{exmpleOp}
\eea
The chain of representations used to define the operator are obtained by dropping the numbered boxes in order. For the example above, we have the following 5 representations defining the restriction from the original Young diagram $R$ to the Young diagram\footnote{The number of primes appearing on $R''''$ tells us how many open string boxes were dropped from $R$.} $R''''$ defining the giant bound state
\bea
R\equiv{\tiny\young({\,}{\,}{\,}{\,}{\,}{\,}4321,{\,}{\,}{\,}{\,})}
\to
{\tiny\young({\,}{\,}{\,}{\,}{\,}{\,}432,{\,}{\,}{\,}{\,})}
\to
{\tiny\young({\,}{\,}{\,}{\,}{\,}{\,}43,{\,}{\,}{\,}{\,})}
\to
{\tiny\young({\,}{\,}{\,}{\,}{\,}{\,}4,{\,}{\,}{\,}{\,})}
\to
{\tiny\young({\,}{\,}{\,}{\,}{\,}{\,},{\,}{\,}{\,}{\,})}\equiv R''''
\nonumber
\eea
Using this set of restrictions the restricted Schur polynomial of \cite{deMelloKoch:2007rqf} is given by
\bea
\chi_{R,R''''}(Z,W^{(i)})&=&{1\over n!}\sum_{\sigma\in S_{14}}\Tr_{R''''}(\Gamma_R(\sigma)) Z^{i_{14}}_{i_{\sigma(14)}}\cdots Z^{i_{5}}_{i_{\sigma(5)}}(W_4)^{i_4}_{i_{\sigma(4)}}(W_3)^{i_3}_{i_{\sigma(3)}}(W_2)^{i_2}_{i_{\sigma(2)}}(W_1)^{i_1}_{i_{\sigma(1)}}\cr\cr
&&\label{rspexmpl}
\eea
In \cite{deMelloKoch:2007rqf}, $\Tr_{R''''}(\Gamma_R(\sigma))$ is called a restricted character because the trace over $R$ to define the character has been restricted to the subspace $R''''$. We could also attach two strings to each giant graviton brane
\bea
\chi_{\tiny\young({\,}{\,}{\,}{\,}{\,}{\,}31,{\,}{\,}{\,}{\,}42)}
\eea
by making use of the following restrictions
\bea
R\equiv {\tiny\young({\,}{\,}{\,}{\,}{\,}{\,}31,{\,}{\,}{\,}{\,}42)}
\to
{\tiny\young({\,}{\,}{\,}{\,}{\,}{\,}3,{\,}{\,}{\,}{\,}42)}
\to
{\tiny\young({\,}{\,}{\,}{\,}{\,}{\,}3,{\,}{\,}{\,}{\,}4)}
\to
{\tiny\young({\,}{\,}{\,}{\,}{\,}{\,},{\,}{\,}{\,}{\,}4)}
\to
{\tiny\young({\,}{\,}{\,}{\,}{\,}{\,},{\,}{\,}{\,}{\,})}
\equiv R''''
\eea
To stretch the strings between the branes we use a different restriction for the row and column index of the restricted character. The row restriction specifies the open string start points while the column restriction specifies the end points.

What is the relation between the operators of \cite{deMelloKoch:2007rqf} and those constructed above? In Appendix \ref{2AdS} we explicitly evaluate the form of the operators constructed in \cite{deMelloKoch:2007rqf} and compare them to the operators which follow from the generating function (\ref{ZAdSW}). Simple examples immediately confirm that
\begin{itemize}
\item[1.] In the case that $m_{22}=0$, $m_{12}=0=m_{21}$ with $m_{11}$ arbitrary, the generating function (\ref{ZAdSW}) precisely reproduces the restricted Schur polynomials constructed in \cite{deMelloKoch:2007rqf}.
\item[2.] In the case that $m_{11}=0$, $m_{12}=0=m_{21}$ with $m_{22}$ arbitrary, the generating function (\ref{ZAdSW}) again precisely reproduces the restricted Schur polynomials constructed in \cite{deMelloKoch:2007rqf}.
\end{itemize}
This agreement is not an accident and we can prove that, for the cases described in 1. and 2. above, the two constructions agree. A central result of \cite{deMelloKoch:2007rqf}, is a rule for the derivative of a restricted Schur polynomial with respect to an open string. The idea behind the proof is to show that the operators constructed from the generating function (\ref{ZAdSW}) obey this derivative rule. 

To state the derivative rule of \cite{deMelloKoch:2007rqf} we need to review a few facts. When strings are added to the giant system, the order in which strings are added matters\footnote{For the specific case that we attach all the open strings to a single giant, as in (\ref{exmpleOp}) it turns out that attaching the strings in different orders gives the same operator. In this case and only this case the order in which strings are attached does not matter.}. Bearing in mind how the operator was constructed, this is not surprising: changing the order in which open strings are attached changes the set of restrictions used to define the restricted character and the restricted character depends sensitively on which subspace is traced. If we differentiate with respect to the last open string added, the open string together with its box is removed from the restricted Schur polynomial and the remaining polynomial is multiplied by the factor $c$ of the box removed. Recall that the factor of a box in row $i$ and column $j$ is $c=N-i+j$. 

The proof of this result, given in \cite{deMelloKoch:2004crq,deMelloKoch:2007rqf}, uses results from the representation theory of symmetric groups. The derivative with respect to the open string word restricts the sum over $S_{n+q}$ to a sum over $S_{n+q-1}$ and its cosets. A simple rewriting of the sum produces the Jucys–Murphy elements in the group algebra $\mathbb {C} [S_{n+q}]$ of the symmetric group $S_{n+q}$. They generate a commutative subalgebra of $\mathbb {C} [S_{n}]$ and they commute with all elements of $\mathbb{C} [S_{n+q-1}]$ with $S_{n+q-1}$ the subgroup we are summing over. The factor that appears multiplying the polynomial reflects the eigenvalues of the Jucys-Murphy elements on the subspace used in the construction of the restricted Schur polynomial. The derivation of this formula is an exercise in the representation theory of finite groups \cite{deMelloKoch:2007rqf}. 

The operators constructed using the generating function obey exactly the same derivative rule. The proof uses only elementary manipulations. Consider a two giant graviton bound state, with $q$ open strings attached to the first row. An example of the operator we consider is given in (\ref{exmpleOp}), which has $q=4$. The box labelled $i$ stands for the open string $W_{A_i}$ for $i=1,2,3,4$. The rule that we are trying to reproduce says
\bea
{\partial\over\partial (W_{A_1})^a_a}\chi_{\tiny\young({\,}{\,}{\,}{\,}{\,}{\,}{\,}{\,}{\,}4321,{\,}{\,}{\,}{\,}{\,}{\,})}=(N+p_1+q-1)\chi_{\tiny\young({\,}{\,}{\,}{\,}{\,}{\,}{\,}{\,}{\,}432,{\,}{\,}{\,}{\,}{\,}{\,})}\label{fstderrule}
\eea
To derive this rule for the operators defined by the generating function, consider ($S_0\equiv-\bar{\varphi}^i\varphi_i-\bar{\phi}^i\phi_i$ and $S_1\equiv t_1\bar{\varphi}^iZ_i^j\varphi_j+t_2^2\bar{\varphi}^i\bar{\phi}^j (Z^k_i Z^l_j-Z^l_i Z^k_j)\varphi_k\phi_l$)
\bea
&&{\partial\over\partial (W_{A_1})^a_a}\int d^N\bar{\varphi}\int d^N\varphi \int d^N\bar{\phi} \int d^N\phi \,\,{e^{S_0}\over \pi^{2N}}\,\,e^{S_1}\,\,\prod_{r=1}^q\bar{\varphi}\cdot W_{A_r}\cdot\varphi\cr\cr\cr
&=&\int d^N\bar{\varphi}\int d^N\varphi\int d^N\bar{\phi} \int d^N\phi \,\,{e^{S_0}\over \pi^{2N}}\,\, e^{S_1}\,\,\prod_{r=2}^q\bar{\varphi}\cdot W_{A_r}\cdot\varphi\,\,\bar{\varphi}\cdot\varphi\cr\cr\cr
&=&-\int d^N\bar{\varphi}\int d^N\varphi \int d^N\bar{\phi} \int d^N\phi \,\,\,\varphi_a {\partial\over\partial\varphi_a}\left({e^{S_0}\over \pi^{2N}}\right)
e^{S_1}\,\,\prod_{r=2}^q\bar{\varphi}\cdot W_{A_r}\cdot\varphi\cr\cr\cr
&=&\int d^N\bar{\varphi}\int d^N\varphi \int d^N\bar{\phi} \int d^N\phi 
\left({e^{S_0}\over \pi^{2N}}\right)(N+\varphi_a {\partial\over\partial\varphi_a})\Big[e^{S_1}\,\,\prod_{r=2}^q\bar{\varphi}\cdot W_{A_r}\cdot \varphi\Big]\nonumber
\eea
The operator $\varphi_a {\partial\over\partial\varphi_a}$ counts how many boxes there are in the first row, after the box for $W_1$ is removed. This follows because boxes in the first row represent fields with their upper index contracted with a $\varphi$ field. Thus, we obtain
\bea
&&{\partial\over\partial (W_{A_1})^a_a}\int d^N\bar{\varphi}\int d^N\varphi \int d^N\bar{\phi} \int d^N\phi \,\,{e^{S_0}\over \pi^{2N}}\,\,e^{S_1}\,\,\prod_{r=1}^q\bar{\varphi}\cdot W_{A_r}\cdot \varphi\cr\cr\cr
&=&(N+p_1+q-1)\int d^N\bar{\varphi}\int d^N\varphi \int d^N\bar{\phi} \int d^N\phi 
\,\,{e^{S_0}\over \pi^{2N}}\,\, e^{S_1}\,\,\prod_{r=2}^q\bar{\varphi}\cdot W_{A_r}\cdot\varphi
\eea
which is indeed the correct rule. Using derivatives which remove all open strings we obtain a Schur polynomial. Given the discussion of Section \ref{TwoRowSchur}, we know that the generating function reproduces this operator correctly. Alternatively, since there is no term in the operators considered that is independent of the open string word $W_{A_1}$, the equality of these derivatives implies the equality of the operators.

We can extend this result to operators corresponding to a bound state of two giant gravitons with all strings attached to the second brane. In this case the derivative rule is
\bea
{\partial\over\partial (W_{B_1})^a_a}\chi_{\tiny\young({\,}{\,}{\,}{\,}{\,}{\,}{\,}{\,}{\,}{\,}{\,}{\,}{\,},{\,}{\,}{\,}{\,}{\,}{\,}4321)}=(N+p_2+q-2)\chi_{\tiny\young({\,}{\,}{\,}{\,}{\,}{\,}{\,}{\,}{\,}{\,}{\,}{\,}{\,},{\,}{\,}{\,}{\,}{\,}{\,}432)}
\eea
Note that the coefficient on the right hand side is the coefficient of (\ref{fstderrule}) with $p_1$ replaced by $p_2$ minus one. The extra minus one is because the open string now occupies the second row. To derive this rule for the operators derived from the generating function, consider
\bea
&&{\partial\over\partial (W_{B_1})^a_a}\int d^N\bar{\varphi}\int d^N\varphi \int d^N\bar{\phi} \int d^N\phi\,\,\,{e^{S_0}\over \pi^{2N}}e^{S_1}\cr\cr
&&\times
\prod_{s=1}^q(\bar{\varphi}\cdot Z\cdot \varphi\,\bar{\phi}\cdot W_{B_s}\cdot \phi-\bar{\varphi}\cdot Z\cdot\phi\, \bar{\phi}\cdot W_{B_s}\cdot\varphi)\cr\cr
&=&\int d^N\bar{\varphi}\int d^N\varphi \int d^N\bar{\phi} \int d^N\phi\,\,\, {e^{S_0}\over \pi^{2N}}e^{S_1}(\bar{\varphi}\cdot Z\cdot \varphi\,\bar{\phi}\cdot \phi-\bar{\varphi}\cdot Z\cdot\phi\, \bar{\phi}\cdot \varphi)\cr\cr
&&\times
\prod_{s=2}^q(\bar{\varphi}\cdot Z\cdot \varphi\,\bar{\phi}\cdot W_{B_s}\cdot \phi-\bar{\varphi}\cdot Z\cdot\phi\, \bar{\phi}\cdot W_{B_s}\cdot\varphi)\cr\cr
&=&(N+p_2+q-1)\int d^N\bar{\varphi}\int d^N\varphi \int d^N\bar{\phi} \int d^N\phi \,\,\,{e^{S_0}\over \pi^{2N}}e^{S_1}\cr\cr
&&\times\,\, \bar{\varphi}\cdot Z\cdot \varphi\,\,\prod_{s=2}^q(\bar{\varphi}\cdot Z\cdot \varphi\,\bar{\phi}\cdot W_{B_s}\cdot \phi-\bar{\varphi}\cdot Z\cdot\phi\, \bar{\phi}\cdot W_{B_s}\cdot\varphi)\cr\cr
&&-\int d^N\bar{\varphi}\int d^N\varphi \int d^N\bar{\phi} \int d^N\phi\,\,\, 
\left({e^{S_0}\over \pi^{2N}}\right)\cr\cr
&&\qquad \varphi_a{\partial\over\partial\phi_a}\Big[\bar{\varphi}\cdot Z\cdot\phi\,e^{S_1}\prod_{s=2}^q(\bar{\varphi}\cdot Z\cdot \varphi\,\bar{\phi}\cdot W_{B_s}\cdot \phi-\bar{\varphi}\cdot Z\cdot\phi\, \bar{\phi}\cdot W_{B_s}\cdot\varphi)\Big]
\eea
Using the obvious facts
\bea
\varphi_a{\partial\over\partial\phi_a}\Big[\bar{\varphi}\cdot Z\cdot \varphi\,\bar{\phi}\cdot W_{B_s}\cdot \phi-\bar{\varphi}\cdot Z\cdot\phi\, \bar{\phi}\cdot W_{B_s}\cdot\varphi\Big]=0
\eea
\bea
\varphi_a{\partial\over\partial\phi_a}\Big[t_2^2(\bar{\varphi}\cdot Z\cdot \varphi\,\bar{\phi}\cdot Z\cdot \phi-\bar{\varphi}\cdot Z\cdot\phi\, \bar{\phi}\cdot Z\cdot\varphi)\Big]=0
\eea
we find
\bea
&&{\partial\over\partial (W_{B_1})^a_a}\int d^N\bar{\varphi}\int d^N\varphi \int d^N\bar{\phi} \int d^N\phi \,\, {e^{S_0}\over \pi^{2N}}e^{S_1}
\prod_{s=1}^q(\bar{\varphi}\cdot Z\cdot \varphi\,\bar{\phi}\cdot W_{B_s}\cdot \phi-\bar{\varphi}\cdot Z\cdot\phi\, \bar{\phi}\cdot W_{B_s}\cdot\varphi)\cr\cr
&=&(N+p_2+q-2)\int d^N\bar{\varphi} \int d^N\varphi \int d^N\bar{\phi} \int d^N\phi \,\, {e^{S_0}\over \pi^{2N}}e^{S_1}\cr\cr
&&\times\,\, \bar{\varphi}\cdot Z\cdot \varphi\,\,\prod_{s=2}^q(\bar{\varphi}\cdot Z\cdot \varphi\,\bar{\phi}\cdot W_{B_s}\cdot \phi-\bar{\varphi}\cdot Z\cdot\phi\, \bar{\phi}\cdot W_{B_s}\cdot\varphi)
\eea
The extra factor of $\bar{\varphi}\cdot Z\cdot \varphi$ is simply adding one box to the first row of the Young diagram labelling the operator. The open string $W_{B_1}$ was located in a column of length 2, with the other box in the column occupied by a $Z$ field. After we differentiate with respect to $W_{B_1}$ the second box in the column is left over and this is precisely the extra factor $\bar{\varphi}\cdot Z\cdot \varphi$. This reproduces the correct derivative rule in complete detail. As above, since there is no term in the operators considered that is independent of the open string word $W_{B_1}$, the equality of these derivatives implies the equality of the operators. Thus, as long as we attach strings to a specific giant, the generating function we have written down is exactly reproducing the restricted Schur polynomials constructed in \cite{deMelloKoch:2007rqf}. 

This agreement between the operators obtained from the generating function (\ref{ZAdSW}) and those constructed in \cite{deMelloKoch:2007rqf} does not hold in general. Indeed, it is simple to verify that
\begin{itemize}
\item[3.] In the case that $m_{12}=0=m_{21}$ with both $m_{11}\ne 0$ and $m_{22}\ne 0$ but otherwise arbitrary, the operators defined by the generating function (\ref{ZAdSW}) do not agree with the restricted Schur polynomials constructed in \cite{deMelloKoch:2007rqf}.
\end{itemize}
As a simple example, the operator with $p_1=1$ and $p_2=0$, with a single  string attached to each row, is given by
\bea
&&\int d^N\bar{\varphi}\int d^N\varphi \int d^N\bar{\phi} \int d^N\phi\,\,{e^{S_0}\over \pi^{2N}}\,\,\bar{\varphi}\cdot W_A\cdot\varphi\,\,(\bar{\varphi}\cdot Z\cdot \varphi\,\bar{\phi}\cdot W_B\cdot \phi-\bar{\varphi}\cdot Z\cdot\phi\, \bar{\phi}\cdot W_B\cdot\varphi)\cr\cr\cr
&=&\Tr(W_A)\Tr(W_B)\Tr(Z)+\Tr(ZW_A)\Tr(W_B)-\Tr(W_A)\Tr(W_B Z)-\Tr(W_AZW_B)
\eea
while the construction of  \cite{deMelloKoch:2007rqf} gives
\bea
\chi_{\tiny\young({\,}A,B)}&=&\Tr(W_A)\Tr(W_B)\Tr(Z)+{1\over 2}\Tr(W_AW_B)\Tr(Z)-\Tr(W_A)\Tr(W_B Z)\cr\cr
&+&{1\over 2}\Tr(ZW_A)\Tr(W_B)-{1\over 2}\Tr(W_AZW_B)-{1\over 2}\Tr(W_AW_BZ)
\eea
This proves that the restricted Schur construction of \cite{deMelloKoch:2007rqf} and the construction using the generating function (\ref{ZAdSW}) are not the same. There is a simple and general way to see that these two sets of operators are different. The operators produced using the generating function (\ref{ZAdSW}) treats all open strings with the same orientation identically. Thus, we can permute any of the strings of the same letter\footnote{By this we mean that we can permute the $W_{A_i}$'s among themselves, the $W_{B_i}$'s among themselves and so on.} and this is a symmetry of the operator. This symmetry is not present in the restricted Schur polynomial construction of \cite{deMelloKoch:2007rqf}.

Describing open string excitations of the giant graviton system in terms of open strings has the advantage that we can make contact with a worldsheet description of the open string \cite{Berenstein:2005fa}. Another possibility is that we excite of the half BPS giant graviton bound state by including a second field $Y$ in the description of the operator\footnote{We can include the complete set of fields in ${\cal N}=4$ super Yang Mills theory. This restriction to a single field $Y$ is for simplicity only.}. This description is naturally related to the restricted Schur polynomial construction developed in \cite{Bhattacharyya:2008rb,Bhattacharyya:2008xy}. Consider operators constructed using $n$ $Z$ fields and $m$ $Y$ fields. The restricted Schur polynomial $\chi_{R,(r,s)\alpha\beta}(Z,Y)$ construction of \cite{Bhattacharyya:2008rb,Bhattacharyya:2008xy} uses a representation $R\vdash m+n$ to organize the complete set of fields, as well as a representation $r\vdash n$ for the $Z$ fields and a representation $s\vdash m$ for the $Y$ fields. In addition, there are multiplicity labels $\alpha,\beta$ needed\footnote{To define the relevant restricted character, we need one multiplicity label for the row index and one for the column index, that we restrict to.} because in general there is more than one way in which the $S_n\times S_m$ representation $(r,s)$ can be obtained  after restricting the $S_{n+m}$ representation $R$ to the $S_n\times S_m$ subgroup. This description of the local operators does not immediately have a visible open string interpretation. By diagonalizing the one loop dilatation operator, the studies \cite{Carlson:2011hy,deMelloKoch:2011wah,deMelloKoch:2011ci,deMelloKoch:2012ck} discovered an alternative basis, known as the Gauss graph basis. The operators of this basis, called Gauss graph operators, are eigenstates of the one loop dilatation operator. These eigenstates are labelled by graphs, which provide a diagrammatic picture of the open strings stretching between the different giant graviton branes. The complete basis of operators manifestly obey the Gauss Law of the brane world volume theory, motivating the terminology ``Gauss graph basis''. We will now argue that it is the Gauss graph operators that correspond to the operators constructed from our generating function. Towards this end, consider the generating function 
\bea
Z_{\rm AdS}(t_1,t_2,Y) &=&{1\over \pi^{2N}}\int d^N\bar{\varphi}\int d^N\varphi 
\int d^N\bar{\phi} \int d^N\phi \,\, e^{-\bar{\varphi}\cdot\varphi-\bar{\phi}\cdot\phi}\cr\cr
&&\times
e^{t_1\bar{\varphi}\cdot Z\cdot\varphi
+t_2^2\bar{\varphi}\cdot Z\cdot\varphi \bar{\phi}\cdot Z\cdot\phi
-t_2^2\bar{\varphi}\cdot Z\cdot\phi \bar{\phi}\cdot Z\cdot\varphi}
(\bar{\varphi}\cdot Y\cdot\varphi)^{m_{11}}\cr\cr
&&\times(\bar{\varphi}\cdot Z\cdot \varphi\,\bar{\phi}\cdot Y\cdot \phi-\bar{\varphi}\cdot Z\cdot\phi\, \bar{\phi}\cdot Y\cdot\varphi)^{m_{22}}\cr\cr
&&\times(\bar{\varphi}\cdot Z\cdot\varphi\,\bar{\phi}\cdot Y\cdot \varphi\,\bar{\varphi}\cdot Y\cdot \phi-\bar{\varphi}\cdot Z\cdot \phi\,\bar{\phi}\cdot Y \cdot \varphi\,\bar{\varphi}\cdot Y\cdot\varphi)^{m_{12}}\cr\cr
&&\label{AdSGGgen}
\eea
which is obtained by replacing $W_*\to Y$ in (\ref{ZAdSW}). The bound states described by this generating function have $m_{11}$ strings attached to the brane corresponding to the first row, $m_{22}$ strings attached to the brane corresponding to the second row, $m_{12}$ strings stretching from the first to the second branes and $m_{21}=m_{12}$ strings stretched from the second to the first. The Gauss Law is clearly manifest in this description since the above expression is only defined for $m_{12}=m_{21}$. In addition, the construction is symmetric under swaps of any strings that have the same orientation. This is a symmetry that the Gauss graph operators enjoy and this agreement of the symmetry enjoyed by the two constructions is a key piece of evidence suggesting that the operators constructed from the generating function (\ref{AdSGGgen}) are the Gauss graph operators. The Gauss graph operators with all strings attached to a single brane agree with the restricted Schur polynomials \cite{deMelloKoch:2011wah,deMelloKoch:2012ck} with all open strings attached in a given row. Consequently, the results we have obtained above already establish that restricted Schur polynomials and Gauss graph operators coincide when the open strings are all attached to a single brane, i.e. when there are no strings stretching between branes.

For a detailed comparison of operators that involve stretched strings, study the simpler expression obtained for $m_{11}=0=m_{22}$ and $m_{12}=1=m_{21}$
\bea
G_{\rm AdS}(t_1,t_2) &=& \int d^N\bar{\varphi}\int d^N\varphi \int d^N\bar{\phi} \int d^N\phi \,\,{e^{S_0}\over \pi^{2N}}\,\, e^{S_1}\cr\cr
&&\times\,\, (\bar{\varphi}\cdot Z\cdot\varphi\,\bar{\phi}\cdot Y\cdot \varphi\,\bar{\varphi}\cdot Y\cdot \phi-\bar{\varphi}\cdot Z\cdot \phi\,\bar{\phi}\cdot Y\cdot \varphi\,\bar{\varphi}\cdot Y\cdot\varphi)
\eea
The dependence on $t_1$ and $t_2$ continues to keep track of the momentum carried by each brane, exactly as above. A simple computation proves that
\bea
G_{\rm AdS}(t_1,t_2)&=& \Tr (Z)\Tr(Y^2)-\Tr(ZY)\Tr(Y)+
t_1\left(\Tr(Z)^2\Tr(Y^2)+\Tr(Z^2)\Tr(Y^2)+\Tr(Z)\Tr(ZY^2)\right.\cr\cr
&&\left.+\Tr(Z^2Y^2)-\Tr(ZY)\Tr(Z)\Tr(Y)-\Tr(Z^2Y)\Tr(Y)-\Tr(ZY)^2-\Tr(ZYZY)\right)\cr\cr
&+&t_2\Big(\Tr(Z)^3 \Tr(Y^2)+2\Tr(Z^2)\Tr(Y^2Z)-\Tr(Z)^2\Tr(Y)\Tr(YZ)+2\Tr(Z)^2\Tr(Y^2Z)\cr\cr
&-&\Tr(YZ)\Tr(YZ^2)-\Tr(Y^2)\Tr(Z^3)-2\Tr(Z)\Tr(YZ)^2-2\Tr(Z)\Tr(Y)\Tr(YZ^2)\cr\cr
&-&2\Tr(Z) \Tr(Y^2Z^2)+\Tr(Z)\Tr(YZYZ)+2\Tr(Y)\Tr(YZ)\Tr(Z^2)+\Tr(Y)\Tr(YZ^3)\cr\cr
&-&2\Tr(Y^2Z^3)+2\Tr(YZYZ^2)\Big)+\cdots
\label{SSanswers}
\eea
In the Appendices we prove that the Gauss graph operators are given by
\bea
\left({\tiny\yng(1)}\,\,,
\begin{gathered}\includegraphics[scale=0.4]{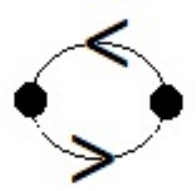}\end{gathered}\right)
&=&\Tr (Z)\Tr(Y^2)-\Tr(ZY)\Tr(Y)\cr\cr
\left({\tiny \yng(2)}\,\,,
\begin{gathered}\includegraphics[scale=0.4]{GG3}\end{gathered}\right)
&=&\Tr(Z)^2\Tr(Y^2)+\Tr(Z^2)\Tr(Y^2)+\Tr(Z)\Tr(ZY^2)+\Tr(Z^2Y^2)\cr\cr
&&-\Tr(ZY)\Tr(Z)\Tr(Y)-\Tr(Z^2Y)\Tr(Y)-\Tr(ZY)^2-\Tr(ZYZY)\cr\cr
\left({\tiny \yng(2,1)}\,\,,
\begin{gathered}\includegraphics[scale=0.4]{GG3}\end{gathered}\right)
&=&\Tr(Z)^3 \Tr(Y^2)+2\Tr(Z^2)\Tr(Y^2Z)-\Tr(Z)^2\Tr(Y)\Tr(YZ)+2\Tr(Z)^2\Tr(Y^2Z)\cr\cr
&-&\Tr(YZ)\Tr(YZ^2)-\Tr(Y^2)\Tr(Z^3)-2\Tr(Z)\Tr(YZ)^2-2\Tr(Z)\Tr(Y)\Tr(YZ^2)\cr\cr
&-&2\Tr(Z) \Tr(Y^2Z^2)+\Tr(Z)\Tr(YZYZ)+2\Tr(Y)\Tr(YZ)\Tr(Z^2)+\Tr(Y)\Tr(YZ^3)\cr\cr
&-&2\Tr(Y^2Z^3)+2\Tr(YZYZ^2)
\eea
in complete agreement with (\ref{SSanswers}).

With this encouraging evidence in hand, we would now like to prove that the operators generated by (\ref{AdSGGgen}) are indeed identical to the Gauss graph operators. 
First in Appendix \ref{ggequality} we prove that, at large $N$, operators derived from the generating function for arbitrary $m_{11}$ and $m_{22}$ agree with the Gauss graph operators. Thus, all that is left is to consider operators with non-zero $m_{12}$. The proof introduces a graph changing operator: acting on a given Gauss graph, we can easily generate any other Gauss graph operator. Then, given the equality between our operators and the Gauss graph operators when all strings are attached to a definite brane (i.e. there are no stretched string states), and the fact that we can generate any boundstate with any number of strings stretched between the giants with the graph changing operator, the proof is complete.

\begin{figure}[h]
\begin{center}
\resizebox{!}{6cm}{\includegraphics{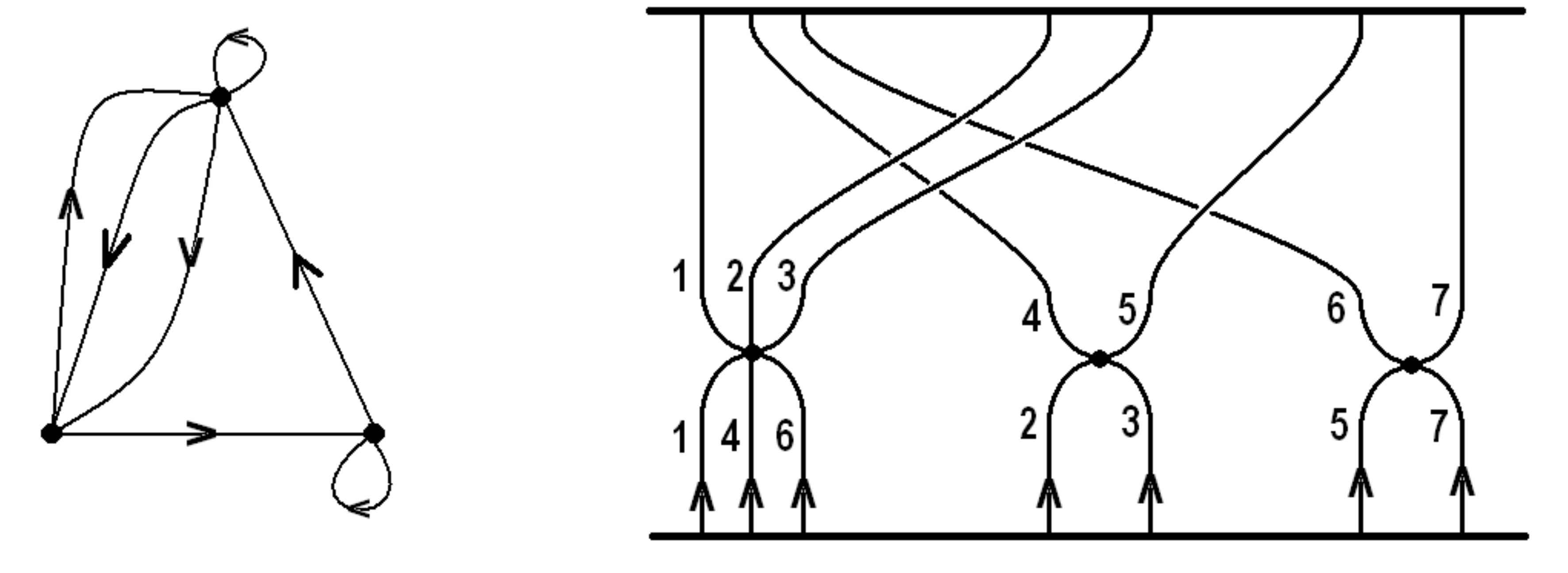}}
\caption{Any open string configuration can be mapped to a labelled graph as shown. The two bold horizontal lines in the Figure on the right are identified. Rows of $R$ form nodes in the graph. For the graph shown, $R$ has three rows. There are 7 open strings, so the operator is constructed using 7 $Y$ fields. The graph determines a permutation, so each open string configuration maps to a permutation. For the graph shown the permutation, in cycle notation, is $\sigma= (24)(536)$. A nice example is the configuration in which all open strings loop back to the brane they start from, described by the identity permutation.}
 \label{fig:gaussgraph}
\end{center}
\end{figure}

For the proof that follows, it is necessary to briefly review the construction of the Gauss graph operators. The restricted Schur polynomial $\chi_{R,(r,s)\alpha\beta}(Z,Y)$ constructed using $n$ $Z$ fields and $m$ $Y$ fields is labelled by three Young diagrams $R\vdash m+n$, $r\vdash n$ and $s\vdash m$ and two multiplicity labels $\alpha,\beta$. The Gauss graph operators $O_{R,r}(\sigma)$ have the same $R$ and $r$ labels, but the $s$ label and the multiplicity labels are replaced by a permutation $\sigma$. If we think about each individual $Y$ field as an open string, to specify a state of the $Y$ fields we need to specify a configuration of the open strings. This can be achieved with a permutation, as illustrated in Figure \ref{fig:gaussgraph}. This is the permutation $\sigma$ labelling the Gauss graph operator. From the Figure it is clear that swapping endpoints of open strings terminating on the brane and swapping endpoints of open strings originating from the brane is a symmetry of the open string configuration. To correctly enumerate distinct physical states we must divide out this symmetry so that each configuration corresponds to an element of a double coset. For the configuration shown in Figure \ref{fig:gaussgraph} the relevant double coset is $H \setminus S_7/H$ where $H$ is the group of symmetries for the end points, given by $H=S_3\times S_2\times S_2$.

The transformation to the Gauss graph basis is given by
\bea
O_{R,r}(\sigma)=\sum_{s\vdash m}\sum_{\alpha,\beta}C_{\alpha\beta}^s (\sigma)\chi_{R,(r,s)\alpha\beta}\label{ggo}
\eea
where
\bea 
C_{\alpha\beta}^s (\tau)=|H|\sqrt {d_s\over m!}\,\,\sum_{k,l=1}^{d_s}\,\,\Gamma^{(s)}_{kl}(\tau)\,\, B^{s\rightarrow 1_H} _{k\alpha} B^{s\rightarrow 1_H} _{l\beta} 
\eea 
This last formula makes use of branching coefficients, which are defined as follows
\bea
{1\over |H|}\sum_{ \gamma \in H } \Gamma^{(s)}_{ik} (\gamma)
=\sum_\mu B^{s \rightarrow 1_H}_{ i \mu}  B^{s \rightarrow 1_H}_{ k \mu}
\eea
where the index $\mu$ on the right hand side is a multiplicity label.

As we have just reviewed, the Gauss graph operator is labelled by a permutation $\sigma$ that specifies how the open strings are stretched between the giant gravitons. We can easily rewrite the formula (\ref{AdSGGgen}) in terms of $\sigma$ as follows ($m=m_{11}+m_{12}+m_{21}+m_{22}$)
\bea
Z_{\rm AdS}(t_1,t_2,Y,\sigma) &=&{1\over \pi^{2N}}\int d^N\bar{\varphi}\int d^N\varphi 
\int d^N\bar{\phi} \int d^N\phi \,\, 
e^{-\bar{\varphi}\cdot\varphi-\bar{\phi}\cdot\phi}\cr\cr
&&\times
e^{t_1\bar{\varphi}\cdot Z\cdot\varphi+t_2^2\bar{\varphi}^i\bar{\phi}^j (Z^k_i Z^l_j-Z^l_i Z^k_j)\varphi_k\phi_l}\,\,\prod_{i=1}^{m_{11}+m_{12}}
\bar{\varphi}^{b_i}Y_{b_i}^{a_i}\varphi_{a_{\sigma(i)}}\cr\cr
&&\prod_{j=m_{11}+m_{12}+1}^{m} 
(\bar{\varphi}\cdot Z\cdot \varphi\,\bar{\phi}^{b_j} Y^{a_j}_{b_j}\phi_{a_{\sigma(j)}}-\bar{\varphi}\cdot Z\cdot\phi\, \bar{\phi}^{b_j} Y^{a_j}_{b_j}\varphi_{a_{\sigma(j)}})\nonumber
\eea
It is clear that $\sigma$ specifies the row to which the upper indices of the $Y$ fields belong i.e. $\sigma$ specifies how the $Y$ excitations connect to the giants. The graph changing operator, acting on a system with a total of $m$ $Y$ excitations, has the form
\bea
\hat{G}(\sigma)=Y^{i_1}_{j_{\sigma(1)}}\cdots Y^{i_m}_{j_{\sigma(m)}}
{\partial\over\partial Y^{i_1}_{j_1}}\cdots {\partial\over\partial Y^{i_m}_{j_m}}
\eea
It is simple to verify that
\bea
\hat{G}(\sigma_1)Z_{\rm AdS}(t_1,t_2,Y,\sigma_2)=Z_{\rm AdS}(t_1,t_2,Y,\sigma_1\sigma_2)
\eea

We will now argue that this graph changing operator has exactly the same action on the Gauss graph operators. When $R$ has only two rows we have no multiplicity labels so that, up to a normalization factor of ${|H|\over m!}$ which we drop, we have\footnote{For a detailed discussion and derivation of this formula, see \cite{deMelloKoch:2012ck}.}
\bea
  O(\sigma)&=&\sum_{s\vdash m}\sqrt{d_s}\Gamma^{(s)}_{jk}(\sigma )B^{s \rightarrow 1_H }_{j}B^{s \rightarrow 1_H }_{k} O_{R,(r,s)}\cr
           &=&{1\over\prod_i m_i!}\sum_{s\vdash m}
            \sum_{\alpha\in \prod_i S_{m_i}}\sqrt{d_s}
            \Gamma^{(s)}_{jk}(\sigma )\Gamma^{(s)}_{kj}(\alpha) O_{R,(r,s)}\cr
  &=&{1\over\prod_i m_i!}\sum_{s\vdash m}\sum_{\alpha\in \prod_i S_{m_i}}\sqrt{d_s}\chi_{s}(\sigma \alpha) O_{R,(r,s)}
\label{GGmanip}
\eea
where $\chi_{s}(\sigma )$ is the character of $\sigma\in S_m$ in irrep $s$. Now, again up to a normalization that plays no role in this proof, we have
\bea
O_{R,(r,s)}=\sum_{\rho\in S_{n+m}}\sum_{\sigma_1\in S_n}\sum_{\sigma_2\in S_m}
{1\over\sqrt{d_s}}\chi_r(\sigma_1)\chi_s(\sigma_2)\chi_R(\sigma_1\circ\sigma_2\,\rho)\Tr(\rho Z^{\otimes n} Y^{\otimes m})
\eea
In writing this formula we have explicitly used the crucial fact that restricted Schur polynomials with only two rows carry no multiplicity labels\footnote{In the case that there are no multiplicities we can write the projection operator, which projects to the $(r,s)$ irreducible representation of $S_n\times S_m$, within the carrier space of $R$ as
\bea P_{R\to (r,s)}={d_r d_s\over n!m!}\sum_{\sigma_1\in S_n}\sum_{\sigma_2\in S_m}\chi_r(\sigma_1)\chi_s(\sigma_2)\Gamma_R(\sigma_1\circ\sigma_2)\eea}. Using this formula, as well as the character orthogonality identity, which says
\bea
\sum_{s\vdash m}\chi_s(g)\chi_s(h)=|[g]|\delta_{[g][h]}
\eea
we have
\bea
O(\sigma)=\sum_{\rho\in S_{n+m}}\sum_{\sigma_1\in S_n}\sum_{\alpha\in S_{m_1}\times S_{m_2}}\chi_r(\sigma_1)\chi_R(\sigma_1\circ\sigma\alpha\,\rho)\Tr(\rho Z^{\otimes n} Y^{\otimes m})\label{niceggformula}
\eea
Using this last formula it is simple to verify that
\bea
\hat{G}(\sigma_1)O(\sigma_2)=O(\sigma_1\sigma_2)
\eea
In Appendix \ref{ggequality} we argue that
\bea
\sum_{p_1,p_2} t_1^{p_1-p_2}t_2^{p_2}O_{(p_1,p_2)}({\bf 1})=Z_{\rm AdS}(t_1,t_2,Y,{\bf 1})
\eea
and since the graph changing operator acts in the same way on both sides of this equation, the proof is complete.

\subsection{Bound states of two Sphere Giants}

A bound state of two sphere giants, one of momentum $p_1$ and a second of momentum $p_2$ with $p_2\le p_1$ corresponds to a Schur polynomial labeled by a Young diagram with two columns. There are $p_1$ boxes in the first column and $p_2$ boxes in the second. To describe giant gravitons, both $p_1$ and $p_2$ must be order $N$. The Young symmetrizer must antisymmetrize over all indices in column 1, and all indices in column 2, as well as symmetrize the pair of indices belonging to each row. In contrast to the AdS giants, it is now the antisymmetrization to be performed that represents the bulk of the work. This antisymmetrization of indices can again be accomplished with the help of a Gaussian integral, but now we integrate over Grassmann variables. Again it is a simple application of Wick's theorem to learn that
\bea
\int d^N\bar{\xi} \int d^N\xi 
\,\, e^{-\bar{\xi}^a\xi_a} \,\,
\bar{\xi}^{i_1}\xi_{j_1}\bar{\xi}^{i_2}\xi_{j_2}
\cdots \bar{\xi}^{i_k}\xi_{j_k}=\sum_{\sigma\in S_k} {\rm sgn}(\sigma)
\delta^{i_1}_{j_{\sigma(1)}}\delta^{i_2}_{j_{\sigma(2)}}\cdots 
\delta^{i_k}_{j_{\sigma(k)}}
\label{fermionicresult}
\eea
\bea
d^N\xi \equiv  d\xi_1 d\xi_2 \cdots d\xi_N \qquad\qquad
d^N\bar{\xi} \equiv d\bar{\xi}^N\cdots d\bar{\xi}^2 d\bar{\xi}^1
\eea
To write down the generating function for Young diagrams $R=(2^{p_2},1^{p_1-p_2})$ that have two columns, one of length $p_1$ and one of length $p_2$, we perform the symmetrization in each row by hand and use $(\psi_i,\bar{\psi}^i)$ to antisymmetrize indices in the first column and $(\xi_i,\bar{\xi}^i)$ to perform the antisymmetrization of indices in the second column. The two giant graviton boundstate generating function can then be written as follows
\bea
Z_{\rm S}(t_1,t_2)&=&\sum_{p_1\ge p_2}^\infty\sum_{p_2=1}^\infty t_2^{p_2}t_1^{p_1-p_2}\,\,
{(p_1+p_2)!\over d_{(2^{p_2}1^{p_1-p_2})}p_2!(p_1-p_2)!}\,\,
\chi_{(2^{p_2},1^{p_1-p_2})}(Z)\cr\cr
&=& \int d^N\bar{\psi}\int d^N\psi \int d^N\bar{\xi} \int d^N\xi\cr\cr
&&\qquad\qquad 
e^{-\bar{\psi}^i\psi_i-\bar{\xi}^i\xi_i-t_1\bar{\psi}^iZ_i^j\psi_j
+t_2^2\bar{\psi}^i\bar{\xi}^j (Z^k_i Z^l_j+Z^l_i Z^k_j)\psi_k\xi_l}
\label{SZ}
\eea
In Appendix \ref{2AdS} we check this result by performing the integral to second order in $t_1$ and $t_2$ and show that we reproduce the correct formulas for the relevant Schur polynomials. 

As was the case as for the AdS giants, this generating function can also be interpreted as a generating function for characters of the symmetric group $S_{p_1+p_2}$ in the representation labelled by Young diagram $R=(2^{p_2},1^{p_1-p_2})$. The generating function is
\bea
Z_{\rm S}(t_1,t_2)&=&
\sum_{p_1=p_2}^\infty \sum_{p_2=1}^\infty\sum_{\{n_i\}}\,\,
{t_2^{2p_2}t_1^{p_1-p_2}\over 1^{n_1}n_1!2^{n_2}n_2!\cdots (p_1+p_2)^{n_{p_1+p_2}}n_{(p_1+p_2)!}}
{(p_1+1)!\over (p_1-p_2+1)(p_1-p_2)!}\cr\cr
&&\qquad\qquad\qquad \times\chi_{(2^{p_2},1^{p_1-p_2})}(\{i^{n_i}\}) {\rm Tr}(Z)^{n_1}{\rm Tr}(Z^2)^{n_2}\cdots {\rm Tr}(Z^{p_1+p_2})^{n_{p_1+p_2}}\cr\cr
\label{charlongcolumns}&&
\eea
where $\sum_{\{n_i\}}$ denotes a sum over the conjugacy classes of $S_{p_1+p_2}$.
To build some intuition for this formula, expand to linear order in $t_1$ and $t_2$ to find
\bea
Z_{\rm S}(t_1,t_2)&=&
1+t_1\chi_{\tiny\yng(1)}((\cdot))\Tr(Z)+t_2^2\chi_{\tiny\yng(2)}((\cdot)(\cdot))\Tr(Z)^2+t_2^2\chi_{\tiny\yng(2)}((\cdot\cdot))\Tr(Z^2)+\cdots\cr\cr
&=&1+t_1\Tr(Z)+t_2^2\Tr(Z)^2+t_2^2\Tr(Z^2)+\cdots
\eea
reproducing the correct characters for $S_1$ and $S_2$
\bea
\chi_{\tiny\yng(1)}((\cdot))=1\qquad \chi_{\tiny\yng(2)}((\cdot)(\cdot))=1\qquad 
\chi_{\tiny\yng(2)}((\cdot\cdot))=1
\eea

Following our discussion for characters with long rows, we could easily use the result (\ref{charlongcolumns}) to develop a formula for characters in representations labelled by Young diagrams with two columns.

\subsection{Excited bound states of two Sphere Giants}

In this section we extend the results obtained above, constructing generating functions for operators dual to excited sphere giant gravitons. Again consider a boundstate of two sphere giant gravitons, the first with momentum $p_1$ and the second with momentum $p_2$. The boundstate is excited by attaching open strings. The generating function for this class of bound states is
\bea
Z_{\rm S}(t_1,t_2,\{W\})&=& \int d^N\bar{\psi}\int d^N\psi 
\int d^N\bar{\xi} \int d^N\xi\,\, e^{-\bar{\psi}\cdot\psi-\bar{\xi}\cdot\xi}\cr\cr
&&\quad\times\,\, e^{-t_1\bar{\psi}\cdot Z\cdot \psi
+t_2^2\bar{\psi}^i\bar{\xi}^j (Z^k_i Z^l_j+Z^l_i Z^k_j)\psi_k\xi_l}\,\,
\prod_{r=1}^{m_{11}}\bar{\psi}\cdot W_{A_r}\cdot\psi\cr\cr
&&\times \prod_{s=1}^{m_{22}}(\bar{\psi}\cdot Z\cdot \psi\,\bar{\xi}\cdot W_{B_s}\cdot \xi-\bar{\psi}\cdot Z\cdot\xi\, \bar{\xi}\cdot W_{B_s}\cdot\psi)\cr\cr
&&\times\prod_{t=1}^{m_{12}}(\bar{\psi}\cdot Z\cdot\psi\,\bar{\xi}\cdot W_{D_t}\cdot \psi\,\bar{\psi}\cdot W_{C_t}\cdot \xi
+\bar{\psi}\cdot Z\cdot \xi\,\bar{\xi}\cdot W_{D_t}\cdot \psi\,\bar{\psi}W_{C_t}\,\psi)\cr\cr
&&\label{ZSW}
\eea
The signs in the final two lines would both have been positive if we had kept all barred variables to the left with $\bar{\psi}$'s before $\bar{\xi}$'s and all unbarred variables to the right with $\psi$'s before $\xi$'s. . The dependence on $t_1$ and $t_2$ again tells us about the quantum numbers of the brane boundstate with no open strings attached: the coefficient of the term $t_1^{p_1-p_2}t_2^{2p_2}$ describes the case that the first brane has momentum $p_1$ and the second momentum $p_2$. We again focus mainly on the su(2) sector in which case the open string words are constructed using two complex adjoint scalars, $Z$ and $Y$, with the last and first letter of the word both $Y$s. The Gauss Law is again respected because the open strings are added to the bound state in pairs, with one string ($W_{C_t}$) stretching from brane 1 to brane 2 and another ($W_{D_t}$) stretching from 2 to 1. In a discussion which closely parallels the discussion for the AdS giant bound state, we can again prove that the operators with strings attached only to brane 1 or only to brane 2 agree with the construction given in \cite{deMelloKoch:2007rqf}. The proof again proceeds by verifying that derivatives with respect to open strings words agree with the rule derived in \cite{deMelloKoch:2007rqf}. We will sketch the case that strings are attached to the giant of momentum $p_1$ since it is interesting to see the role of the Grassman nature of the integration variables. Recall that in the case of the AdS giant (a long row), the factor of the box containing $W_{A_1}$ is $N+(q-1)$, since each time we move along the Young diagram, to the right, we add 1 to the factor. In the case of the sphere giant (a long column), the factor of the box containing $W_{A_1}$ is $N-(q-1)$, since each time we move down the Young diagram we subtract 1 from the factor. It is the change in the sign of the $q-1$ contribution to the factor that we want to demonstrate. Towards this end, consider ($S_0=-\bar{\psi}\cdot\psi-\bar{\xi}\cdot\xi$, $S_1=-t_1\bar{\psi}\cdot Z\cdot \psi+t_2^2\bar{\psi}^i\bar{\xi}^j (Z^k_i Z^l_j+Z^l_i Z^k_j)\psi_k\xi_l$)
\bea
&&{\partial\over\partial (W_{A_1})^a_a} \int d^N\bar{\psi}\int d^N\psi \int d^N\bar{\xi} \int d^N\xi e^{S_0}e^{S_1}\,\,\prod_{r=1}^q\bar{\psi}\cdot W_{A_r}\cdot\psi\cr\cr
&=&\int d^N\bar{\psi}\int d^N\psi \int d^N\bar{\xi} \int d^N\xi e^{S_0}e^{S_1}\,\,\bar{\psi}\cdot\psi\,\,\prod_{r=2}^q\bar{\psi}\cdot W_{A_r}\cdot\psi\cr\cr
&=&-\int d^N\bar{\psi}\int d^N\psi \int d^N\bar{\xi} \int d^N\xi \left(\bar{\psi}^i{\partial\over\partial\bar\psi^i}e^{S_0}\right) e^{S_1}\,\,\prod_{r=2}^q\bar{\psi}\cdot W_{A_r}\cdot\psi\cr\cr
&=&\left(N-\bar{\psi}^i{\partial\over\partial\bar\psi^i}\right)\int d^N\bar{\psi}\int d^N\psi \int d^N\bar{\xi} \int d^N\xi \,\,e^{S_0+S_1}\,\,\prod_{r=2}^q\bar{\psi}\cdot W_{A_r}\cdot\psi
\eea
where the negative sign in the expression in brackets on the last line above is exactly what is required to get the correct factor. This sign comes from the Grassman nature of $\bar\psi^i$ since
\bea
{\partial\over\partial\bar\psi^i}\bar\psi^i=N-\bar\psi^i{\partial\over\partial\bar\psi^i}
\eea
Again, exactly as we found for the excited AdS giant bound states, by constructing and comparing operators we learn that the restricted Schur construction of \cite{deMelloKoch:2007rqf} is in general not the same as the construction using the generating function (\ref{ZSW}).

The operators constructed using the generating function (\ref{ZSW}) are again symmetric under swapping strings of the same orientation. This is a strong hint that they are closely related to the Gauss graph operators. This expectation turns out to be correct. The relevant generating function is
\bea
Z_{\rm S}(t_1,t_2,Y) &=& \int d^N\bar{\psi}\int d^N\psi 
\int d^N\bar{\xi} \int d^N\xi \,\, e^{-\bar{\psi}\cdot\psi-\bar{\xi}\cdot\xi}\cr\cr
&&\quad\times \,\, e^{-t_1\bar{\psi}\cdot Z\cdot \psi
+t_2^2\bar{\psi}^i\bar{\xi}^j (Z^k_i Z^l_j+Z^l_i Z^k_j)\psi_k\xi_l}\,\,
\prod_{r=1}^{m_{11}}\bar{\psi}\cdot Y\cdot\psi\cr\cr
&&\times \prod_{s=1}^{m_{22}}(\bar{\psi}\cdot Z\cdot \psi\,\bar{\xi}\cdot Y\cdot \xi-\bar{\psi}\cdot Z\cdot\xi\, \bar{\xi}\cdot Y\cdot\psi)\cr\cr
&&\times\prod_{t=1}^{m_{12}}(\bar{\psi}\cdot Z\cdot\psi\,\bar{\xi}\cdot Y\cdot \psi\,\bar{\psi}\cdot Y\cdot \xi+\bar{\psi}\cdot Z\cdot \xi\,\bar{\xi}\cdot Y\cdot \psi\,\bar{\psi}Y\,\psi)\cr\cr
&&\label{ggfors}
\eea
The bound states described by this generating function have $m_{11}$ strings attached to the brane corresponding to the first column, $m_{22}$ strings attached to the brane corresponding to the second column, $m_{12}$ strings stretching from the first to the second branes and $m_{21}=m_{12}$ strings stretched from the second to the first. The construction is manifestly symmetric under swaps of any strings that have the same orientation. In Appendix \ref{2S} we give detailed checks of the construction of operators coming from (\ref{ggfors}), as well as detailed examples demonstrating their equality to the Gauss graph operators. It is possible to generalize the proof given in Appendix \ref{ggequality} to the case of operators labelled by Young diagrams with long columns, and then to repeat the argument that worked for the AdS giant graviton, making use of the graph changing operator. The details are an obvious generalization of the discussion for the AdS gaint graviton.

\section{Action of the dilatation operator}\label{actionofD}

In the previous section we have obtained an integral representation of the operators dual to excited giant graviton bound states. Our goal in this section is study the action of the dilatation operator on these integral representations. We will restrict to the SU(2) sector, which corresponds to studying open string words constructed using two complex matrices $Y$ and $Z$. The one loop dilatation operator in the SU(2) sector (up to overall normalization) is \cite{Beisert:2003tq}
\bea
D=-g_{YM}^2{\rm Tr}\left([Z,Y][\partial_Z,\partial_Y]\right)\label{DilOprtr}
\eea 
When the open strings are described using words, one can derive a Hamiltonian \cite{Berenstein:2005fa,Berenstein:2006qk} for particles (representing the $Z$ fields) hopping on a lattice (defined by the $Y$ fields). The $Z$ fields can hop between the open string world sheet and the giant graviton bound state. A key physical question is the description of this hopping, relevant for the dynamics of the string endpoints. We take this up in Section \ref{osbc}. For the description in which the open strings are described using the $Y$ field, we have argued that the operators we construct are the Gauss graph operators. The Gauss graph operators are eigenstates of the one loop dilatation operator at large $N$. We will define an integral representation for the action of the dilatation operator $D$ on the generating function (\ref{AdSGGgen}) in Section \ref{ggtrue}. This integral representation is a natural starting point for a systematic ${1\over N}$ expansion.

\subsection{Open string boundaries}\label{osbc}

Our goal is to work out the action of the dilatation operator, in the large $N$ limit. Concretely, this amounts to constructing a lattice Hamiltonian for each of the four types of open strings (attaching to brane 1 or 2, and stretching from 1 to 2 or from 2 to 1) we can consider. What is of most interest is the boundary conditions for the strings. There are two derivatives in $D$, so it is possible that the two derivatives will act on different open strings. However, these effects are subleading at large $N$ and we ignore them\footnote{Here we are simply alluding to the fact that at large $N$ string interactions are suppressed and hence open string words do not mix. In the closed string sector the analogous statement is that in the planar limit, different trace structures do not mix.}. In this case, we are not losing any information by simplifying the discussion to a single (for $W_A$ and $W_B$) or pair (for $W_C$ and $W_D$) of open strings. 

The number of fields in the giant graviton is $O(N)$, the planar approximation fails and to obtain the correct large $N$ limit we need to sum all possible contractions of fields in the two giant gravitons. The number of fields in each open string is $\sim L$ with $L\sim O(\sqrt{N})$. If we take ${L^2\over N}\ll 1$, then at large $N$ it is accurate to contract the open string words planarly \cite{deMelloKoch:2007rqf}. 

We restrict to the su(2) sector which amounts to constructing all open string words $W_*$ from the pair $Z,Y$ of matrices. To obtain the world sheet description of the open strings, following \cite{Berenstein:2006qk}, we interpret the $Y$ fields in $W_*$ as a ``lattice" which can be populated by inserting excitations (in this case $Z$'s) into the lattice (gaps between the $Y$'s). We will refer to the $Z$'s as impurities. For a word containing $L$ $Y$ fields, there are $L-1$ sites in the lattice
\bea
W_*=YZ^{n_1}YZ^{n_2}Y\cdots YZ^{n_{L-1}}Y
\eea
The $\{n_i\}$ are lattice site occupation numbers. The one loop dilatation operator preserves the number of $Y$'s (the lattice is not dynamical) and allows impurities to hop between neighbouring sites ($Z$'s and $Y$'s can swap their order in $W_*$). This interpretation maps the problem of determining the anomalous dimensions of operators in the super Yang-Mills theory into the dynamics of a Cuntz oscillator chain \cite{Berenstein:2006qk}. The bulk interactions are described by the Hamiltonian
\begin{equation}
H_{\rm bulk} = 2\lambda\sum_{l=1}^L \hat{a}_l^\dagger \hat{a}_l -\lambda\sum_{l=1}^{L-1}(\hat{a}_l^\dagger \hat{a}_{l+1}
+\hat{a}_l \hat{a}^\dagger_{l+1}),
\label{bulk}
\end{equation}
where $\lambda=g_{YM}^2 N$ is the 't Hooft coupling and the Cuntz oscillators obey
$$ \hat{a}_i \hat{a}_i^\dagger = I,\qquad \hat{a}^\dagger_i \hat{a}_i = I-|0\rangle\langle 0|.$$
To obtain the full Hamiltonian, we need to include boundary interactions arising from the string/brane system interaction. This interaction, which introduces sources and sinks for the impurities at the boundaries of the lattice, was first derived in \cite{Berenstein:2006qk} for a string attached to a single sphere giant graviton, and then derived in \cite{deMelloKoch:2007nbd,Bekker:2007ea} for strings attached to an arbitrary bound state. One such interaction allows a $Z$ to hop from the first or last site of the string onto the giant, or from the giant into the first or last site of the string. In this process the string exchanges momentum with the giant graviton, so that the giant graviton is exerting a force on the string. There is also a boundary interaction in which a $Z$ belonging to the giant ``kisses" the first (or last) $Y$ in the open string word. In this case, no momentum is exchanged, and the string does not feel a force. The logic developed in \cite{deMelloKoch:2007nbd} derives the interaction for the ``hop off" process, in which a $Z$ hops off the string and onto the giant. Requiring that the Hamiltonian is hermitian then fixes the ``hop on" interaction. The momentum conserving boundary interaction follows by expressing the kiss as a hop on followed by a hop off.

The hop off boundary term for open string $W_*$ is given by \cite{deMelloKoch:2007nbd,Bekker:2007ea}
\bea
H_{\rm hop\,\,off}=\lambda\sqrt{c\over N} \hat{A}^\dagger \hat{a}_1 +\lambda\sqrt{c\over N} \hat{A}^\dagger \hat{a}_L\label{hopoffH}
\eea
where $c$ is the factor of the box occupied by the open string word $W_*$, $\hat{a}_1$ and $\hat{a}_L$ annihilate a $Z$ from the last and first sites of the open string word $W_*$ and $\hat{A}^\dagger$ adds a $Z$ to the giant graviton i.e. it adds a box in the Young diagram label of the operator. The key step in the derivation of this hop off boundary term is an identity obeyed by the restricted Schur polynomials. When a string hops out of the first (or last) site we obtain an open string word of the form $ZW_*$ (or $W_*Z$). The identity expresses the restricted Schur polynomial with words $ZW_*$ or $W_*Z$ in terms of restricted Schur polynomials with word $W_*$. This identity is derived by writing the sum over $\sigma$ in (\ref{rspexmpl}) as a sum over a subgroup and its cosets and again using known results about the eigenvalues of the Jucys-Murphy elements. This is how the factor of the box appears in (\ref{hopoffH}). In terms of the generating function the derivation of this identity is a trivial exercise. Indeed, the identity of \cite{deMelloKoch:2007nbd} is given by ($S_0=-\bar{\varphi}\cdot\varphi-\bar{\phi}\cdot\phi$ and $S_1=t_1\bar{\varphi}\cdot Z\cdot \varphi+t_2^2(\bar{\varphi}\cdot Z\cdot \varphi \bar{\phi}\cdot Z\cdot\phi -\bar{\varphi}\cdot Z\cdot \phi \bar{\phi}\cdot Z\cdot\varphi)$)
\bea
0&=&{1\over \pi^{2N}}\int d^N\bar{\varphi}\int d^N\varphi\int d^N\bar{\phi}\int d^N\phi {\partial\over\partial\bar{\varphi}^j}\,\,\left(e^{S_0+S_1}\bar{\varphi}^i W^j_i\right)\cr\cr
&=&{1\over \pi^{2N}}\int d^N\bar{\varphi}\int d^N\varphi\int d^N\bar{\phi}\int d^N\phi\,\, e^{S_0+S_1}\left(\Tr (W)-\bar{\varphi}\cdot W\cdot\varphi+t_1\bar{\varphi}\cdot W\cdot Z\cdot\varphi\right.\cr\cr\cr
&&\qquad\qquad\left.
+t_2^2(\bar{\varphi}\cdot W\cdot Z\cdot \varphi\,\bar{\phi}\cdot Z\cdot \phi-\bar{\varphi}\cdot W\cdot Z\cdot\phi\, \bar{\phi}\cdot Z\cdot\varphi)\right)
\eea
This is identical to the identities given in Appendix C.1 of \cite{Bekker:2007ea}. The term proportional to $\Tr (W)$ describes gravitational radiation, whereby the open string $W$ is emitted as a closed string $\Tr(W)$ from the excited giant bound state. It represents a subleading contribution at large $N$. The term proportional to $\bar{\varphi}\cdot W\cdot\varphi$ produces restricted Schur polynomials constructed with open string word $W$. The remaining terms produce restricted Schur polynomials constructed with open string word $W\cdot Z$. The generating function provides a remarkably simple derivation of this formula.

The identity above describes hopping out of the last site of the open string word. To obtain the identity needed for hopping out of the first site we consider
\bea
0&=&{1\over \pi^{2N}}\int d^N\bar{\varphi}\int d^N\varphi\int d^N\bar{\phi}\int d^N\phi \,\,{\partial\over\partial \varphi_i}\left(e^{S_0+S_1}W^j_i\varphi_j\right)
\eea
The identities needed for the sphere giant are equally easy to derive.

\subsection{Gauss graph operators}\label{ggtrue}

We have already argued that when each open string word is replaced by a single letter, the operators we generate are the Gauss graph operators. From the results obtained in \cite{deMelloKoch:2012ck} we know that these are eigenstates of $D$, in the large $N$ limit. In \cite{deMelloKoch:2011wah,deMelloKoch:2012ck} the simplifications of large $N$ were implementing by exploiting simplifications in the action of the symmetric group. This analysis, known as the distant corners approximation, is clumsy. In particular, it is not obvious exactly what has been discarded in this approximation and, consequently, it is not easy to systematically correct and improve the approximation. It is interesting to reconsider this computation using the generating function where we should have better control over the large $N$ expansion. Indeed, a systematic large $N$ expansion should follow from a saddle point approximation of the integral representations we have obtained. To carry out this analysis we need to apply $D$ to the operators defined by (\ref{AdSGGgen}). In this subsection we study some illustrative examples. Return to (\ref{AdSGGgen}) which we write as
\bea
Z_{\rm AdS}(t_1,t_2,Y)&=& {1\over \pi^{2N}}\int d^N\bar{\varphi}\int d^N\varphi\int d^N\bar{\phi}\int d^N\phi\,\,e^{-\bar{\varphi}\cdot\varphi-\bar{\phi}\cdot\phi}\,\,
e^{t_1\Phi_1+t_2^2\Phi_2}
\Phi_3^{m_{11}}\Phi_4^{m_{22}}\Phi_5^{m_{12}}\nonumber
\eea
where we have introduced the new variables
\bea
\Phi_1&=&\bar{\varphi}\cdot Z\cdot \varphi\cr\cr
\Phi_2&=&\bar{\varphi}\cdot Z\cdot \varphi \, \bar{\phi}\cdot Z\cdot\phi 
-\bar{\varphi}\cdot Z\cdot \phi\, \bar{\phi}\cdot Z\cdot\varphi\cr\cr
\Phi_3&=&\bar{\varphi}\cdot Y\cdot \varphi\cr\cr
\Phi_4&=&\bar{\varphi}\cdot Z\cdot \varphi\,\bar{\phi}\cdot Y\cdot \phi-\bar{\varphi}\cdot Z\cdot\phi\, \bar{\phi}\cdot Y\cdot\varphi\cr\cr
\Phi_5&=&\bar{\varphi}\cdot Z\cdot\varphi\,\bar{\phi}\cdot Y\cdot \varphi\,\bar{\varphi}\cdot Y\cdot \phi-\bar{\varphi}\cdot Z\cdot \phi\,\bar{\phi}\cdot Y \cdot \varphi\,\bar{\varphi}\cdot Y\cdot\varphi
\eea
Our operators are constructed using a large number of $Z$ fields (order $N$) and very few $Y$ fields (order 1). It proves helpful to remove the $Y$ fields, as much as possible, from the manipulations that follow. With this in mind, introduce sources for the fields $\varphi$, $\bar{\varphi}$, $\phi$ and $\bar{\phi}$ and write
\bea
Z_{\rm AdS}(t_1,t_2,Y)&=&\widehat\Phi_3^{m_{11}}\widehat\Phi_4^{m_{22}}\widehat\Phi_5^{m_{12}} \Big({1\over \pi^{2N}}\int d^N\bar{\varphi}\int d^N\varphi\int d^N\bar{\phi}\int d^N\phi \,\,\times\cr\cr\cr
&&\qquad\qquad\times\, e^{-\bar{\varphi}\cdot\varphi-\bar{\phi}\cdot\phi}\,\,
e^{t_1\Phi_1+t_2^2\Phi_2}e^{\bar{j}^{\varphi}\cdot\varphi+\bar{\varphi}\cdot j^{\varphi}+\bar{j}^{\phi}\cdot\phi+\bar{\phi}\cdot j^{\phi}}\Big)\Big|_{j=0}\cr\cr
&\equiv& \widehat\Phi_3^{m_{11}}\widehat\Phi_4^{m_{22}}\widehat\Phi_5^{m_{12}}
Z_{\rm AdS,reduced}(t_1,t_2)|_{j=0}
\label{endresulttt}
\eea
where $|_{j=0}$ is an instruction to set all sources to zero and
\bea
\widehat\Phi_3&=&{\partial\over\partial j^{\varphi}}\cdot Y\cdot {\partial\over\partial \bar{j}^\varphi}\cr\cr
\widehat\Phi_4&=&{\partial\over\partial j^{\varphi}}\cdot Z\cdot {\partial\over\partial \bar{j}^\varphi}\,{\partial\over\partial j^{\phi}}\cdot Y\cdot {\partial\over\partial \bar{j}^{\phi}}-{\partial\over\partial j^{\varphi}}\cdot Z\cdot{\partial\over\partial \bar{j}^{\phi}}\, {\partial\over\partial j^{\phi}}\cdot Y\cdot{\partial\over\partial \bar{j}^\varphi}\cr\cr
\widehat\Phi_5&=&{\partial\over\partial j^{\varphi}}\cdot Z\cdot{\partial\over\partial \bar{j}^\varphi}\,{\partial\over\partial j^{\phi}}\cdot Y\cdot {\partial\over\partial \bar{j}^\varphi}\,{\partial\over\partial j^{\varphi}}\cdot Y\cdot {\partial\over\partial \bar{j}^{\phi}}\cr\cr\cr
&&-{\partial\over\partial j^{\varphi}}\cdot Z\cdot {\partial\over\partial \bar{j}^{\phi}}\,{\partial\over\partial j^{\phi}}\cdot Y \cdot {\partial\over\partial \bar{j}^\varphi}\,{\partial\over\partial j^{\varphi}}\cdot Y\cdot{\partial\over\partial \bar{j}^\varphi}
\label{openstringoperators}
\eea
The integration is not a Gaussian integral because $\Phi_2$ introduces terms that are quartic in the integration variables. To proceed perform a Hubbard-Stratanovich like transformation, introducing two new complex fields, and rewrite $Z_{\rm AdS,reduced}(t_1,t_2)$ as follows
\bea
&&Z_{\rm AdS,reduced}(t_1,t_2)\,\,\, =\,\,\,{1\over \pi^{2N}}\int d^N\bar{\varphi}\int d^N\varphi\int d^N\bar{\phi}\int d^N\phi \int dx\int d\bar{x}\int dy\int d\bar{y}\cr\cr\cr
&&\,\, \times\,e^{-\bar{\varphi}\cdot\varphi-\bar{\phi}\cdot\phi-x\bar{x}-y\bar{y}+t_1\bar{\varphi}\cdot Z\cdot \varphi+t_2x\bar{\varphi}\cdot Z\cdot \varphi+t_2\bar{x}\bar{\phi}\cdot Z\cdot\phi+t_2 y\bar{\varphi}\cdot Z\cdot \phi-t_2\bar{y}\bar{\phi}\cdot Z\cdot\varphi+\bar{j}^{\varphi}\cdot\varphi+\bar{\varphi}\cdot j^{\varphi}+\bar{j}^{\phi}\cdot\phi+\bar{\phi}\cdot j^{\phi}}\nonumber
\eea
We can streamline this expression as follows
\bea
Z_{\rm AdS,reduced}(t_1,t_2)&=&{1\over \pi^{2N+2}}\int d^N\bar{\varphi}\int d^N\varphi 
\int d^N\bar{\phi}\int d^N\phi\int dx\int d\bar{x}\int dy\int d\bar{y}\cr\cr\cr
&&\,\,\times\, e^{-v\cdot M\cdot \bar{v} -x\bar{x}-y\bar{y}+v \cdot\bar{j}+j\cdot\bar{v}}
\eea
where
\bea
v_i&=&\left[\begin{array}{cc} \varphi_i &\phi_i \end{array}\right] \qquad
\bar{v}^i\,\,=\,\,\left[\begin{array}{c} \bar{\varphi}^i \\ \bar{\phi}^i \end{array}\right] \qquad j_i\,\,=\,\,\left[\begin{array}{cc} j^{\varphi}_i &j^\phi_i \end{array}\right] 
\qquad
\bar{j}^i\,\,=\,\,\left[\begin{array}{c} \bar{j}^{\varphi\,i} \\ \bar{j}^{\phi\, i} \end{array}\right] \cr\cr\cr
M &=& \left[\begin{array}{cc} {\bf 1}_N-t_1 Z-t_2 x Z &-t_2 y Z\\ t_2 \bar{y} Z &{\bf 1}_N-t_2\bar{x} Z\end{array}\right]\,\,=\,\,
{\bf 1}_2\otimes {\bf 1}_N-\left[\begin{array}{cc} t_1+t_2 x &t_2 y\\ -t_2\bar{y} & t_2\bar{x} \end{array}\right]\otimes Z\cr\cr
&\equiv&{\bf 1}_2\otimes {\bf 1}_N- {\cal M} \otimes Z
\eea
The integrals over $\varphi$, $\bar{\varphi}$, $\phi$ and $\bar{\phi}$ are Gaussian and can be done exactly. The result is
\bea
Z_{\rm AdS,reduced}(t_1,t_2)&=&{1\over \pi^2}\int dx\int d\bar{x}\int dy\int d\bar{y}
\,\,\,{e^{j\cdot (M^{-1})\cdot \bar{j} -x\bar{x}-y\bar{y}}\over\det M}\label{GIdone}
\eea
The eigenvalues of $Z$ are $z_i$. The eigenvalues of ${\cal M}$ are given in (\ref{eigens}). In terms of these eigenvalues we can write
\bea
{1\over \det (M)}=\prod_{a=1}^2\prod_{i=1}^N {1\over 1-\lambda_a z_i}
\eea
The Cauchy-Littlewood identity \cite{macdonald} is
\bea
\prod_{i=1}^L\prod_{j=1}^M {1\over 1-x_i y_j}=\sum_{\Lambda}
\chi_\Lambda (x_1,\cdots,x_L)\chi_\Lambda (y_1,\cdots,y_M)
\eea
where the sum on the right hand side only includes Young diagrams $\Lambda$ with less than the smaller of $L,M$ rows, and $\chi_\Lambda$ is a Schur polynomial. Applying this formula to our problem we find
\bea
{1\over \det (M)}=\sum_{\Lambda}
\chi_\Lambda ({\cal M})\chi_\Lambda (Z)
\eea
so that
\bea
Z_{\rm AdS,reduced}(t_1,t_2)&=&\sum_{\Lambda}f_\Lambda(j,\bar{j})\chi_\Lambda (Z)
\label{SchurIdent}
\eea
\bea
f_\Lambda(j,\bar{j})&=&{1\over\pi^2}\int dx\int d\bar{x}\int dy\int d\bar{y}\,\,\,e^{j\cdot (M^{-1})\cdot \bar{j} -x\bar{x}-y\bar{y}}\,\,\chi_\Lambda ({\cal M})
\eea
The sum over $\Lambda$ is only over Young diagrams with at most 2 rows.

We can perform a convincing check of (\ref{SchurIdent}) after we set $j=\bar{j}=0$. When the sources vanish we know that the generating function produces the Schur polynomials with two rows. Further, the Schur polynomial labelled by a Young diagram with $r_1$ boxes in the first row and $r_2$ boxes in the second is multiplied by a monomial $t_1^{r_1-r_2}t_2^{2r_2}$. Using the usual formula for the Schur polynomials, we find
\bea
\chi_{\tiny\yng(1)}({\cal M})&=&t_1 + t_2 x + t_2 \bar{x}
\eea
\bea
\chi_{\tiny\yng(2)}({\cal M})&=&t_1^2+2t_1t_2x+t_2^2x^2+t_1t_2\bar{x}+t_2^2x\bar{x}+t_2^2\bar{x}^2-t_2^2y\bar{y}
\eea
\bea
\chi_{\tiny\yng(1,1)}({\cal M})&=&t_1t_2\bar{x}+t_2^2x\bar{x}+t_2^2y\bar{y}
\eea
\bea
\chi_{\tiny\yng(3)}({\cal M})&=&t_1^3+3t_1^2t_2x+3t_1t_2^2x^2+t_2^3x^3+t_1^2t_2\bar{x}+2t_1t_2^2x\bar{x}+t_2^3x^2\bar{x}\cr\cr
&&\quad +t_1t_2^2\bar{x}^2+t_2^3x\bar{x}^2+t_2^3\bar{x}^3-2t_1t_2^2y\bar{y}-2 t_2^3xy\bar{y}-2 t_2^3 \bar{x} y \bar{y}
\eea
\bea
\chi_{\tiny\yng(2,1)}({\cal M})&=&t_1^2t_2\bar{x}+2 t_1 t_2^2 x \bar{x} + t_2^3 x^2 \bar{x}+ t_1 t_2^2 \bar{x}^2 + t_2^3 x \bar{x}^2+ t_1 t_2^2 y \bar{y} \cr\cr
&&\qquad + t_2^3 x y \bar{y} + t_2^3 \bar{x} y \bar{y}
\eea
We then easily obtain
\bea
f_{\tiny\yng(1)}&=&t_1\qquad
f_{\tiny\yng(2)}\,\,\,=\,\,\,t_1^2\qquad
f_{\tiny\yng(1,1)}\,\,\,=\,\,\,2 t_2^2\qquad
f_{\tiny\yng(3)}\,\,\,=\,\,\,t_1^3\qquad
f_{\tiny\yng(2,1)}\,\,\,=\,\,\,3t_1t_2^2
\eea
which are the correct results. We have also tested that the integer coefficient of each monomial is correct.

If we work in a basis in which $Z$ is diagonal, after a little work we can rewrite (\ref{GIdone}) as
\bea
&&Z_{\rm AdS,reduced}(t_1,t_2)\,\,\,=\,\,\,{1\over \pi^2}\int dx\int d\bar{x}\int dy\int d\bar{y}\,\,\,\exp \Bigg[-x\bar{x}-y\bar{y}+\sum_{n=1}^\infty f_n \Tr(Z^n)\cr\cr
&&\quad +j_{\varphi}\cdot\left(\sum_{n=0}^\infty X^n(1-t_2\bar{x}Z)\right)\cdot\bar{j}_\varphi-j_{\varphi}\cdot\left(\sum_{n=0}^\infty X^n t_2yZ\right)\cdot\bar{j}_\phi+j_{\phi}\cdot\left(\sum_{n=0}^\infty X^n t_2\bar{y}Z\right)\cdot\bar{j}_\varphi\cr\cr
&&\qquad\qquad +j_{\phi}\cdot\left(\sum_{n=0}^\infty X^n (1-t_2\bar{x}Z-t_1Z)\right)\cdot\bar{j}_\phi\Bigg]
\label{twointegralstodo}
\eea
where $f_n$ has been defined in (\ref{definefn}) and
\bea
X&=& (t_1+t_2(x+\bar{x}))Z-(t_1t_2\bar{x}+t_2^2(x\bar{x}+y\bar{y}))Z^2
\eea
This formula is exact. We are now ready to apply the operators defined in (\ref{openstringoperators}) to this expression. We evaluate the result at $j=\bar{j}=0$. One derivative acting produces a term linear in the sources. To get a non-zero result we need to differentiate this again, with respect to the source that appears. Consequently, a pair of derivatives must act before we get a non-zero result. The combinatorics is therefore identical to the combinatorics from Wick's theorem and we only need to know what the basic pairings are. It is easy to read off from the above expression what each possible pairing gives.

It is instructive to consider the case that $t_2=0$. This corresponds to a single giant system, which we know is a ${1\over 2}$-BPS operator. A simple computation shows that
\bea
Z_{\rm AdS,reduced}(t_1,0)\,\,\,=\,\,\,\exp \Bigg(\sum_{n=1}^\infty f_n \Tr(Z^n)+j_{\varphi}\cdot\left(\sum_{n=0}^\infty t_1^n Z^n\right)\cdot\bar{j}_\varphi+j_{\phi}\cdot\bar{j}_\phi\Bigg)\cr
\eea
For two strings attached to the giant graviton, the generating function is
\bea
Z_{\rm AdS}(t_1,0,Y)&=&\left({\partial\over\partial j^\varphi}\cdot Y\cdot{\partial\over\partial\bar{j}^\varphi}\right)^2Z_{\rm AdS,reduced}(t_1,0)\cr\cr
&=&e^{\sum_{n=1}^\infty {t_1^n\over n}\Tr(Z^n)}\sum_{n_1,n_2=1}^\infty t_1^{n_1+n_2}\Big(\Tr(YZ^{n_1})\Tr(YZ^{n_2})+\Tr(YZ^{n_1}YZ^{n_2})\Big)\cr\cr
&=&\sum_{q=1}^\infty {t_1^q\over q!} O_q^{(2)}
\eea
For $k$ strings attached to the giant graviton, the generating function is
\bea
Z_{\rm AdS}(t_1,0,Y)&=&e^{\sum_{n=1}^\infty {t_1^n\over n}\Tr(Z^n)}\sum_{n_1,\cdots,n_k=1}^\infty t_1^{n_1+n_2+\cdots+n_k}\cr\cr
&&\qquad \sum_{\sigma\in S_k}(YZ^{n_1})^{i_1}_{i_{\sigma(1)}}(YZ^{n_2})^{i_2}_{i_{\sigma(2)}}\cdots (YZ^{n_k})^{i_k}_{i_{\sigma(k)}}\cr\cr
&=&\sum_{q=1}^\infty {t_1^q\over q!} O_q^{(k)}
\eea
This gives a very explicit description of the Gauss graph operator, for a single giant graviton with $k$ strings attached. It is straight forward to verify that
\bea
O^{(k)}_q&=&\sum_{\sigma\in S_{k+q}}Z^{i_1}_{i_{\sigma(1)}}\cdots Z^{i_q}_{i_{\sigma(q)}}Y^{i_{q+1}}_{i_{\sigma(q+1)}}\cdots Y^{i_{q+k}}_{i_{\sigma(q+k)}}\cr\cr &=& k!q!\,\,\chi_{(k+q),(k)(q)}(Z,Y)
\eea
where $\chi_{(k+q),(k)(q)}(Z,Y)$ is a restricted Schur polynomial. The corresponding projection operator $P_{(k+q),(k)(q)}$ needed to define the restricted trace is trivial since the carrier space of representation $(k+q)$ is one dimensional.  The results of \cite{DeComarmond:2010ie} then imply that
\bea
DO_q^{(k)}=0
\eea
to all orders in ${1\over N}$.

As an example of a non-trivial eigenstate of the dilatation operator $D$, consider a pair of stretched strings, i.e. $m_{11}=0=m_{22}$ and $m_{12}=1=m_{21}$. To obtain a manageable problem, consider the coefficient of $t_1^n$ in the expansion of (\ref{endresulttt}), which is given by
\bea
{\cal O}&=&\sum_{\sigma\in S_{n+2}}Z^{i_1}_{i_{\sigma(1)}}\cdots Z^{i_{n+1}}_{i_{\sigma(n+1)}}(Y^2)^{i_{n+2}}_{i_{\sigma (n+2)}}\cr\cr
&&-\sum_{\sigma\in S_{n+2}}Z^{i_1}_{i_{\sigma(1)}}\cdots Z^{i_n}_{i_{\sigma(n)}}Y^{i_{n+1}}_{i_{\sigma(n+1)}}(ZY)^{i_{n+2}}_{i_{\sigma (n+2)}}\cr\cr
&=&{\cal O}_1-{\cal O}_2
\eea
We work at the leading order in a large $N$ expansion and we take $n$ to be of order $N$. In this limit ${\cal O}$ simplifies. Recall that in any conformal field theory, we have the operator/state correspondence, which associates a state to every operator. Thus associated to the two operators above, we have the states $|{\cal O}_1\rangle$ and $|{\cal O}_2\rangle$. Norms of states map to correlation functions of operators
\bea
\langle {\cal O}_1|{\cal O}_1\rangle = \langle {\cal O}_1{\cal O}_1^\dagger\rangle\qquad 
\langle {\cal O}_2|{\cal O}_2\rangle = \langle {\cal O}_2{\cal O}_2^\dagger\rangle
\eea
In Appendix \ref{correlators} we show that
\bea
{\langle {\cal O}_2|{\cal O}_2\rangle\over \langle {\cal O}_1|{\cal O}_1\rangle}=O(N^{-1})
\eea
so that at large $N$ we can simplify
\bea
{\cal O}&=&{\cal O}_1+O(N^{-1})\,\,=\,\,\sum_{\sigma\in S_{n+2}}Z^{i_1}_{i_{\sigma(1)}}\cdots Z^{i_{n+1}}_{i_{\sigma(n+1)}}(Y^2)^{i_{n+2}}_{i_{\sigma (n+2)}}
\eea
It is interesting to trace this simplification back to the generating function $Z_{\rm AdS}(t_1,t_2,Y)$. This simplification amounts to dropping the second term in
\bea
\Phi_5&=&\bar{\varphi}\cdot Z\cdot\varphi\,\bar{\phi}\cdot Y\cdot \varphi\,\bar{\varphi}\cdot Y\cdot \phi-\bar{\varphi}\cdot Z\cdot \phi\,\bar{\phi}\cdot Y \cdot \varphi\,\bar{\varphi}\cdot Y\cdot\varphi
\eea
which is intuitively appealing. Indeed the factor $\bar{\phi}\cdot Y\cdot \varphi\,\bar{\varphi}\cdot Y\cdot \phi$ has the very natural interpretation that the two $Y$ fields have endpoints on different branes. Acting on ${\cal O}$ with $D$ given in (\ref{DilOprtr}) we obtain
\bea
D{\cal O}&=&-g_{YM}^2(n+1)\sum_{\sigma\in S_{n+2}}Z^{i_2}_{i_{\sigma(2)}}\cdots Z^{i_{n+1}}_{i_{\sigma(n+1)}}\Big(Y^{i_1}_{i_{\sigma(n+2)}}[Z,Y]^{i_{n+2}}_{i_{\sigma(1)}}
+Y^{i_{n+2}}_i[Z,Y]^i_{i_{\sigma(1)}}\delta^{i_1}_{i_{\sigma(n+2)}}\cr\cr
&&\qquad\qquad-[Z,Y]^{i_1}_jY^j_{i_{\sigma(n+2)}}\delta^{i_{n+2}}_{i_{\sigma(1)}}
-Y^{i_{n+2}}_{i_{\sigma(1)}}[Z,Y]^{i_1}_{i_{\sigma(n+2)}}\Big)
\eea
To simplify this expression we need a few identities. The most basic of these follow by simply changing variables in the sum. For example
\bea
\sum_{\sigma\in S_{n+2}}Z^{i_2}_{i_{\sigma(2)}}\cdots Z^{i_{n+1}}_{i_{\sigma(n+1)}}Y^{i_1}_{i_{\sigma(n+2)}}[Z,Y]^{i_{n+2}}_{i_{\sigma(1)}}&=&\sum_{\rho\in S_{n+2}}Z^{i_2}_{i_{\rho(2)}}\cdots Z^{i_{n+1}}_{i_{\rho(n+1)}}Y^{i_1}_{i_{\rho(1)}}[Z,Y]^{i_{n+2}}_{i_{\rho(n+2)}}\cr\cr
&=&\sum_{\psi\in S_{n+2}}Z^{i_1}_{i_{\psi(1)}}\cdots Z^{i_{n}}_{i_{\psi(n)}}Y^{i_{n+1}}_{i_{\psi(n+1)}}[Z,Y]^{i_{n+2}}_{i_{\psi(n+2)}}\nonumber
\eea
where for the first equality $\sigma=\rho (1,n+2)$ and for the second $\rho=(1,n+1)\psi(1,n+1)$. Another useful trick is to write sums in terms of cosets of a subgroup. For example, examine the sum
\bea
{\cal S}&=&\sum_{\sigma\in S_{n+3}}Z^{i_1}_{i_{\sigma(1)}}\cdots Z^{i_{n+1}}_{i_{\sigma(n+1)}}Y^{i_{n+2}}_{i_{\sigma(n+2)}}Y^{i_{n+3}}_{i_{\sigma(n+3)}}
\eea
Consider the subgroup $S_{n+2}$ defined by $\rho(n+3)=n+3$ for $\rho\in S_{n+2}$, i.e. the subgroup leaves $n+3$ inert. Now rewrite the above sum in terms of the right cosets, labelled by the integer $i=1,2,\cdots,n+3$ defined by
\bea
\sigma&=&\rho (i,n+3)\qquad \sigma\in S_{n+3}\quad \rho\in S_{n+2}
\eea
The sum over $\sigma$ is equivalent to separately summing over $\rho$ and $i$. After  rewriting we find
\bea
{\cal S}&=&\sum_{\rho\in S_{n+2}}Z^{i_1}_{i_{\rho(1)}}\cdots Z^{i_{n+1}}_{i_{\rho(n+1)}}Y^{i_{n+2}}_{i_{\rho(n+2)}}\Tr(Y)
+\sum_{\rho\in S_{n+2}}Z^{i_1}_{i_{\rho(1)}}\cdots Z^{i_{n+1}}_{i_{\rho(n+1)}}(Y^2)^{i_{n+2}}_{i_{\sigma(n+2)}}\cr\cr
&&+(n+1)\sum_{\rho\in S_{n+2}}Z^{i_1}_{i_{\sigma(1)}}\cdots Y^{i_{n+1}}_{i_{\sigma(n+1)}}(ZY)^{i_{n+2}}_{i_{\sigma(n+2)}}\label{frstresforsum}
\eea
If we use left cosets we find
\bea
{\cal S}&=&\sum_{\rho\in S_{n+2}}Z^{i_1}_{i_{\rho(1)}}\cdots Z^{i_{n+1}}_{i_{\rho(n+1)}}Y^{i_{n+2}}_{i_{\rho(n+2)}}\Tr(Y)
+\sum_{\rho\in S_{n+2}}Z^{i_1}_{i_{\rho(1)}}\cdots Z^{i_{n+1}}_{i_{\rho(n+1)}}(Y^2)^{i_{n+2}}_{i_{\sigma(n+2)}}\cr\cr
&&+(n+1)\sum_{\rho\in S_{n+2}}Z^{i_1}_{i_{\sigma(1)}}\cdots Y^{i_{n+1}}_{i_{\sigma(n+1)}}(YZ)^{i_{n+2}}_{i_{\sigma(n+2)}}\label{secresforsum}
\eea
Comparing (\ref{frstresforsum}) and (\ref{secresforsum}) we have
\bea
\sum_{\rho\in S_{n+2}}Z^{i_1}_{i_{\sigma(1)}}\cdots Y^{i_{n+1}}_{i_{\sigma(n+1)}}(ZY)^{i_{n+2}}_{i_{\sigma(n+2)}}&=&\sum_{\rho\in S_{n+2}}Z^{i_1}_{i_{\sigma(1)}}\cdots Y^{i_{n+1}}_{i_{\sigma(n+1)}}(YZ)^{i_{n+2}}_{i_{\sigma(n+2)}}
\eea
Making use of these identities we find
\bea
D{\cal O}&=&-g_{YM}^2(n+1)\sum_{\sigma\in S_{n+2}}Z^{i_1}_{i_{\sigma(1)}}\cdots Z^{i_{n}}_{i_{\sigma(n)}}(2YZY-Y^2Z-ZY^2)^{i_{n+1}}_{i_{\sigma(n+1)}}\delta^{i_{n+2}}_{i_{\sigma(n+2)}}\nonumber
\eea
Next, introduce the $S_{n+1}$ subgroup defined by $\rho(n+2)=n+2$. By making use of right cosets exactly as above, we find
\bea
D{\cal O}&=&-g_{YM}^2(N+n+1)(n+1)\sum_{\sigma\in S_{n+1}}Z^{i_1}_{i_{\sigma(1)}}\cdots Z^{i_{n}}_{i_{\sigma(n)}}(2YZY-Y^2Z-ZY^2)^{i_{n+1}}_{i_{\sigma(n+1)}}\nonumber
\eea
The next identity we need is obtained by starting from the sum which defines ${\cal O}$
\bea
{\cal O}=\sum_{\sigma\in S_{n+2}}Z^{i_1}_{i_{\sigma(1)}}\cdots Z^{i_n}_{i_{\sigma(n)}}Z^{i_{n+1}}_{i_{\sigma(n+1)}}(Y^2)^{i_{n+2}}_{i_{\sigma(n+2)}}
\eea
and writing in terms of cosets of the subgroup $S_{n+1}$ which leaves $n+2$ inert. We learn that
\bea
{\cal O}&=&\sum_{\sigma\in S_{n+1}}Z^{i_1}_{i_{\sigma(1)}}\cdots Z^{i_n}_{i_{\sigma(n)}}Z^{i_{n+1}}_{i_{\sigma(n+1)}}\Tr(Y^2)
+(n+1)\sum_{\sigma\in S_{n+1}}Z^{i_1}_{i_{\sigma(1)}}\cdots Z^{i_n}_{i_{\sigma(n)}}(ZY^2)^{i_{n+1}}_{i_{\sigma(n+1)}}\nonumber
\eea
when using right cosets, and
\bea
{\cal O}&=&\sum_{\sigma\in S_{n+1}}Z^{i_1}_{i_{\sigma(1)}}\cdots Z^{i_n}_{i_{\sigma(n)}}Z^{i_{n+1}}_{i_{\sigma(n+1)}}\Tr(Y^2)
+(n+1)\sum_{\sigma\in S_{n+1}}Z^{i_1}_{i_{\sigma(1)}}\cdots Z^{i_n}_{i_{\sigma(n)}}(Y^2Z)^{i_{n+1}}_{i_{\sigma(n+1)}}\nonumber
\eea
when using left cosets. Recalling that $n$ is of order $N$, it is clear that the first term on the right hand side of these last two formulas can be dropped at large $N$. We then find
\bea
D{\cal O}&=&2g_{YM}^2(N+n+1){\cal O}+2{\cal O}'
\eea
where
\bea
{\cal O}'&=&\sum_{\sigma\in S_{n+1}}Z^{i_1}_{i_{\sigma(1)}}\cdots Z^{i_{n}}_{i_{\sigma(n)}}
(YZY)^{i_{n+1}}_{i_{\sigma(n+1)}}
\eea
In Appendix \ref{correlators} we show that
\bea
{\langle {\cal O}'|{\cal O}'\rangle\over \langle {\cal O}|{\cal O}\rangle}=O(N^{-1})
\eea
so that the ${\cal O}'$ contribution can be dropped at large $N$. Thus, in the end we have
\bea
D{\cal O}&=&2g_{YM}^2(N+n+1){\cal O}
\eea
proving that our operators are indeed eigenstates of $D$ at large $N$. There are a few points worth stressing:
\begin{itemize} 
\item[1.] The above calculation has summed a lot more than just the planar diagrams.
\item[2.] The operator ${\cal O}$ is a sum of many different trace structures.
\item[3.] It is not the usual 't Hooft coupling $g_{YM}^2 N$ that appears, but rather $g_{YM}^2 (N+n+1)$. This is also fixed in the large $N$ limit so that we have a finite renormalization of the 't Hooft coupling. Notice that $N+n+1$ is the factor of the last box in the first row of the $R$ label of the Gauss graph operator. The same renormalization is evident in the results of \cite{DeComarmond:2010ie,deMelloKoch:2020agz}, obtained using very different methods.
\item[4.] The above result is another nice check of the equivalence of the operators defined by the generating function and the Gauss graph operators.
\end{itemize}

\section{Discussion and Conclusions}\label{conclusions}

The spectrum of type IIB string theory on AdS$_5\times$S$^5$ includes giant graviton branes. These are ${1\over 2}$-BPS solutions. Superposing a collection of giant graviton branes gives a ${1\over 2}$-BPS bound state of giant gravitons. The CFT operators dual to these bound states are given by Schur polynomials \cite{Corley:2001zk}. Excitations of these bound states, which are in general not BPS states, are obtained by attaching open strings to the branes. CFT operators describing these excited giant graviton bound states, known as Gauss graph operators, were described in \cite{deMelloKoch:2011ci,deMelloKoch:2012ck}. Their construction employs group representation theory and a Fourier transform on a double coset. The construction is technical which obscures many details. For example, it is not easy to see how the description simplifies at large $N$ or how systematic ${1\over N}$ corrections can be computed. The discussion is general in the sense that it applies to arbitrary bound states of giant gravitons. An important assumption is that the number of open strings that are attached is limited so that we are considering a small deformation of the original ${1\over 2}$-BPS operator.

The question of developing a systematic ${1\over N}$ expansion for excited giant graviton systems is the central motivation for our study. A standard way of generating a ${1\over N}$ expansion is through a saddle point evaluation of the path integral. Consequently, a promising starting point is an integral representations for the relevant operators. This philosophy was recently employed, in a very similar setting, in \cite{Berenstein:2022srd}. In this article we have obtained these integral representations, for operators dual to a boundstate of two sphere giant gravitons or two AdS giant gravitons. These representations provide a simple description of the relevant operators.

The starting point for the construction is the observation that Schur polynomials can be written using Young symmetrizers. These Young symmetrizers symmetrize and anti-symmetrize indices of matrix fields from which our operator is composed. Motivated by the results of \cite{Jiang:2019xdz,Jiang:2019zig,Yang:2021kot,Chen:2019gsb,Chen:2019kgc} these symmetrization and anti-symmetrizations can be implemented as integrals over ordinary (for symmetrization) and Grassman (for anti-symmetrization) variables which carry a single $U(N)$ index. In this way one obtains an integral representation for Schur polynomials with two rows or two columns. One consequence of this representation is a simple integral formula for characters of the symmetric group, in irreps labelled by Young diagrams with two rows, or two columns. There is a natural proposal for operators dual to excited giant graviton boundstates. One of our most important results has been the demonstration that this gives an integral representation for the Gauss graph operators themselves. 

We have considered the action of the dilatation operator on these novel generating functions. In the case that the open strings are described using an open string word, a key physical question is the boundary condition imposed on the string endpoints. The dilatation operator translates into a world sheet Hamiltonian as usual. The open string endpoints are represented by sources and sinks in this Hamiltonian, allowing the brane and open string to exchange momentum, i.e. to exert a force on each other. The form of this interaction was first obtained in \cite{Berenstein:2005fa,Berenstein:2006qk} for a single brane, and in \cite{deMelloKoch:2007nbd,Bekker:2007ea} for a general brane boundstate. In the case of the general brane boundstate, the discussion makes use of identities obeyed by restricted characters and these identities were derived using sums over cosets of the group and properties of Jucys-Murphy elements. We have recovered these known results, but with much less effort: the restricted character identity follows from the statement that the integral of a total derivative vanishes! In the case that each open string is described by a single $Y$ field, we are considering the Gauss graph operators which are eigenstates of $D$ \cite{deMelloKoch:2011ci,deMelloKoch:2012ck}. By introducing a set of sources we have written down a set of differential operators (\ref{openstringoperators}) which create the open strings when applied to the generating function (\ref{twointegralstodo}). In (\ref{twointegralstodo}) there are only 4 integrals left to do and consequently, this is a promising starting point to generate a systematic ${1\over N}$ expansion through a saddle point evaluation.

Our analysis has been specialized to the case of a boundstate of two giant gravitons. A natural extension is to boundstates of more giant gravitons. The extension is straight forward, if a little tedious. To sketch what is involved, we will give some of the details relevant to describe bound states of three AdS giant gravitons. Since there are three rows, and the indices belonging to each row must be symmetrized, the description will involve three sets of fields, $\varphi_i$, $\bar\varphi^i$, $\phi_i$, $\bar\phi^i$ as well as $\eta_i$, $\bar\eta^i$. There is also a new antisymmetrization, associated to the new columns of length three. As usual, we list row lengths so that $R=(p_1,p_2,p_3)$ denotes a Young diagram with $p_1$ boxes in the first row, $p_2$ in the second and $p_3$ in the third. The generating function for Schur polynomials with three rows is given by
\bea
Z_{\rm AdS}(t_1,t_2,t_3)&=&\sum^\infty_{p_1=p_2+p_3}\sum_{p_2=p_3}^\infty
\sum_{p_3=1}^\infty 
{t_3^{p_3}t_2^{p_2-p_3}t_1^{p_1-p_2}(p_1+p_2+p_3)!\over d_{(p_1,p_2,p_3)}p_3!(p_2-p_3)!(p_1-p_2)!}\chi_{(p_1,p_2,p_3)}(Z)\cr\cr\cr
&=& {1\over \pi^{3N}}\int d^N\varphi\int d^N\bar{\varphi}\int d^N\phi\int d^N\bar{\phi}\int d^N\eta\int d^N\bar{\eta}\cr\cr
&&\qquad\qquad 
e^{-\bar{\varphi}^i\varphi_i-\bar{\phi}^i\phi_i-\bar{\eta}^i\eta_i+t_1S_1
+t_2^2S_2+t_3^3S_3}
\label{threeAdSgiants}
\eea
where
\bea
S_1&=&\bar{\varphi}^iZ_i^j\varphi_j\cr\cr
S_2&=&\bar{\varphi}^i\bar{\phi}^j (Z^k_i Z^l_j-Z^l_i Z^k_j)\varphi_k\phi_l\cr\cr
S_3&=&\bar{\varphi}^i\bar{\phi}^j\bar{\eta}^k 
(Z^l_i Z^m_jZ^n_k-Z^m_i Z^l_jZ^n_k-Z^n_i Z^m_jZ^l_k-Z^l_i Z^n_jZ^m_k\cr\cr
&&\qquad +Z^n_i Z^l_jZ^m_k+Z^m_i Z^n_jZ^l_k)\varphi_l\phi_m\eta_n
\eea
When considering excited bound states, we now have many different possibilities for the open strings that can be attached, since the open string can start on any brane and terminate on any brane. Factors that attach strings to brane 1, brane 2 and brane 3 respectively, are given by
\bea
&&\bar{\varphi}^i Y_i^j\varphi_j\qquad
\bar{\varphi}^i\bar{\phi}^j (Z^k_i Y^l_j-Z^l_i Y^k_j)\varphi_k\phi_l\cr\cr
&&\bar{\varphi}^i\bar{\phi}^j\bar{\eta}^k 
(Z^l_i Z^m_jY^n_k-Z^m_i Z^l_jY^n_k-Z^n_i Z^m_jY^l_k-Z^l_i Z^n_jY^m_k\cr\cr
&&\qquad\qquad +Z^n_i Z^l_jY^m_k+Z^m_i Z^n_jY^l_k)\varphi_l\phi_m\eta_n
\eea
To describe strings that stretch between branes, we again use a product of the above factors and simply swap column indices. It is no longer necessary to attach pairs of strings. We can for example define a factor with three $Y$ fields, corresponding to a configuration in which a $Y$ stretches from the first to the second brane, another from the second to the third and the last from the third to the first. 

We can also attach strings that stretch between the two branes. To do this, we start with the product of a factor which attaches a string to the first brane and a factor which attaches a string to the second brane. From this we obtain a pair of strings stretched between the branes by permuting column indices, thereby swapping the endpoints of the two strings. The resulting generating function is again consistent with the Gauss Law arising from the world volume theory of the giant gravitons. Further, it is manifestly invariant under swapping strings of the same orientation. This again suggests that these operators are identical to the corresponding Gauss graph operators, something that we have checked in some examples. It is equally easy to consider bound states with even more AdS giant gravitons, or to consider bound states of more than two sphere giant gravitons.

The world volume dynamics of a bound state of $p$ giant gravitons is described in terms of the open string excitations of the branes \cite{Dai:1989ua}. At low energy this should become a non-Abelian Yang-Mills theory with gauge group $U(p)$ \cite{Witten:1995im}. Attempts to construct this emergent Yang-Mills theory starting from a description of the giant graviton bound state in conformal field theory have been given in \cite{Balasubramanian:2004nb,deCarvalho:2020pdp,deMelloKoch:2020agz,Su:2022ntc}. Although these initial results are encouraging, they are frustrated by the technical difficulty of the combinatorics. Perhaps the simpler description of giant graviton bound states obtained in this paper will allow progress. Specifically, a systematic ${1\over N}$ expansion, through a saddle point evaluation of (\ref{twointegralstodo}), should lead to ${1\over N}$ interactions which allow open strings to split and join. These interactions should then match those of the expected emergent Yang-Mills theory.

\begin{center} 
{\bf Acknowledgements}
\end{center}
RdMK is supported by the South African Research Chairs initiative of the Department of Science and Technology and the National Research Foundation. MK is supported by the National Research Foundation of Korea Grants funded by the Korea government (MSIT),  NRF-2020R1A6A1A03047877 (Center for Quantum Space Time) and NRF-2022R1I1A1A010. This work was initiated while RdMK was a participant of the KITP program ``Integrability in String, Field, and Condensed Matter Theory.'' This research was also supported in part by the National Science Foundation under Grant No. NSF PHY-1748958. We thank David Berenstein and Adolfo Holguin for useful discussions.

\begin{appendix}

\section{Check of the generating function for a bound state of two AdS giants}\label{2AdS}

In this Appendix we expand the generating function for the bound states of two AdS giant gravitons, in a power series in $t_1$ and $t_2$. This allows us to test their correctness since, for a small number of fields, the construction of the relevant Schur polynomials is entirely straightforward. 

\subsection{Operators constructed from the generating function}

\noindent
{\bf No open strings attached:} To start, consider the two AdS giant graviton bound state generating function, which is given by
\bea
Z_{\rm AdS}(t_1,t_2)&=&\sum^\infty_{p_1=p_2}\sum_{p_2=1}^\infty t_2^{2p_2}t_1^{p_1-p_2}{(p_1+p_2)!\over d_{(p_1,p_2)}p_2!(p_1-p_2)!}\chi_{(p_1,p_2)}(Z)\cr\cr
&=& {1\over \pi^{2N}}\int d^N\bar{\varphi}\int d^N\varphi\int d^N\bar{\phi}\int d^N\phi \cr\cr\cr
&&\qquad\qquad 
e^{-\bar{\varphi}^i\varphi_i-\bar{\phi}^i\phi_i+t_1\bar{\varphi}^iZ_i^j\varphi_j
+t_2^2\bar{\varphi}^i\bar{\phi}^j (Z^k_i Z^l_j-Z^l_i Z^k_j)\varphi_k\phi_l}
\label{formulatocheck}
\eea
where $d_{(p_1,p_2)}$ is the dimension of the symmetric group $S_{p_1+p_2}$ representation labelled by a Young diagram with $p_1$ boxes in the first row and $p_2$ boxes in the second row, $p_1\ge p_2$. After a simple computation we have
\bea
d_{(p_1,p_2)}&=&{(p_1+p_2)!(p_1-p_2+1)\over(p_1+1)!p_2!}\cr\cr
{\cal M}_{p_1,p_2}&=&{(p_1+p_2)!\over d_{(p_1,p_2)}p_2!(p_1-p_2)!}
\,\,\,=\,\,\,{(p_1+1)!\over (p_1-p_2+1)!}
\eea
Power series expanding the integrand of (\ref{formulatocheck}) in the parameters $t_1$ and $t_2$, we generate a sequence of polynomials in the variables $\bar{\varphi}^i$, $\varphi_i$, $\bar{\phi}^i$ and $\phi_i$. The resulting integrals are all easily performed using (\ref{bosonicresult}) and the obvious analogous formula for integrals over $\bar{\varphi}^i$ and $\varphi_i$. The result is
\bea
Z(t_1,t_2) &=& 1
+ t_1 \Tr Z
+ t_2^2 \left( \Tr(Z)^{2} - \Tr(Z^2) \right)
+ t_1 t_2^2 \left( \Tr(Z)^3 - \Tr(Z^3) \right) \cr \cr
&&+ {t_1^2\over 2} \left( \Tr(Z)^2 + \Tr(Z^2) \right)
+ {t_2^4 \over 2} \left( \Tr(Z)^4+3\Tr(Z^2)^2 - 4\Tr Z \Tr(Z^3) \right)
+ \dots\cr\cr
&&\label{resultofintegral}
\eea
In the first column below we show the value of ${\cal M}_{p_1,p_2}$, in the second we show the Schur polynomial $\chi_{(p_1,p_2)}(Z)$ and in the third we show the coefficient of $t_1^{p_1-p_2}t_2^{2p_2}$. The fact that the product of the first two entries on any row reproduces the third entry shows that the result (\ref{resultofintegral}) has indeed confirmed the formula (\ref{formulatocheck}).
\bea
\begin{array}{ccccc}
{\cal M}_{1,0}\,\,=\,\,1 & &\chi_{\tiny\yng(1)}=\Tr Z
& &\Tr Z\\
{\cal M}_{1,1}\,\,=\,\,2 & &\chi_{\tiny\yng(1,1)}={1\over 2}((\Tr Z)^2-\Tr Z^2)
& &\left( \Tr(Z)^{2} - \Tr(Z^2) \right)\\
{\cal M}_{2,1}\,\,=\,\,3 & &\chi_{\tiny\yng(2,1)}={1\over 3}((\Tr Z)^3-\Tr Z^3)
& &\left( \Tr(Z)^3 - \Tr(Z^3) \right)\\
{\cal M}_{2,0}\,\,=\,\,1 & &\chi_{\tiny\yng(2)}={1\over 2}((\Tr Z)^2+\Tr Z^2)
& &{1\over 2} \left( \Tr(Z)^2 + \Tr(Z^2) \right)\\
{\cal M}_{2,2}\,\,=\,\,6 & &\chi_{\tiny\yng(2,2)}={((\Tr Z)^4-4\Tr Z\Tr Z^3+3(\Tr Z^2)^2)\over 12}
& &{1\over 2} \left( \Tr(Z)^4+3\Tr(Z^2)^2 - 4\Tr Z \Tr(Z^3) \right)
\end{array}
\eea

This example nicely illustrates how the checks that follow will be performed. We will not however quote the explicit formulas for all of the operators we produce, since these become rather lengthy expressions that are not very illuminating.

{\vskip 0.5cm}

\noindent
{\bf Two open strings attached, to the giant of momentum $p_1$:} Next consider the generating function obtained when we attach two open strings to the giant graviton with momentum $p_1$. We employ the description in which each open string is described with a word. The relevant generating function is
\bea
Z(t_1,t_2) &=& {1\over \pi^{2N}}\int d^N\bar{\varphi}\int d^N\varphi\int d^N\bar{\phi}\int d^N\phi\,\,e^{-\bar{\varphi}^i\varphi_i-\bar{\phi}^i\phi_i+t_2^2\bar{\varphi}^i\bar{\phi}^j (Z^k_i Z^l_j-Z^l_i Z^k_j)\varphi_k\phi_l}\cr\cr
&&\quad\times\,\,e^{t_1\bar{\varphi}^iZ_i^j\varphi_j}\,\,\bar{\varphi}^{b_1}(W_{A_1})^{a_1}_{b_1}\varphi_{a_1}\,\,\bar{\varphi}^{b_2}(W_{A_2})^{a_2}_{b_2}\varphi_{a_2}
\eea
Expanding to second order in $t_1$ and $t_2^2$ we obtain
\bea
Z(t_1,t_2)&=&a_0+a_1t_1+a_2t_2^2+t_1^2a_3+t_1t_2^2 a_4+\frac{t_2^4}{2}a_5+ \dots
\label{fstexp}
\eea
where\footnote{When writing $Z^i_j$ the upper $i$ index is a row index and the lower $j$ index is a column index.}, for example
\bea
a_0 &=&\Tr(W_{A_1}) \Tr(W_{A_2})+\Tr(W_{A_1}W_{A_2})\cr\cr
a_1&=&\Tr(Z)\Tr(W_{A_1}) \Tr(W_{A_2}) +\Tr(Z)\Tr(W_{A_1}W_{A_2})+\Tr(ZW_{A_1})\Tr(W_{A_2})\cr\cr    
&+&\Tr(W_{A_1}) \Tr(Z W_{A_2})+\Tr(W_{A_1}W_{A_2}Z)+\Tr(W_{A_1}Z W_{A_2})\label{frstbsgf}
\eea

\noindent
{\bf Two open strings attached, to the giant of momentum $p_2$:} Next consider the generating function obtained when we attach two open strings to the giant graviton with momentum $p_2$. The relevant generating function is
\bea
Z(t_1,t_2) &=& {1\over \pi^{2N}}\int d^N\bar{\varphi}\int d^N\varphi\int d^N\bar{\phi}\int d^N\phi\,\,e^{-\bar{\varphi}^i\varphi_i-\bar{\phi}^i\phi_i+t_2^2\bar{\varphi}^i\bar{\phi}^j (Z^k_i Z^l_j-Z^l_i Z^k_j)\varphi_k\phi_l}\cr\cr
&&\quad\times\,\,e^{t_1\bar{\varphi}^iZ_i^j\varphi_j}\,\,(\bar{\varphi}\cdot Z\cdot \varphi\,\bar{\phi}\cdot W_{B_1}\cdot \phi-\bar{\varphi}\cdot Z\cdot\phi\, \bar{\phi}\cdot W_{B_1}\cdot\varphi)\cr\cr
&&\quad\times\,\, (\bar{\varphi}\cdot Z\cdot \varphi\,\bar{\phi}\cdot W_{B_2}\cdot \phi-\bar{\varphi}\cdot Z\cdot\phi\, \bar{\phi}\cdot W_{B_2}\cdot\varphi)
\eea
Expanding to second order in $t_1$ and $t_2^2$ we obtain
\bea
Z(t_1,t_2)&=&b_0+b_1t_1+b_2t_2^2+t_1^2b_3+t_1t_2^2 b_4+\frac{t_2^4}{2}b_5+ \dots
\label{scndexp}
\eea
where, for example
\bea
b_0&=&\Tr(W_{B_1}) \Tr(W_{B_2}) \Tr(Z)^2 + \Tr(Z)^2 \Tr(W_{B_1}W_{B_2}) - \Tr(W_{B_2}) \Tr(Z) \Tr(W_{B_1}Z)\cr\cr
&&- \Tr(W_{B_1}) \Tr(Z) \Tr(W_{B_2}Z) + 2 \Tr(W_{B_1}Z) \Tr(W_{B_2}Z) + \Tr(W_{B_1}) \Tr(W_{B_2}) \Tr(Z^2)\cr\cr
&&+\Tr(W_{B_1}W_{B_2}) \Tr(Z^2) - \Tr(Z) \Tr(W_{B_1}W_{B_2}Z) - \Tr(W_{B_2}) \Tr(W_{B_1}Z^2)- \Tr(W_{B_2}W_{B_1}Z^2)\cr\cr 
&&-\Tr(Z) \Tr(W_{B_2}W_{B_1}Z) - \Tr(W_{B_1}) \Tr(W_{B_2}Z^2) - \Tr(W_{B_1}W_{B_2}Z^2) + 2 \Tr(W_{B_1}ZW_{B_2}Z)\cr\cr
b_1&=&\Tr(W_{B_1}) \Tr(W_{B_2}) \Tr(Z)^3 + \Tr(Z)^3 \Tr(W_{B_1}W_{B_2}) - \Tr(W_{B_2}) \Tr(Z)^2 \Tr(W_{B_1}Z)\cr\cr 
&&-\Tr(W_{B_1}) \Tr(Z)^2 \Tr(W_{B_2}Z) + 2 \Tr(Z) \Tr(W_{B_1}Z) \Tr(W_{B_2}Z) + 
 3 \Tr(W_{B_1}) \Tr(W_{B_2}) \Tr(Z) \Tr(Z^2)\cr\cr 
&&+ 3 \Tr(Z) \Tr(W_{B_1}W_{B_2}) \Tr(Z^2) - 
 \Tr(W_{B_2}) \Tr(W_{B_1}Z) \Tr(Z^2) - \Tr(W_{B_1}) \Tr(W_{B_2}Z) \Tr(Z^2)\cr\cr 
&&-\Tr(Z)^2 \Tr(W_{B_1}W_{B_2}Z) - \Tr(Z^2) \Tr(W_{B_1}W_{B_2}Z) - 
 2 \Tr(W_{B_2}) \Tr(Z) \Tr(W_{B_1}Z^2)\cr\cr 
&&+ 2 \Tr(W_{B_2}Z) \Tr(W_{B_1}Z^2) -  \Tr(Z)^2 \Tr(W_{B_2}W_{B_1}Z) - \Tr(Z^2) \Tr(W_{B_2}W_{B_1}Z)\cr\cr 
&&- 2 \Tr(W_{B_1}) \Tr(Z) \Tr(W_{B_2}Z^2) + 2 \Tr(W_{B_1}Z) \Tr(W_{B_2}Z^2) + 
 2 \Tr(W_{B_1}) \Tr(W_{B_2}) \Tr(Z^3)\cr\cr 
&&+ 2 \Tr(W_{B_1}W_{B_2}) \Tr(Z^3) - 2 \Tr(Z) \Tr(W_{B_1}W_{B_2}Z^2) + 2 \Tr(Z) \Tr(W_{B_1}ZW_{B_2}Z)\cr\cr 
&&- 2 \Tr(W_{B_2}) \Tr(W_{B_1}Z^3) - 2 \Tr(Z) \Tr(W_{B_2}W_{B_1}Z^2) - 
 2 \Tr(W_{B_1}) \Tr(W_{B_2}Z^3)\cr\cr 
&&- 2 \Tr(W_{B_1}W_{B_2}Z^3) + 2 \Tr(W_{B_1}ZW_{B_2}Z^2) - 
 2 \Tr(W_{B_2}W_{B_1}Z^3)\cr\cr 
&&+ 2 \Tr(W_{B_2}ZW_{B_1}Z^2)
\eea

\noindent
{\bf A pair of strings attached, one to each giant:} The generating function for two strings attached, one to each of the two giants in the bound state, again employing a description in which each open string is described with a word, is given by
\bea
Z(t_1,t_2) &=& {1\over \pi^{2N}}\int d^N\bar{\varphi}\int d^N\varphi\int d^N\bar{\phi}\int d^N\phi\,\,e^{-\bar{\varphi}^i\varphi_i-\bar{\phi}^i\phi_i+t_2^2\bar{\varphi}^i\bar{\phi}^j (Z^k_i Z^l_j-Z^l_i Z^k_j)\varphi_k\phi_l+t_1\bar{\varphi}^iZ_i^j\varphi_j}\cr\cr
&&\times\,\,\,\bar{\varphi}\cdot W_A\cdot \varphi\,\,\,(\bar{\varphi}\cdot Z\cdot \varphi\,\bar{\phi}\cdot W_B\cdot \phi-\bar{\varphi}\cdot Z\cdot\phi\, \bar{\phi}\cdot W_B\cdot\varphi)
\eea
Again expanding to second order in $t_1$ and $t_2$ we obtain
\bea
Z(t_1,t_2)&=& c_0+t_1 c_1+t_2^2 c_2+t_1^2 c_3+t_1 t_2^2 c_4+t_2^4 c_5+\cdots
\label{oneoneach}
\eea
where, for example
\bea
c_0&=&\Tr(W_A)\Tr(W_B)\Tr(Z)-\Tr(W_B)\Tr(W_A Z)+\Tr(W_A)\Tr(W_B Z)-\Tr(W_AW_BZ)\cr\cr
c_1&=&\Tr(W_A)\Tr(W_B)\Tr(Z)^2-\Tr(W_B)\Tr(Z)\Tr(W_AZ)+2\Tr(W_A)\Tr(Z)\Tr(W_BZ)\cr\cr
&&-\Tr(W_AZ)\Tr(W_BZ) + \Tr(W_A)\Tr(W_B)\Tr(Z^2) -\Tr(Z)\Tr(W_AW_BZ)\cr\cr
&&-\Tr(W_B) \Tr(W_AZ^2)+2\Tr(W_A)\Tr(W_BZ^2)-\Tr(W_AW_BZ^2)-\Tr(W_AZW_BZ)\cr\cr
&&\label{diffgntstringresults}
\eea

\noindent
{\bf A pair of strings stretched between the giants:} Finally, consider the generating function for two strings stretched between the two giants in the bound state, again employing a description in which each open string is described with a word
\bea
Z(t_1,t_2) &=& {1\over \pi^{2N}}\int d^N\bar{\varphi}\int d^N\varphi\int d^N\bar{\phi}\int d^N\phi\,\,e^{-\bar{\varphi}^i\varphi_i-\bar{\phi}^i\phi_i+t_2^2\bar{\varphi}^i\bar{\phi}^j (Z^k_i Z^l_j-Z^l_i Z^k_j)\varphi_k\phi_l+t_1\bar{\varphi}^iZ_i^j\varphi_j}\cr\cr
&&\times\,\,(\bar{\varphi}\cdot Z\cdot \varphi\,\bar{\phi}\cdot W_C\cdot \varphi \bar{\varphi}\cdot W_D\cdot \phi-\bar{\varphi}\cdot Z\cdot\phi\, \bar{\phi}\cdot W_C\cdot\varphi\bar{\varphi}\cdot W_D\cdot \varphi)
\eea
Again expanding to second order in $t_1$ and $t_2$ we obtain
\bea
Z(t_1,t_2)&=& s_0+t_1 s_1+t_2^2 s_2+t_1^2 s_3+t_1 t_2^2 s_4+t_2^4 s_5+\cdots
\eea
where, for example
\bea
s_0&=&\Tr(Z)\Tr(W_CW_D)-\Tr(W_D)\Tr(W_CZ)-\Tr(W_CW_DZ) +\Tr(W_DW_CZ)\cr\cr
s_1&=&\Tr(Z)^2 \Tr(W_CW_D)-\Tr(W_D)\Tr(Z)\Tr(W_CZ)-\Tr(W_CZ)\Tr(W_DZ)+ 
 \Tr(W_CW_D)\Tr(Z^2)\cr\cr 
&&- \Tr(Z) \Tr(W_C W_D Z) - \Tr(W_D) \Tr(W_C Z^2) + 
 2 \Tr(Z) \Tr(W_D W_C Z) - \Tr(W_C W_DZ^2)\cr\cr 
&&-\Tr(W_C Z W_D Z) + 2 \Tr(W_D W_C Z^2)
\label{strecthedstringresults}
\eea

\subsection{Comparison to restricted Schur polynomial and Gauss graph operators}

The formulas derived above, in terms of open string words $W_*$, are extremely useful. As written, they describe the operators obtained when each open string excitation is described using a word. These are the operators we use in Section \ref{actionofD} to determine the lattice Hamiltonians describing the dynamics of open strings attached to giant gravitons. Alternatively, we can set all open string words equal to a single letter $W_*=Y$ and compare to the Gauss graph operators. It is this second exercise that we will carry out in this Appendix. 

To construct the Gauss graph operators, we need to construct the restricted Schur polynomials. There are a number of simplifications that arise when we consider restricted Schur polynomial operators labeled by a single row (this is the case for the coefficients $a_0$, $a_1$ and $a_3$ in (\ref{fstexp})). In this case the Gauss graph operators and the restricted Schur polynomial bases coincide. Further, since the representation $R$ is one dimensional, the subspace we restrict to when defining the restricted characters is not a proper subspace and the restricted characters simply reduce to ordinary characters. In this case it is simple to check the following equalities
\bea
\begin{array}{ccc}
a_0 &=\,\,2\chi_{\tiny \yng(2) \big(\cdot,\yng(2)\big)}(Z,Y) &= \left(\cdot \,\,,
\begin{gathered}\includegraphics[scale=0.4]{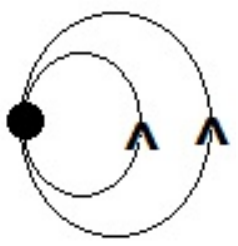}\end{gathered}\right)\cr\cr
a_1 &=\,\,2\chi_{\tiny \yng(3) \big(\yng(1),\yng(2)\big)}(Z,Y) &=\left({\tiny\yng(1)}\,\,, \begin{gathered}\includegraphics[scale=0.4]{GG1}\end{gathered}\right)\cr\cr
a_3 &=\,\,2\chi_{\tiny \yng(4) \big(\yng(2),\yng(2)\big)}(Z,Y) &=\left({\tiny\yng(2)}\,\,,\begin{gathered}\includegraphics[scale=0.4]{GG1}\end{gathered}\right)
\end{array}\label{firstcomp}
\eea
The equality above is between (1) the coefficient $a_*$ appearing in (\ref{frstbsgf})), (2) the restricted Schur polynomial with three Young diagram labels and (3) the Gauss graph operator which has both a Young diagram label and a graph label. The graph is the same in all three cases because in all three we have a single brane with two strings attached to it. The equalities above are explicit checks of the equality between Gauss graph operators with all strings attached to a single brane, and our generating function for excited bound states with all excitations associated to a given row. Here we are, of course, using the fact that these equalities are meant to be exact and not just large $N$ identities.

The remaining terms in (\ref{fstexp}) should be compared to Gauss graph operators constructed from restricted Schur polynomials that are labelled by Young diagrams that have two rows. The two row restricted Schur polynomial is still simpler than the general case, since with only two rows we do not need to introduce multiplicity labels. However, the representation $R$ is no longer one dimensional and the subspace we restrict to when defining the restricted characters is usually a proper subspace. Thus, the restricted characters are not equal to ordinary characters, and they must be computed. Since we are attaching both strings to a single brane, it is again true that, for these operators, the Gauss graph operators and the restricted Schur polynomial bases coincide. This is again an exact statement. The restricted Schur polynomials that we need are given by
\bea
\chi_{\tiny \yng(3,1) \big(\yng(1,1),\yng(2)\big)}(Z,Y)&=&
{1\over 4}\Big(\Tr(Z)^2\Tr(Y)^2+\Tr(Y^2)\Tr(Z)^2+2\Tr(YZ)\Tr(Z)\Tr(Y)\cr
&&-\Tr(Z^2)\Tr(Y)^2-\Tr(Y^2)\Tr(Z^2)-2\Tr(YZ^2)\Tr(Y)\cr
&&+2\Tr(ZY^2)\Tr(Z)-2\Tr(Y^2Z^2)\Big)\cr\cr
\chi_{\tiny \yng(4,1) \big(\yng(2,1),\yng(2)\big)}(Z,Y)&=&
{1\over 6}\Big(\Tr(Z)^3\Tr(Y)^2+4\Tr(Z)^2\Tr(Y)\Tr(ZY)
+4\Tr(Z)^2\Tr(ZY^2)\cr\cr
&&+2\Tr(Z)\Tr(Y)\Tr(Z^2Y)+\Tr(Z)^3\Tr(Y^2)+2\Tr(Z)\Tr(ZY)^2\cr\cr
&&+2\Tr(Z)\Tr(Z^2Y^2)+2\Tr(Z)\Tr(ZYZY)-2\Tr(YZ)\Tr(Y)\Tr(Z^2)\cr\cr
&&-\Tr(Y)^2\Tr(Z^3)-\Tr(Y^2)\Tr(Z^3)-2\Tr(ZY)\Tr(Z^2Y)\cr\cr
&&-2\Tr(Z^2)\Tr(ZY^2)-4\Tr(Z^3Y)\Tr(Y)-4\Tr(Y^2Z^3)\cr\cr
&&-2\Tr(YZYZ^2)\Big)\cr\cr
\chi_{\tiny \yng(4,2) \big(\yng(2,2),\yng(2)\big)}(Z,Y) &=&{1\over 24} \Tr (Y)^2 \Tr (Z)^4 + {1\over 24} \Tr (Z)^4 \Tr (Y^2)+{1\over 6} \Tr (Y)\Tr (Z)^3 \Tr (YZ)\cr\cr
&&+{1\over 12} \Tr (Z)^2 \Tr (YZ)^2 + {1\over 12} \Tr (YZ)^2 \Tr (Z^2)+{1\over 8} \Tr (Y)^2 \Tr (Z^2)^2\cr\cr 
&&+ {1\over 8} \Tr (Y^2) \Tr (Z^2)^2+{1\over 6} \Tr (Z)^3 \Tr (Y^2Z) - 
{1\over 6} \Tr (Z) \Tr (YZ) \Tr (YZ^2) \cr\cr
&&+ {1\over 2} \Tr (Y) \Tr (Z^2) \Tr (YZ^2) + 
 {1\over 6} \Tr (YZ^2)^2 - {1\over 6} \Tr (Y)^2 \Tr (Z) \Tr (Z^3)\cr\cr 
&&- {1\over 6} \Tr (Z) \Tr (Y^2) \Tr (Z^3) - {1\over 6} \Tr (Y) \Tr (YZ) \Tr (Z^3) - 
 {1\over 6} \Tr (Y^2Z) \Tr (Z^3)\cr\cr 
&&+ {1\over 2} \Tr (Z^2) \Tr (Y^2Z^2) + 
 {1\over 12} \Tr (Z)^2 \Tr (YZYZ) + {1\over 12} \Tr (Z^2) \Tr (YZYZ)\cr\cr 
&&-{1\over 2} \Tr (Y) \Tr (Z) \Tr (YZ^3) - {1\over 6} \Tr (YZ) \Tr (YZ^3) - 
 1/2 \Tr (Z) \Tr (Y^2Z^3)\cr\cr 
&&- {1\over 6} \Tr (Z) \Tr (YZYZ^2)-{1\over 6}\Tr(YZYZ^3) 
+ {1\over 6} \Tr (YZ^2YZ^2)
\eea
It is now a simple exercise to check the following equalities
\bea
\begin{array}{ccc}
a_2 &=\,\,\,\,4\chi_{\tiny \yng(3,1) \big(\yng(1,1),\yng(2)\big)}(Z,Y) &=\,\,\,\, \left({\tiny \yng(1,1)}\,\,,
\begin{gathered}\includegraphics[scale=0.4]{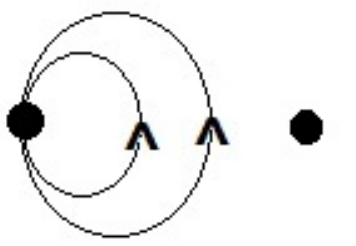}\end{gathered}\right)\cr\cr
a_4 &=\,\,\,\,6\chi_{\tiny \yng(4,1) \big(\yng(2,1),\yng(2)\big)}(Z,Y) &=\,\,\,\,\left({\tiny\yng(2,1)}\,\,, \begin{gathered}\includegraphics[scale=0.4]{GG2}\end{gathered}\right)\cr\cr
a_5 &=\,\,\,\,12\chi_{\tiny \yng(4,2) \big(\yng(2,2),\yng(2)\big)}(Z,Y) &=\,\,\,\,\left({\tiny\yng(2,2)}\,\,,\begin{gathered}\includegraphics[scale=0.4]{GG2}\end{gathered}\right)
\end{array}\label{secondcomp}
\eea
after setting $W_1=W_2=Y$ in $a_2$, $a_4$ and $a_5$.

It is equally easy to test the results for strings attached to the second giant graviton, given in (\ref{scndexp}). In this case, again because all strings are attached to a single giant, there is an exact equality between the restricted Schur polynomials and the Gauss graph operators. We have tested the equalities
\bea
\begin{array}{ccc}
b_0 &=\,\,\,\,4\,\chi_{\tiny \yng(2,2) \big(\yng(2),\yng(2)\big)}(Z,Y) &=\,\,\,\, \left({\tiny \yng(2)}\,\,,
\begin{gathered}\includegraphics[scale=0.4]{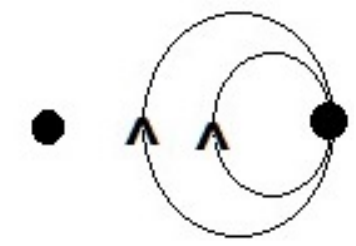}\end{gathered}\right)\cr\cr
b_1 &=\,\,\,\,12\,\chi_{\tiny \yng(3,2) \big(\yng(3),\yng(2)\big)}(Z,Y) &=\,\,\,\,\left({\tiny\yng(3)}\,\,, \begin{gathered}\includegraphics[scale=0.4]{GG4}\end{gathered}\right)\cr\cr
b_2 &=\,\,\,\,16\,\chi_{\tiny \yng(3,3) \big(\yng(3,1),\yng(2)\big)}(Z,Y) &=\,\,\,\,\left({\tiny\yng(3,1)}\,\,,\begin{gathered}\includegraphics[scale=0.4]{GG4}\end{gathered}\right)\cr\cr
b_3 &=\,\,\,\,24\,\chi_{\tiny \yng(4,2) \big(\yng(4),\yng(2)\big)}(Z,Y) &=\,\,\,\,\left({\tiny\yng(4)}\,\,,\begin{gathered}\includegraphics[scale=0.4]{GG4}\end{gathered}\right)
\end{array}
\eea
after setting $W_1=W_2=Y$ in $b_0$, $b_1$, $b_2$ and $b_3$.

Testing the results when there is a string attached to each giant graviton is also interesting. In this case we gave a proof that the two sets of operators agree at large $N$, but have not proved an exact statement. These results are given in (\ref{oneoneach}). For this class of operators it is no longer true that the Gauss graph operators and the restricted Schur polynomials coincide. Recall that from (\ref{GGmanip}) we have
\bea
O(\sigma)&=&{1\over\prod_i m_i!}\sum_{s\vdash m}\sum_{\alpha\in \prod_i S_{m_i}}\sqrt{d_s}\chi_{s}(\sigma \alpha) O_{R,(r,s)}
\eea
where $\chi_{s}(\sigma )$ is the character of $\sigma\in S_m$ in irrep $s$. Since we only have a single $Y$ box in each row, the groups $S_{m_i}$ are trivial we have the lovely simple formula
\bea
  O(\sigma)  &=&\sum_{s\vdash m}\sqrt{d_s}\chi_{s}(\sigma) O_{R,(r,s)}
\label{finalGG}
\eea
To obtain the state with a string attached to each of the two branes we need to choose $\sigma$ equals the identity in (\ref{finalGG}). We then have $\chi_{s}({\bf 1})=d_s$ and hence 
\bea
  O( {\bf 1} )&=&O_{R,(r,{\tiny\yng(2)})}+O_{R,(r,{\tiny\yng(1,1)})}
\label{finalfinalGG}
\eea
Using this formula, we can test the polynomials defined in (\ref{oneoneach}). A subset of the restricted Schur polynomials that we need are
\bea
\chi_{{\tiny\yng(2,1),(\yng(1),\yng(2))}}&=&{1\over 2!}\Big(
\Tr(Y)^2\Tr(Z)+\Tr(Y^2)\Tr(Z)-\Tr(Y)\Tr(YZ)-\Tr(Y^2Z)\Big)\cr\cr
\chi_{{\tiny\yng(2,1),(\yng(1),\yng(1,1))}}&=&{1\over 2!}\Big(
\Tr(Y)^2\Tr(Z)-\Tr(Y^2)\Tr(Z)+\Tr(Y)\Tr(YZ)-\Tr(Y^2Z)\Big)\cr\cr
\chi_{{\tiny\yng(3,1),(\yng(2),\yng(2))}}&=&{1\over 4}\Big(
\Tr(Z^2)\Tr(Y^2)+\Tr(Z^2)\Tr(Y)^2+\Tr(Y^2)\Tr(Z)^2\cr\cr
&&-2\Tr(YZYZ)+\Tr(Y^2)\Tr(Z^2)- 2\Tr(YZ)^2\Big)\cr\cr
\chi_{{\tiny\yng(3,1),(\yng(2),\yng(1,1))}}&=&{1\over 4}\Big(
\Tr(Z)^2\Tr(Y)^2+\Tr(Z^2)\Tr(Y)^2-\Tr(Y^2)\Tr(Z)^2\cr\cr
&&-2\Tr(Y^2Z^2)-\Tr(Y^2)\Tr(Z^2)+2\Tr(YZ^2)\Tr(Y)\cr\cr
&&- 2\Tr(ZY^2)\Tr(Z)+2\Tr(YZ)\Tr(Z)\Tr(Y)\Big)\cr\cr
&&
\eea

We have tested the equalities
\bea
\begin{array}{ccc}
c_0 &=\,\,\,\,\chi_{\tiny \yng(2,1) \big(\yng(1),\yng(2)\big)}(Z,Y)+\chi_{\tiny \yng(2,1) \big(\yng(1),\yng(1,1)\big)}(Z,Y) &=\,\,\,\, \left({\tiny \yng(1)}\,\,,
\begin{gathered}\includegraphics[scale=0.4]{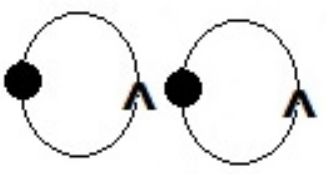}\end{gathered}\right)\cr\cr
c_1 &=\,\,\,\,\chi_{\tiny \yng(3,1) \big(\yng(2),\yng(2)\big)}(Z,Y)+\chi_{\tiny \yng(3,1) \big(\yng(2),\yng(1,1)\big)}(Z,Y) &=\,\,\,\, \left({\tiny \yng(2)}\,\,,
\begin{gathered}\includegraphics[scale=0.4]{GG5}\end{gathered}\right)\cr\cr
c_2 &=\,\,\,\,3\,\chi_{\tiny \yng(3,2) \big(\yng(2,1),\yng(2)\big)}(Z,Y)+3\,\chi_{\tiny \yng(3,2) \big(\yng(2,1),\yng(1,1)\big)}(Z,Y) &=\,\,\,\, \left({\tiny \yng(2,1)}\,\,,
\begin{gathered}\includegraphics[scale=0.4]{GG5}\end{gathered}\right)\cr\cr
c_3 &=\,\,\,\,{3\over 2}\,\chi_{\tiny \yng(4,1) \big(\yng(3),\yng(2)\big)}(Z,Y)+{3\over 2}\,\chi_{\tiny \yng(4,1) \big(\yng(3),\yng(1,1)\big)}(Z,Y) &=\,\,\,\, \left({\tiny \yng(3)}\,\,,
\begin{gathered}\includegraphics[scale=0.4]{GG5}\end{gathered}\right)\cr\cr
c_4 &=\,\,\,\,8\,\chi_{\tiny \yng(4,2) \big(\yng(3,1),\yng(2)\big)}(Z,Y)+8\,\chi_{\tiny \yng(4,2) \big(\yng(3,1),\yng(1,1)\big)}(Z,Y) &=\,\,\,\, \left({\tiny \yng(3,1)}\,\,,
\begin{gathered}\includegraphics[scale=0.4]{GG5}\end{gathered}\right)\end{array}
\label{CoolMatch}
\eea
after setting $W_1=W_2=Y$ in $c_0$, $c_1$, $c_2$, $c_3$ and $c_4$. The fact that we obtain exact agreement is not expected, suggesting we should attempt to extend the proof given in Appendix \ref{ggequality} to finite $N$.

We would now like to test the results for the stretched string operators, summarized in (\ref{strecthedstringresults}) above. Since the operators with a string on each brane are related to the stretched string operators by the action of the graph changing operator, the results (\ref{CoolMatch}) suggest we should again find exact agreement. To obtain the state with strings stretching between the two branes we need to choose $\sigma=(1,2)$ in (\ref{finalGG}). Thus, in the end we have
\bea
  O( (1,2) )&=&O_{R,(r,{\tiny\yng(2)})}-O_{R,(r,{\tiny\yng(1,1)})}
\label{finalfinalGG2}
\eea
Using this formula, we have argued for and tested the equalities
\bea
\begin{array}{ccc}
s_0 &=\,\,\,\,\chi_{\tiny \yng(2,1) \big(\yng(1),\yng(2)\big)}(Z,Y)-\chi_{\tiny \yng(2,1) \big(\yng(1),\yng(1,1)\big)}(Z,Y) &=\,\,\,\, \left({\tiny \yng(1)}\,\,,
\begin{gathered}\includegraphics[scale=0.4]{GG3}\end{gathered}\right)\cr\cr
s_1 &=\,\,\,\,\chi_{\tiny \yng(3,1) \big(\yng(2),\yng(2)\big)}(Z,Y)-\chi_{\tiny \yng(3,1) \big(\yng(2),\yng(1,1)\big)}(Z,Y) &=\,\,\,\, \left({\tiny \yng(2)}\,\,,
\begin{gathered}\includegraphics[scale=0.4]{GG3}\end{gathered}\right)\cr\cr
s_2 &=\,\,\,\,3\,\chi_{\tiny \yng(3,2) \big(\yng(2,1),\yng(2)\big)}(Z,Y)-3\,\chi_{\tiny \yng(3,2) \big(\yng(2,1),\yng(1,1)\big)}(Z,Y) &=\,\,\,\, \left({\tiny \yng(2,1)}\,\,,
\begin{gathered}\includegraphics[scale=0.4]{GG3}\end{gathered}\right)\cr\cr
s_3 &=\,\,\,\,{3\over 2}\,\chi_{\tiny \yng(4,1) \big(\yng(3),\yng(2)\big)}(Z,Y)-{3\over 2}\,\chi_{\tiny \yng(4,1) \big(\yng(3),\yng(1,1)\big)}(Z,Y) &=\,\,\,\, \left({\tiny \yng(3)}\,\,,
\begin{gathered}\includegraphics[scale=0.4]{GG3}\end{gathered}\right)\cr\cr
s_4 &=\,\,\,\,8\,\chi_{\tiny \yng(4,2) \big(\yng(3,1),\yng(2)\big)}(Z,Y)-8\,\chi_{\tiny \yng(4,2) \big(\yng(3,1),\yng(1,1)\big)}(Z,Y) &=\,\,\,\, \left({\tiny \yng(3,1)}\,\,,
\begin{gathered}\includegraphics[scale=0.4]{GG3}\end{gathered}\right)\end{array}
\eea
after setting $W_1=W_2=Y$ in $s_0$, $s_1$, $s_2$, $s_3$ and $s_4$.

Our final comparison is between operators for which the open string excitations are described using open string words. A simple example is given by attaching two strings to the first brane and setting $p_1=p_2=1$. As above, each row of $R$ is, of course, not a giant graviton, but this simple example allows us to compare the construction of \cite{deMelloKoch:2007rqf} with the construction developed above. From the above construction this operator is given as the coefficient of $t_2^2$ in the power series expansion of the expression
\bea
&&\int d^N\bar{\varphi}\int d^N\varphi\int d^N\bar{\phi}\int d^N\phi\,\,{e^{-\bar{\varphi}\cdot\varphi-\bar{\phi}\cdot\phi}\over \pi^{2N}}\,\,e^{t_1\bar{\varphi}\cdot Z\cdot\varphi+t_2^2\bar{\varphi}\cdot Z\cdot\varphi \bar{\phi}\cdot Z\cdot\phi - t_2^2\bar{\varphi}\cdot Z\cdot\phi \bar{\phi}\cdot Z\cdot\varphi}\prod_{r=1}^2\bar{\varphi}\cdot W_{A_r}\cdot\varphi\cr\cr
&&
\eea
The result is
\bea
O_{11}(Z,W_{A_1},W_{A_2})&=&
\Tr (Z)^2\Tr (W_{A_1})\Tr (W_{A_2})-\Tr (Z^2)\Tr (W_{A_1})\Tr (W_{A_2})
+\Tr (ZW_{A_1})\Tr (Z)\Tr (W_{A_2})\cr\cr
&&+\Tr (ZW_{A_2})\Tr (Z)\Tr (W_{A_1})+\Tr (Z)^2\Tr (W_{A_1}W_{A_2})
-\Tr (Z^2)\Tr (W_{A_1}W_{A_2})\cr\cr
&&-\Tr (Z^2W_{A_1})\Tr (W_{A_2})-\Tr (Z^2W_{A_2})\Tr (W_{A_1})
+\Tr (ZW_{A_1}W^{A_2})\Tr (Z)\cr\cr
&&+\Tr (Z W_{A_2}W^{A_1})\Tr (Z)-\Tr (Z^2W^{A_1}W^{A_2})
-\Tr (Z^2 W^{A_2}W_{A_1})\label{2stringsattached}
\eea
We want to compare this to the construction given in \cite{deMelloKoch:2007rqf}, which makes use of projection operators. For the operator that we are studying, we define a representation $R_1''$ by removing two boxes from representation $R$, as follows
\bea
R={\small \yng(3,1)}\quad\longrightarrow\quad {\small \yng(2,1)}
\quad\longrightarrow\quad R''_1={\small \yng(1,1)}
\eea
We use the notation $R''_1$ to match the notation of Appendix C1 of \cite{deMelloKoch:2007rqf}. The projection operator that acts within the carrier space of $R$ and projects from $R$ to $R_1''$ is denoted by $P_{R,R_1''}$. In terms of this projector, the giant bound state with two strings attached is given by
\bea
\chi_{R,R_1''}(Z,W_{A_1},W_{A_2})={1\over 2}\sum_{\sigma\in S_4}Z^{i_1}_{i_{\sigma(1)}}Z^{i_2}_{i_{\sigma(2)}}
(W_{A_1})^{i_3}_{i_{\sigma(3)}}(W_{A_2})^{i_4}_{i_{\sigma(4)}}{\rm Tr} (P_{R,R_1''}\Gamma_R(\sigma) )
\eea
Using the results of Appendix C1 of \cite{deMelloKoch:2007rqf} and the explicit expression (\ref{2stringsattached}) it is simple to verify that
\bea
O(Z,W_{A_1},W_{A_2})&=&2\chi_{R,R_1''}(Z,W_{A_1},W_{A_2})
\eea
so that our generating function has correctly reproduced the restricted polynomial constructed in \cite{deMelloKoch:2007rqf}. In Section \ref{excitedAdS} our arguments for the equality of restricted Schur polynomials and the operators obtained from the generating function made extensive use of the derivative rule proved in \cite{deMelloKoch:2007rqf}. It is worth reviewing this rule with the formula (\ref{2stringsattached}) in hand. Towards this end, we employ a convenient graphical notation (for this example, when defining the restrictions, string $W_{A_1}$ is removed first and then $W_{A_2}$ is removed) as follows
\bea
\chi_{R,R_1''}(Z,W_{A_1},W_{A_2})=
\chi_{\tiny\young({\,}21,{\,})}
\eea
More details about this notation are available in \cite{deMelloKoch:2007rqf}. For this example, the derivative rule of \cite{deMelloKoch:2007rqf} states that
\bea
{\partial\over\partial (W_{A_1})^a_a}\chi_{\tiny\young({\,}21,{\,})}
=(N+2)\chi_{\tiny\young({\,}2,{\,})}\qquad\qquad
{\partial\over\partial (W_{A_2})^a_a}\chi_{\tiny\young({\,}2,{\,})}
=(N+1)\chi_{\tiny\young({\,},{\,})}
\eea
It is a simple matter to verify these formulas using (\ref{2stringsattached}).

Finally, consider the case that a pair of strings are stretched between the two giant gravitons. Giant graviton bound states with a single pair of stretched strings are generated by the integral
\bea
&&\int d^N\bar{\varphi}\int d^N\varphi\int d^N\bar{\phi}\int d^N\phi\,\,{e^{S_0}\over \pi^{2N}}\,\,e^{S_1}\,\,(\bar{\varphi}\cdot Z\cdot\varphi\,\bar{\phi}\cdot W_D\cdot \varphi\,\bar{\varphi}\cdot W_C\cdot \phi\cr\cr
&&\qquad\qquad\qquad -\bar{\varphi}\cdot Z\cdot \phi\,\bar{\phi}\cdot W_D\cdot \varphi\,\bar{\varphi}\cdot W_C\cdot\varphi)
\nonumber
\eea
The term independent of $t_1$ and $t_2$ in the power series expansion of integral is
\bea
O_{ss}=\Tr (Z)\Tr(W_CW_D)+\Tr(ZW_{C}W_{D})-\Tr(ZW_D)\Tr(W_C)-\Tr(ZW_DW_C)\label{strstr}
\eea
which is in precise agreement with the construction of \cite{deMelloKoch:2007rqf}. To see this, note that the restricted character needed to construct the restricted Schur polynomial is given by a single matrix element (we use the Young's orthonormal basis)
\bea
\chi_{ss}(\sigma)=\langle{\tiny\young(31,2)}|\Gamma_{\tiny\yng(2,1)}(\sigma)|{\tiny\young(32,1)}\rangle
\eea
The complete set of restricted characters are now easily evaluated
\bea
\chi_{ss}((1)(2)(3))&=&0\qquad \chi_{ss}((12))\,\,\,=\,\,\,{\sqrt{3}\over 2}
\qquad\chi_{ss}((13))\,\,\,=\,\,\,-{\sqrt{3}\over 2}\cr\cr
 \chi_{ss}((23))&=&0\qquad \chi_{ss}((123))\,\,\,=\,\,\,-{\sqrt{3}\over 2}
\qquad \chi_{ss}((132))\,\,\,=\,\,\,{\sqrt{3}\over 2}
\eea
It is now trivial to verify that
\bea
\chi_{ss}(Z,W_C,W_D)\equiv\sum_{\sigma\in S_3}\chi_{ss}(\sigma)
(W_D)^{i_1}_{i_{\sigma(1)}}(W_C)^{i_2}_{i_{\sigma(2)}}Z^{i_3}_{i_{\sigma(3)}}
={\sqrt{3}\over 2}O_{ss}
\eea
It is again interesting to explore the derivative of this operator. The derivative rule of \cite{deMelloKoch:2007rqf} states that the derivative with respect to string $D$, stretching between the two branes, vanishes\footnote{The thing that is special about string $D$ is that it occupies slot 1, i.e. its upper index is $i_1$.}. It is a simple exercise to verify that this is indeed the case
\bea
{\partial\over\partial (W_D)^a_a}O_{ss}&=&0
\eea

\section{Check of the generating function for a bound state of two sphere giants}\label{2S}

In this Appendix we expand the generating functions for the bound states of two sphere giant gravitons, in a power series in $t_1$ and $t_2$, in order to test their correctness. 

\subsection{Operators constructed from the generating function}

\noindent
{\bf No open strings attached:} The two sphere giant graviton bound state generating function is given by
\bea
Z(t_1,t_2)&=&\sum_{p_1\ge p_2}^\infty\sum_{p_2=1}^\infty t_2^{2p_2}t_1^{p_1-p_2}
{(p_1+p_2)!\over d_{(2^{p_2}1^{p_1-p_2})}p_2!(p_1-p_2)!}\chi_{(2^{p_2},1^{p_1-p_2})}(Z)\cr\cr
&=& \int d^N\bar{\psi}\int d^N\psi\int d^N\bar{\xi}\int d^N\xi\cr\cr
&&\qquad\qquad 
e^{-\bar{\psi}^i\psi_i-\bar{\xi}^i\xi_i-t_1\bar{\psi}^iZ_i^j\psi_j
-t_2^2\bar{\psi}^i\bar{\xi}^j (Z^k_i Z^l_j+Z^l_i Z^k_j)\psi_k\xi_l}
\label{sphgntchck}
\eea
where $d_{(2^{p_2}1^{p_1-p_2})}$ is the dimension of the symmetric group $S_{p_1+p_2}$ representation labelled by a Young diagram with $p_1$ boxes in the first column and $p_2\le p_1$ boxes in the second column. After a simple computation we have
\bea
d_{(2^{p_2}1^{p_1-p_2})}&=&{(p_1+p_2)!(p_1-p_2+1)\over (p_1+1)!p_2!}\cr\cr
{\cal M}_{p_1,p_2}&=&{(p_1+p_2)!\over d_{(2^{p_2}1^{p_1-p_2})}p_2!(p_1-p_2)!}
\,\,\,=\,\,\,{(p_1+1)!\over (p_1-p_2+1)!}
\eea
Power series expanding this integral in the parameters $t_1$ and $t_2^2$ we find a sequence of polynomials in the variables $\bar{\psi}^i$, $\psi_i$, $\bar{\xi}^i$ and $\xi_i$. These are all easily performed using (\ref{fermionicresult}) and the obvious analogous formula for integrals over $\bar{\xi}^i$ and $\xi_i$. The result is
\bea
Z(t_1,t_2,\{W\})
&=&1+t_1 d_1+t_2^2 d_2+t_1^2 d_3+t_1t_2^2 d_4+t_2^4 d_5+\cdots
\eea
where
\bea
d_1&=&\Tr(Z)\,\,\,=\,\,\,{\cal M}_{1,0}\chi_{\tiny\yng(1)}\cr\cr
d_2&=&\Tr(Z)^2+\Tr(Z^2)\,\,\,=\,\,\,{\cal M}_{2,0}\chi_{\tiny\yng(2)}\cr\cr
d_3&=&\Tr(Z)^2-\Tr(Z^2)\,\,\,=\,\,\,{\cal M}_{1,1}\chi_{\tiny\yng(1,1)}\cr\cr
d_4&=&\Tr(Z)^3-\Tr(Z^3)\,\,\,=\,\,\,{\cal M}_{2,1}\chi_{\tiny\yng(2,1)}\cr\cr
d_5&=&\Tr(Z)^4+3\Tr(Z^2)^2-4\Tr(Z)\Tr(Z^3)\,\,\,=\,\,\,{\cal M}_{2,2}
\chi_{\tiny\yng(2,2)}
\eea
confirming (\ref{sphgntchck}).

\noindent
{\bf Two open strings attached, to the giant of momentum $p_1$:} Next consider the generating function describing bound states of two sphere giant gravitons, with two strings attached to the first giant graviton. The generating function is given by
\bea
Z(t_1,t_2,\{W\})&=& \int d^N\bar{\psi}\int d^N\psi 
\int d^N\bar{\xi} \int d^N\xi \,\, e^{-\bar{\psi}^i\psi_i-\bar{\xi}^i\xi_i}\cr\cr
&&\quad\times
e^{-t_1\bar{\psi}^iZ_i^j\psi_j
-t_2^2\bar{\psi}^i\bar{\xi}^j (Z^k_i Z^l_j+Z^l_i Z^k_j)\psi_k\xi_l}
\psi_{a_1}(W_{A_1})^{a_1}_{b_1}\bar{\psi}^{b_1}
\psi_{a_2}(W_{A_2})^{a_2}_{b_2}\bar{\psi}^{b_2}\cr\cr
&=&f_0+t_1 f_1+t_2^2 f_2+t_1^2 f_3+t_1t_2^2 f_4+t_2^4 f_5+\cdots
\label{scndsetofgiantres}
\eea
where, for example
\bea
f_0&=&\Tr(W_{A_1})\Tr(W_{A_2})-\Tr(W_{A_1}W_{A_2})\cr\cr
f_1&=&\Tr(W_{A_1})\Tr(W_{A_2})\Tr(Z)-\Tr(ZW_{A_1})\Tr(W_{A_2})-\Tr(ZW_{A_2})\Tr(W_{A_1})\cr\cr
&&-\Tr(W_{A_1}W_{A_2})\Tr(Z)+\Tr(W_{A_1}W_{A_2}Z)+\Tr(W_{A_2}W_{A_1}Z)
\eea

\noindent
{\bf Two open strings attached, to the giant of momentum $p_2$:} Next consider the generating function describing bound states of two sphere giant gravitons, with two strings attached to the first giant graviton. The generating function is given by
\bea
Z(t_1,t_2,\{W\})&=& \int d^N\bar{\psi}\int d^N\psi 
\int d^N\bar{\xi} \int d^N\xi \,\,e^{-t_1\bar{\psi}^iZ_i^j\psi_j
-t_2^2\bar{\psi}^i\bar{\xi}^j (Z^k_i Z^l_j+Z^l_i Z^k_j)\psi_k\xi_l}\cr\cr
&&\quad\times e^{-\bar{\psi}^i\psi_i-\bar{\xi}^i\xi_i}
(\bar{\psi}\cdot Z\cdot\psi\, \bar{\xi}\cdot W_{B_1}\cdot \xi - \bar{\psi}\cdot Z\cdot\xi\, \bar{\xi}\cdot W_{B_1}\cdot \psi)\cr\cr
&&\quad\times(\bar{\psi}\cdot Z\cdot\psi\, \bar{\xi}\cdot W_{B_2}\cdot \xi - \bar{\psi}\cdot Z\cdot\xi\, \bar{\xi}\cdot W_{B_2}\cdot \psi)\cr\cr
&=&g_0+t_1 g_1+t_2^2 g_2+t_1^2 g_3+t_1t_2^2 g_4+t_2^4 g_5+\cdots
\label{onsecondspheregiant}
\eea
where, for example
\bea
g_0&=&\Tr(W_{B_1})\Tr(W_{B_2})\Tr(Z)^2-\Tr(Z)^2\Tr(W_{B_1}W_{B_2})+\Tr(W_{B_2}) \Tr (Z)\Tr(W_{B_1}Z)\cr\cr
&&+\Tr(W_{B_1}) \Tr (Z)\Tr(W_{B_2}Z)+2\Tr(W_{B_1}Z)\Tr(W_{B_2}Z)-\Tr(W_{B_1})\Tr (W_{B_2})\Tr (Z^2)\cr\cr 
&&+\Tr (W_{B_1}W_{B_2})\Tr (Z^2)- \Tr (Z)\Tr (W_{B_1}W_{B_2}Z)-\Tr(W_{B_2})\Tr (W_{B_1}Z^2) -  \Tr (Z) \Tr (W_{B_2}W_{B_1}Z)\cr\cr 
&&- \Tr (W_{B_1}) \Tr (W_{B_2}Z^2) + \Tr (W_{B_1}W_{B_2}Z^2) -  2 \Tr (W_{B_1}ZW_{B_2}Z) + \Tr (W_{B_2}W_{B_1}Z^2)\cr\cr
g_1&=&\Tr (W_{B_1})\Tr(W_{B_2})\Tr (Z)^3-\Tr(Z)^3\Tr(W_{B_1}W_{B_2})+\Tr(W_{B_2}) \Tr (Z)^2 \Tr (W_{B_1}Z)\cr\cr
&&+\Tr(W_{B_1})\Tr(Z)^2\Tr(W_{B_2}Z)+2\Tr(Z)\Tr (W_{B_1}Z)\Tr(W_{B_2}Z)-3\Tr(W_{B_1})\Tr(W_{B_2})\Tr(Z)\Tr (Z^2)\cr\cr
&&+3\Tr(Z)\Tr (W_{B_1}W_{B_2})\Tr (Z^2)-\Tr(W_{B_2})\Tr(W_{B_1}Z)\Tr (Z^2)-\Tr(W_{B_1})\Tr (W_{B_2}Z)\Tr (Z^2)\cr\cr
&&-\Tr (Z)^2 \Tr (W_{B_1}W_{B_2}Z) + \Tr (Z^2) \Tr (W_{B_1}W_{B_2}Z) - 
 2 \Tr (W_{B_2})\Tr (Z)\Tr (W_{B_1}Z^2)\cr\cr
&& - 2 \Tr (W_{B_2}Z)\Tr(W_{B_1}Z^2)-\Tr(Z)^2 \Tr (W_{B_2}W_{B_1}Z)+\Tr (Z^2)\Tr (W_{B_2}W_{B_1}Z)\cr\cr
&&-2\Tr (W_{B_1})\Tr (Z)\Tr (W_{B_2}Z^2)-2\Tr(W_{B_1}Z)\Tr(W_{B_2}Z^2)+2\Tr(W_{B_1})\Tr (W_{B_2})\Tr (Z^3)\cr\cr
&&-2\Tr(W_{B_1}W_{B_2})\Tr (Z^3)+2\Tr(Z)\Tr(W_{B_1}W_{B_2}Z^2)-2\Tr(Z)\Tr(W_{B_1}ZW_{B_2}Z)\cr\cr
&&+2\Tr(W_{B_2})\Tr(W_{B_1}Z^3)+2\Tr(Z)\Tr(W_{B_2}W_{B_1}Z^2)+2\Tr(W_{B_1})\Tr(W_{B_2}Z^3)\cr\cr
&&-2\Tr(W_{B_1}W_{B_2}Z^3)+2\Tr(W_{B_1}ZW_{B_2}Z^2)-2\Tr (W_{B_2}W_{B_1}Z^3)+2\Tr(W_{B_2}ZW_{B_1}Z^2)
\eea

\noindent
{\bf A pair of strings, one attached to each giant:} The generating function describing bound states of two sphere giant gravitons, with a string attached to each of the two giant gravitons, is given by
\bea
Z(t_1,t_2,\{W\})&=& \int d^N\bar{\psi}\int d^N\psi 
\int d^N\bar{\xi} \int d^N\xi \,\, e^{-\bar{\psi}^i\psi_i-\bar{\xi}^i\xi_i}\cr\cr
&&\quad\times
e^{-t_1\bar{\psi}^iZ_i^j\psi_j
-t_2^2\bar{\psi}^i\bar{\xi}^j (Z^k_i Z^l_j-Z^l_i Z^k_j)\psi_k\xi_l}\,\,
\bar{\psi}\cdot W_A\cdot \psi\cr\cr
&&\quad\times (\bar{\psi}\cdot Z\cdot\psi\, \bar{\xi}\cdot W_{B}\cdot \xi - \bar{\psi}\cdot Z\cdot\xi\, \bar{\xi}\cdot W_{B}\cdot \psi)\cr\cr
&=&h_0+t_1 h_1+t_2^2 h_2+t_1^2 h_3+t_1t_2^2 h_4+t_2^4 h_5+\cdots
\eea
where, for example
\bea
h_0&=&\Tr (W_A)\Tr (W_B)\Tr (Z)+\Tr (W_B)\Tr(W_AZ)-\Tr (W_A)\Tr(W_BZ)-\Tr (W_AW_BZ)\cr\cr
h_1&=&\Tr(W_{A}) \Tr(W_{B}) \Tr(Z)^2
         +    \Tr(W_{B}) \Tr(Z) \Tr(W_{A} Z)
         -  2 \Tr(W_{A}) \Tr(Z) \Tr(W_{B} Z)\cr\cr
         &&-    \Tr(W_{A} Z) \Tr(W_{B} Z) 
        -    \Tr(W_{A}) \Tr(W_{B}) \Tr(Z^2)
         -    \Tr(Z) \Tr(W_{A} . W_{B} Z)+ \Tr(W_{A} W_{B} Z^2 )\cr\cr
         &&-    \Tr(W_{B}) \Tr(W_{A} Z^{2})
         +  2 \Tr(W_{A}) \Tr(W_{B} Z^{2})
         + \Tr(W_{A} Z W_{B} Z)
\label{frthsetofgiantres}
\eea

\noindent
{\bf A pair of strings stretched between the giants:} Finally consider the generating function describing bound states of two sphere giant gravitons, with two strings stretched between the two giant gravitons. The generating function is given by
\bea
Z(t_1,t_2,\{W\})&=& \int d^N\bar{\psi}\int d^N\psi 
\int d^N\bar{\xi} \int d^N\xi \,\, e^{-\bar{\psi}^i\psi_i-\bar{\xi}^i\xi_i}\cr\cr
&&\quad\times
e^{-t_1\bar{\psi}^iZ_i^j\psi_j
-t_2^2\bar{\psi}^i\bar{\xi}^j (Z^k_i Z^l_j-Z^l_i Z^k_j)\psi_k\xi_l}\cr\cr
&&\quad\times (\bar{\psi}\cdot Z\cdot\psi\,\bar{\xi}\cdot W_C\cdot \psi\,\bar{\psi}\cdot W_D\cdot \xi-\bar{\psi}\cdot Z\cdot \xi\,\bar{\xi}\cdot W_C\cdot \psi\,\bar{\psi}W_D\,\psi)\cr\cr
&=&j_0+t_1 j_1+t_2^2 j_2+t_1^2 j_3+t_1t_2^2 j_4+t_2^4 j_5+\cdots
\eea
where, for example
\bea
j_0&=&	  \Tr(Z) \Tr(W_{C} W_{D})
	- \Tr(W_{D}) \Tr(W_{C} Z)
	+ \Tr(W_{C} W_{D} Z)
	- \Tr(W_{D} W_{C} Z)\cr\cr
j_1&=&		\Tr(Z)^2 \Tr(W_{C} W_{D})
	-   \Tr(W_{D}) \Tr(Z) \Tr(W_{C} Z)
	+   \Tr(Z) \Tr( W_{C} W_{D} Z )
	- 2 \Tr(Z) \Tr(W_{D} W_{C} Z) \cr\cr
	&&+   \Tr(W_{C} Z) \Tr(W_{D} Z)
	-   \Tr(Z^2) \Tr(W_{C} W_{D})
	+   \Tr(W_{D}) \Tr(W_{C} Z^2)
	-   \Tr(W_{C} W_{D} Z^2) \cr\cr
	&&-   \Tr(W_{C} Z W_{D} Z)
	+ 2 \Tr(W_{D} W_{C} Z^2)\label{frthsetofgiantres}
\eea

\subsection{Comparison to restricted Schur polynomial and Gauss graph operators}

Our goal is again to compare the operators we have obtained above to the Gauss graph operators, after setting the open string words equal to $Y$. As for the case of the AdS giants, the simplest possibility to consider is when the Young diagram labelling the operator has a single column, corresponding to a single giant graviton. In this case, the restricted Schur polynomial basis and the Gauss graph basis coincide, and restricted characters reduce to ordinary characters. It is simple to verify that 
\bea
\begin{array}{ccc}
f_0 &=\,\,2\,\chi_{\tiny \yng(1,1) \big(\cdot,\yng(1,1)\big)}(Z,Y) &= \left(\cdot \,\,,
\begin{gathered}\includegraphics[scale=0.4]{GG1}\end{gathered}\right)\cr\cr
f_1 &=\,\,2\,\chi_{\tiny \yng(1,1,1) \big(\yng(1),\yng(1,1)\big)}(Z,Y) &=\left({\tiny\yng(1)}\,\,, \begin{gathered}\includegraphics[scale=0.4]{GG1}\end{gathered}\right)\cr\cr
f_3 &=\,\,2\,\chi_{\tiny \yng(1,1,1,1) \big(\yng(1,1),\yng(1,1)\big)}(Z,Y) &=\left({\tiny\yng(1,1)}\,\,,\begin{gathered}\includegraphics[scale=0.4]{GG1}\end{gathered}\right)
\end{array}
\eea

The remaining terms in (\ref{scndsetofgiantres})) correspond to restricted Schur polynomials that are labelled by Young diagrams with two columns. Since we are attaching both strings to a single brane, it is still true that the Gauss graph operators and the restricted Schur polynomial bases coincide. The restricted Schur polynomials that we need to evaluate are given by
\bea
\chi_{\tiny \yng(2,1,1)(\yng(2),\yng(1,1))}&=&{1\over 4}\Big(
\Tr(Z)^2\Tr(Y)^2+2\Tr(Y^2Z^2)-\Tr(Y^2)\Tr(Z^2)-2\Tr(YZ^2)\Tr(Y)\cr\cr
&&+2\Tr(ZY^2)\Tr(Z)-\Tr(Y^2)\Tr(Z)^2
-2\Tr(YZ)\Tr(Z)\Tr(Y)+\Tr(Z^2)\Tr(Y)^2\Big)\cr\cr
\chi_{\tiny \yng(2,1,1,1)(\yng(2,1),\yng(1,1))}&=&{1\over 6}\Big(
\Tr(Z)^3\Tr(Y)^2- 4 \Tr(Z)^2\Tr(Y)\Tr(ZY)- \Tr(Z)^3\Tr(Y^2)\cr\cr
&&- 2\Tr(Z^2)\Tr(ZY)\Tr(Y)+2\Tr(Z^2Y)\Tr(Z)\Tr(Y)-\Tr(Z^3)\Tr(Y)^2\cr\cr
&&+ 4\Tr(Y^2Z)\Tr(Z)^2+2\Tr(ZY)^2\Tr(Z)+\Tr(Y^2)\Tr(Z^3)\cr\cr 
&&+ 2\Tr(ZY)\Tr(Z^2Y)-2\Tr(ZYZY)\Tr(Z)+2\Tr(Z^2)\Tr(ZY^2)\cr\cr 
&&-2\Tr(Z^2Y^2)\Tr(Z)+4\Tr(Z^3Y)\Tr(Y)-2\Tr(Z^2YZY)-4\Tr(Z^3Y^2)\Big)  \cr\cr
\chi_{\tiny \yng(2,2,1,1)(\yng(2,2),\yng(1,1))}&=&{1\over 24} \Tr (Y)^2 \Tr (Z)^4 - {1\over 24} \Tr (Z)^4 \Tr (Y^2)-{1\over 6} \Tr (Y) \Tr (Z)^3 \Tr (YZ) + {1\over 12} \Tr (Z)^2 \Tr (YZ)^2\cr\cr 
&&-{1\over 12}\Tr (YZ)^2\Tr (Z^2)+{1\over 8}\Tr (Y)^2\Tr (Z^2)^2-{1\over 8}\Tr (Y^2) \Tr (Z^2)^2+{1\over 6}\Tr (Z)^3\Tr (Y^2Z)\cr\cr
&&+{1\over 6}\Tr (Z)\Tr (YZ)\Tr(YZ^2)-{1\over 2} \Tr(Y)\Tr (Z^2)\Tr(YZ^2)+{1\over 6} \Tr (YZ^2)^2\cr\cr
&&-{1\over 6}\Tr(Y)^2\Tr (Z)\Tr (Z^3)+{1\over 6}\Tr(Z)\Tr (Y^2)\Tr(Z^3)+{1\over 6} \Tr (Y) \Tr (YZ) \Tr (Z^3)\cr\cr
&&- {1\over 6} \Tr (Y^2Z) \Tr (Z^3)+{1\over 2}\Tr (Z^2)\Tr (Y^2Z^2)-{1\over 12}\Tr (Z)^2\Tr(YZYZ)\cr\cr
&&+{1\over 12}\Tr(Z^2) \Tr (YZYZ)+{1\over 2}\Tr(Y)\Tr (Z)\Tr(YZ^3)-{1\over 6}\Tr(YZ)\Tr(YZ^3)\cr\cr
&&-{1\over 2}\Tr(Z) \Tr(Y^2Z^3)-{1\over 6}\Tr (Z)\Tr(YZYZ^2)+{1\over 6}\Tr(YZYZ^3) - {1\over 6} \Tr (YZ^2YZ^2)\cr
&&
\eea
It is now a simple exercise to check the following equalities
\bea
\begin{array}{ccc}
f_2 &=\,\,\,\,4\,\chi_{\tiny \yng(2,1,1) \big(\yng(2),\yng(1,1)\big)}(Z,Y) &=\,\,\,\, \left({\tiny \yng(2)}\,\,,
\begin{gathered}\includegraphics[scale=0.4]{GG2}\end{gathered}\right)\cr\cr
f_4 &=\,\,\,\,6\,\chi_{\tiny \yng(2,1,1,1) \big(\yng(2,1),\yng(1,1)\big)}(Z,Y) &=\,\,\,\,\left({\tiny\yng(2,1)}\,\,, \begin{gathered}\includegraphics[scale=0.4]{GG2}\end{gathered}\right)\cr\cr
f_5 &=\,\,\,\,12\,\chi_{\tiny \yng(2,2,1,1) \big(\yng(2,2),\yng(1,1)\big)}(Z,Y) &=\,\,\,\,\left({\tiny\yng(2,2)}\,\,,\begin{gathered}\includegraphics[scale=0.4]{GG2}\end{gathered}\right)
\end{array}\label{seconddcomp}
\eea
after setting $W_1=W_2=Y$ in $f_2$, $f_4$ and $f_5$.

Using the results for strings attached to the second giant graviton, given in (\ref{onsecondspheregiant}) we have checked that the identities
\bea
\begin{array}{ccc}
g_0 &=\,\,\,\,4\,\chi_{\tiny \yng(2,2) \big(\yng(1,1),\yng(1,1)\big)}(Z,Y) &=\,\,\,\, \left({\tiny \yng(1,1)}\,\,,
\begin{gathered}\includegraphics[scale=0.4]{GG4}\end{gathered}\right)\cr\cr
g_1 &=\,\,\,\,2\,\chi_{\tiny \yng(2,2,1) \big(\yng(1,1,1),\yng(1,1)\big)}(Z,Y) &=\,\,\,\,\left({\tiny\yng(1,1,1)}\,\,, \begin{gathered}\includegraphics[scale=0.4]{GG4}\end{gathered}\right)\cr\cr
g_2 &=\,\,\,\,16\,\chi_{\tiny \yng(2,2,2) \big(\yng(2,1,1),\yng(1,1)\big)}(Z,Y) &=\,\,\,\,\left({\tiny\yng(2,1,1)}\,\,,\begin{gathered}\includegraphics[scale=0.4]{GG4}\end{gathered}\right)\cr\cr
g_3 &=\,\,\,\,24\,\chi_{\tiny \yng(2,2,1,1) \big(\yng(1,1,1,1),\yng(1,1)\big)}(Z,Y) &=\,\,\,\,\left({\tiny\yng(1,1,1,1)}\,\,,\begin{gathered}\includegraphics[scale=0.4]{GG4}\end{gathered}\right)
\end{array}\label{seconddcomp}
\eea
are obeyed. To test the results when there is one string attached to each giant graviton, we again have to compare to a linear combination of restricted Schur polynomials. The discussion is parallel to the discussion for the AdS giants, so we simply quote the results. We have verified the identities
\bea
\begin{array}{ccc}
h_0 &=\,\,\,\,\chi_{\tiny \yng(2,1) \big(\yng(1),\yng(2)\big)}(Z,Y)+\chi_{\tiny \yng(2,1) \big(\yng(1),\yng(1,1)\big)}(Z,Y) &=\,\,\,\, \left({\tiny \yng(1)}\,\,,
\begin{gathered}\includegraphics[scale=0.4]{GG5}\end{gathered}\right)\cr\cr
h_1 &=\,\,\,\,2\,\chi_{\tiny \yng(2,1,1) \big(\yng(1,1),\yng(2)\big)}(Z,Y)+2\,\chi_{\tiny \yng(2,1,1) \big(\yng(1,1),\yng(1,1)\big)}(Z,Y) &=\,\,\,\,\left({\tiny\yng(1,1,1)}\,\,, \begin{gathered}\includegraphics[scale=0.4]{GG5}\end{gathered}\right)\cr\cr
h_2 &=\,\,\,\,{1\over 2}\,\chi_{\tiny \yng(2,2,1) \big(\yng(2,1),\yng(2)\big)}(Z,Y)+{1\over 2}\,\chi_{\tiny \yng(2,2,1) \big(\yng(2,1),\yng(1,1)\big)}(Z,Y) &=\,\,\,\,\left({\tiny\yng(2,1,1)}\,\,,\begin{gathered}\includegraphics[scale=0.4]{GG5}\end{gathered}\right)\cr\cr
h_3 &=\,\,\,\,3\,\chi_{\tiny \yng(2,1,1,1) \big(\yng(1,1,1),\yng(2)\big)}(Z,Y)+3\,\chi_{\tiny \yng(2,1,1,1) \big(\yng(1,1,1),\yng(1,1)\big)}(Z,Y) &=\,\,\,\,\left({\tiny\yng(1,1,1)}\,\,,\begin{gathered}\includegraphics[scale=0.4]{GG5}\end{gathered}\right)\cr\cr
h_4 &=\,\,\,\,8\,\chi_{\tiny \yng(2,2,1,1) \big(\yng(2,1,1),\yng(2)\big)}(Z,Y)+8\,\chi_{\tiny \yng(2,2,1,1) \big(\yng(2,1,1),\yng(1,1)\big)}(Z,Y) &=\,\,\,\,\left({\tiny\yng(1,1,1,1)}\,\,,\begin{gathered}\includegraphics[scale=0.4]{GG5}\end{gathered}\right)
\end{array}\label{seconddcomp}
\eea
Note once again that we have a complete match, even at finite $N$. Finally, for the operators corresponding to the stretched string states we have verified the identities
\bea
\begin{array}{ccc}
j_0 &=\,\,\,\,\chi_{\tiny \yng(2,1) \big(\yng(1),\yng(2)\big)}(Z,Y)-\chi_{\tiny \yng(2,1) \big(\yng(1),\yng(1,1)\big)}(Z,Y) &=\,\,\,\, \left({\tiny \yng(1)}\,\,,
\begin{gathered}\includegraphics[scale=0.4]{GG3}\end{gathered}\right)\cr\cr
j_1 &=\,\,\,\,2\,\chi_{\tiny \yng(2,1,1) \big(\yng(1,1),\yng(2)\big)}(Z,Y)-2\,\chi_{\tiny \yng(2,1,1) \big(\yng(1,1),\yng(1,1)\big)}(Z,Y) &=\,\,\,\,\left({\tiny\yng(1,1,1)}\,\,, \begin{gathered}\includegraphics[scale=0.4]{GG3}\end{gathered}\right)\cr\cr
j_2 &=\,\,\,\,{1\over 2}\,\chi_{\tiny \yng(2,2,1) \big(\yng(2,1),\yng(2)\big)}(Z,Y)-{1\over 2}\,\chi_{\tiny \yng(2,2,1) \big(\yng(2,1),\yng(1,1)\big)}(Z,Y) &=\,\,\,\,\left({\tiny\yng(2,1,1)}\,\,,\begin{gathered}\includegraphics[scale=0.4]{GG3}\end{gathered}\right)\cr\cr
j_3 &=\,\,\,\,3\,\chi_{\tiny \yng(2,1,1,1) \big(\yng(1,1,1),\yng(2)\big)}(Z,Y)-3\,\chi_{\tiny \yng(2,1,1,1) \big(\yng(1,1,1),\yng(1,1)\big)}(Z,Y) &=\,\,\,\,\left({\tiny\yng(1,1,1)}\,\,,\begin{gathered}\includegraphics[scale=0.4]{GG3}\end{gathered}\right)\cr\cr
j_4 &=\,\,\,\,8\,\chi_{\tiny \yng(2,2,1,1) \big(\yng(2,1,1),\yng(2)\big)}(Z,Y)-8\,\chi_{\tiny \yng(2,2,1,1) \big(\yng(2,1,1),\yng(1,1)\big)}(Z,Y) &=\,\,\,\,\left({\tiny\yng(1,1,1,1)}\,\,,\begin{gathered}\includegraphics[scale=0.4]{GG3}\end{gathered}\right)
\end{array}\label{seconddcomp}
\eea

\section{A Gauss Graph Identity}\label{ggequality}

In this Appendix we will prove an identity for Gauss graph operators which do not involve any stretched strings. This corresponds to setting $\sigma={\bf 1}$ in $O_{R,r}(\sigma)$. We choose $r$ to have $p_1$ boxes in the first row and $p_2$ boxes in the second and take both $p_1$ and $p_2$ to be $O(N)$. $R$ has $p_1+m_1$ boxes in the first row and $m_2+p_2$ boxes in the second row so that $m_1$ of the boxes in the first row and $m_2$ in the second row correspond to $Y$'s. We take $m_1$ and $m_2$ to be $O(1)$. Recall formula (\ref{niceggformula}) which, after setting $\sigma$ to the identity, says
\bea
O_{R,r}({\bf 1})=\sum_{\rho\in S_{n+m}}\sum_{\sigma_1\in S_n}\sum_{\alpha\in S_{m_1}\times S_{m_2}}\chi_r(\sigma_1)\chi_R(\sigma_1\circ\alpha\,\rho)\Tr(\rho Z^{\otimes n} Y^{\otimes m})
\eea
There is a nice interpretation of this formula that follows after recalling that the projection operator onto representation $q$, within a vector space $V$, is given by
\bea
P_q={d_q\over |{\cal G}|}\sum_{\sigma\in{\cal G}}\chi_q(\sigma)\sigma
\eea
Here $\sigma$ is the action of group ${\cal G}$ on the space $V$, $\chi_q(\sigma)$ is the character of group element $\sigma$ in representation $q$, $d_q$ is the dimension of representation $q$ and $|{\cal G}|$ is the order of the group ${\cal G}$. If there is more than one subspace of $V$ that carries representation $q$, $P_q$ will project to their direct sum. Using $(m)$ to denote the trivial representation of $S_m$, labelled by a Young diagram with a single row of length $m$, we can write (up to an unimportant constant which we drop)
\bea
O_{R,r}({\bf 1})=\sum_{\rho\in S_{n+m}}\chi_R(P_r P_{(m_1)}P_{(m_2)}\,\rho)\Tr(\rho Z^{\otimes n}\otimes Y^{\otimes m})
\eea
We are going to work out a formula for the derivative of this operator with respect to $Y^a_a$. This analysis is motivated by Appendix C of \cite{deMelloKoch:2007rqf} and we recommend that the reader consult this source for further details that we do not spell out here. It is useful to generalize the operators we consider by including a third field
\bea
O^{(3)}_{R,r}({\bf 1})=\sum_{\rho\in S_{n+m}}\chi_R(P_r P_{(m_1)}P_{(m_2)}\,\rho)\Tr(\rho Z^{\otimes n}\otimes X^{\otimes m_2}\otimes Y^{\otimes m_1})
\eea
Describing the states in the carrier space of $R\vdash m+n$ using Young-Yammanouchi patterns, the indices of the fields $Y$ correspond to boxes with labels from 1 to $m_1$, the indices of the fields $X$ to boxes with labels from $m_1+1$ to $m_1+m_2=m$ and $Z$ to labels from $m+1$ to $m+n$. The $Y$ fields are in representation $(m_1)$ and the $X$ fields in representation $(m_2)$. The $Y$ fields populate boxes in the first row, while the $X$ fields populate boxes in the second row. Now, consider the derivative
\bea
{d\over dY^a_a}O^{(3)}_{R,r}({\bf 1})=m_1\sum_{\rho\in S_{n+m}}\chi_R(P_r P_{(m_1)}P_{(m_2)}\,\rho)\Tr(\rho Z^{\otimes n}\otimes X^{\otimes m_2}\otimes Y^{\otimes m_1-1}\otimes {\bf 1})
\eea
where $a$ is summed on the left hand side. Here we have chosen the index of ${\bf 1}$ to correspond to the box in $R$ labelled 1 and we have used the fact that we can differentiate any of the $Y$ fields and then cycle the identity ${\bf 1}$ into slot 1, possible because the $Y$ fields are in representation $(m_1)$. Writing the indices more explicitly we have
\bea
{d\over dY^a_a}O^{(3)}_{R,r}({\bf 1})&=&m_1\sum_{\rho\in S_{n+m}}\chi_R(P_r P_{(m_1)}P_{(m_2)}\,\rho) Z^{i_{n+m}}_{i_{\rho(n+m)}}\cdots Z^{i_{n+1}}_{i_{\rho(n+1)}}
X^{i_{m_1+m_2}}_{i_{\rho(m_1+m_2)}} \dots X^{i_{m_1+1}}_{i_{\rho(m_1+1)}}\cr\cr
&&\times Y^{i_{m_1}}_{i_{\rho(m_1)}}\cdots Y^{i_2}_{i_{\rho(2)}}\delta^{i_1}_{i_{\rho(1)}}
\eea
To proceed, write the sum over $S_{n+m}$ as a sum over the subgroup $S_{n+m-1}$ which permutes $2,3,\cdots,m+n$ but leaves 1 inert. The result is
\bea
{d\over dY^a_a}O^{(3)}_{R,r}({\bf 1})&=&m_1\sum_{\rho\in S_{n+m-1}}\chi_R(P_r P_{(m_1)}P_{(m_2)}\,\rho C) Z^{i_{n+m}}_{i_{\rho(n+m)}}\cdots Z^{i_{n+1}}_{i_{\rho(n+1)}}
X^{i_{m_1+m_2}}_{i_{\rho(m_1+m_2)}} \dots X^{i_{m_1+1}}_{i_{\rho(m_1+1)}}\cr\cr
&&\times Y^{i_{m_1}}_{i_{\rho(m_1)}}\cdots Y^{i_2}_{i_{\rho(2)}}
\eea
where
\bea
C\,\,=\,\,N+\sum_{i=2}^{n+m}(1,i)\,\,=\,\,N+\hat{j}
\eea
In this formula, $\hat{j}$ is a Jucys-Murphy element. Denote the representation obtained after dropping the box labelled 1 from $R$ by $R'$. The box that is dropped is from the first row of $R$. Acting on this representation $\hat{j}$ gives the content of the dropped box and we have
\bea
{d\over dY^a_a}O^{(3)}_{R,r}({\bf 1})&=&m_1 c_{RR'}O_{R',r}({\bf 1})
\eea
This is an exact result. Now consider the derivative with respect to $X$. A simple calculation gives
\bea
{d\over dX^a_a}O^{(3)}_{R,r}({\bf 1})&=&m_2\sum_{\rho\in S_{n+m}}\chi_R(P_r P_{(m_1)}P_{(m_2)}\,\rho) Z^{i_{n+m}}_{i_{\rho(n+m)}}\cdots Z^{i_{n+1}}_{i_{\rho(n+1)}}
X^{i_{m_1+m_2}}_{i_{\rho(m_1+m_2)}} \dots \delta^{i_{m_1+1}}_{i_{\rho(m_1+1)}}\cr\cr
&&\times Y^{i_{m_1}}_{i_{\rho(m_1)}}\cdots Y^{i_2}_{i_{\rho(2)}}Y^{i_1}_{i_{\rho(1)}}
\eea
To proceed, we write the sum over $S_{n+m}$ as a sum over the subgroup $S_{n+m-1}$ which permutes $1,2\cdots,m_1,m_1+2,\cdots m+n$ but leaves $m_1+1$ inert. The result is
\bea
{d\over dX^a_a}O^{(3)}_{R,r}({\bf 1})&=&m_2\sum_{\rho\in S_{n+m-1}}\chi_R(P_r P_{(m_1)}P_{(m_2)}\,\rho C) Z^{i_{n+m}}_{i_{\rho(n+m)}}\cdots Z^{i_{n+1}}_{i_{\rho(n+1)}}
X^{i_{m_1+m_2}}_{i_{\rho(m_1+m_2)}} \dots X^{i_{m_1+2}}_{i_{\rho(m_1+2)}}\cr\cr
&&\times Y^{i_{m_1}}_{i_{\rho(m_1)}}\cdots Y^{i_1}_{i_{\rho(1)}}
\eea
where
\bea
C\,\,=\,\,\kappa_1+\kappa_2
\eea
\bea
\kappa_1\,\,=\,\,N+\sum_{i=m_1+2}^{n+m}(m_1+1,i)\qquad 
\kappa_2\,\,=\,\,\sum_{i=1}^{m_1}(m_1+1,i)
\eea
We will now argue that at large $N$, the terms collected in $\kappa_2$ can be ignored. Above we are summing transpositions (2-cycles). Transpositions square to the identity matrix, so that their eigenvalues square to 1. Consequently, the matrix elements of the transpositions are bounded, independent of $N$. In the term $\kappa_2$ we sum only $m_1=O(1)$ terms, while in $\kappa_1$ we sum $m_2+n=O(N)$ terms. In addition, the first term in $\kappa_1$ is $N$. This certainly bounds the eigenvalues of $\kappa_2$ to be $O(1)$ at most. One might worry that there are cancellations between terms summed in $\kappa_1$, which might in the end imply that $\kappa_1$ itself has eigenvalues of order 1. This is not the case: it is clear $\kappa_1=N+\hat{j}$. For the case that we consider here, $\hat{j}$ gives the content of the last box in the second row, which is $p_2-2$. Thus, $\kappa_1$ has all eigenvalues of order $N$ and we can safely ignore $\kappa_2$ at large $N$. In this case we have
\bea
{d\over dX^a_a}O^{(3)}_{R,r}({\bf 1})&=&m_2 c_{RR'}O_{R',r}({\bf 1})
\eea
In this formula the box dropped to obtain $R'$ from $R$ is the right most box in the second row. The formula we want to derive, is for the derivative of $O_{R,r}({\bf 1})$. This can be written as
\bea
{d\over dY^a_a}O_{R,r}({\bf 1})&=&
{d\over dY^a_a}O^{(3)}_{R,r}({\bf 1})|_{X=Y}+
{d\over dX^a_a}O^{(3)}_{R,r}({\bf 1})|_{X=Y}\cr\cr
&=&m_1 c_{RR_1'}O_{R_1',r}({\bf 1})+m_2 c_{RR_2'}O_{R_2',r}({\bf 1})
\label{finalrule}
\eea
where we have used the notation $R_i'$ to denote the representation obtained by dropping a box from row $i$ of $R$.

It is a much simpler exercise to show that this is the rule obeyed by operators obtained from generating function. The relevant generating function is
\bea
Z_{\rm AdS}(t_1,t_2,Y) &=&{1\over \pi^{2N}}\int d^N\bar{\varphi}\int d^N\varphi 
\int d^N\bar{\phi} \int d^N\phi \,\, e^{-\bar{\varphi}\cdot\varphi-\bar{\phi}\cdot\phi}\cr\cr
&&\times
e^{t_1\bar{\varphi}\cdot Z\cdot\varphi
+t_2^2\bar{\varphi}\cdot Z\cdot\varphi \bar{\phi}\cdot Z\cdot\phi
-t_2^2\bar{\varphi}\cdot Z\cdot\phi \bar{\phi}\cdot Z\cdot\varphi}
(\bar{\varphi}\cdot Y\cdot\varphi)^{m_{11}}\cr\cr
&&\times(\bar{\varphi}\cdot Z\cdot \varphi\,\bar{\phi}\cdot Y\cdot \phi-\bar{\varphi}\cdot Z\cdot\phi\, \bar{\phi}\cdot Y\cdot\varphi)^{m_{22}}
\eea
Setting $m_{11}$ equal to $m_1$ and $m_{22}$ equal to $m_2$, and extracting the coefficient of $(t_2^2)^{p_2}t_1^{p_1-p_2}$ gives the operator $O_{R,r}({\bf 1})$ with $m_1$ $Y$ boxes in the first row, $m_2$ $Y$ boxes in the second and $r$ a Young diagram with $p_1$ boxes in the first row and $p_2$ boxes in the second row. Differentiating with respect to $Y^a_a$ and using the arguments of Section \ref{excitedAdS}, we easily recover (\ref{finalrule}).

This proves that the derivative with respect to $Y$ of the Gauss graph operator matches the derivative with respect to $Y$ of the operators defined by our generating function. Since there are no contributions to either operator that is independent of $Y$, this proves their equality.

There is an obvious extension of these arguments to the case of Gauss graph operators for sphere giant gravitons.

\section{Correlation Functions}\label{correlators}

In this section we evaluate the correlation functions that were used in Section \ref{ggtrue}. Consider
\bea
{\cal O}_1&=&\sum_{\sigma\in S_{n+2}}Z^{i_1}_{i_{\sigma(1)}}
\cdots Z^{i_{n+1}}_{i_{\sigma(n+1)}}(Y^2)^{i_{n+2}}_{i_{\sigma(n+2)}}
\eea
We would like to evaluate the correlator $\langle{\cal O}_1{\cal O}_1^\dagger\rangle$. The $Y$ fields only contract with $Y^\dagger$ fields. Note that
\bea
\langle (Y^2)^i_j (Y^{\dagger 2})^k_l\rangle&=&\delta^k_l\delta^i_j+N\delta^i_l\delta^k_j
\eea
At large $N$ we need only keep the second term on the right hand side. This gives
\bea
\langle (Y^2)^i_j(Y^{\dagger 2})^k_l\rangle&=&N\langle Y^i_jY^k_l\rangle (1+O(N^{-1})) 
\eea
We will not display the $O(N^{-1})$ correction in what follows. Using this formula we have (the operator $P_{(n+2)}$ is defined in (\ref{projop}) above)
\bea
\langle{\cal O}_1{\cal O}_1^\dagger\rangle &=&
N\sum_{\sigma_1,\sigma_2\in S_{n+2}}
\langle Z^{i_1}_{i_{\sigma_1(1)}}\cdots Z^{i_{n+1}}_{i_{\sigma_1(n+1)}}Y^{i_{n+2}}_{i_{\sigma_1(n+2)}}Z^{\dagger\,j_1}_{j_{\sigma_2(1)}}\cdots Z^{\dagger\, j_{n+1}}_{j_{\sigma_2 (n+1)}}Y^{\dagger\, j_{n+2}}_{j_{\sigma_2(n+2)}}\rangle\cr\cr
&=&N\sum_{\sigma_1,\sigma_2\in S_{n+2}}\sum_{\rho\in S_{n+1}}
\Tr (\sigma_1\rho\sigma_2\rho^{-1})\cr\cr
&=&N(n+1)!\sum_{\sigma_1,\sigma_2\in S_{n+2}}\Tr (\sigma_1\sigma_2)\cr\cr
&=&N(n+1)!(n+2)!\sum_{\sigma_1\in S_{n+2}}\Tr (\sigma_1)\cr\cr
&=&N(n+1)! ((n+2)!)^2\Tr (P_{(n+2)})\cr\cr
&=&N(n+1)!(n+2)! \prod_{i=0}^{n+1}(N+i)
\eea
Next, consider
\bea
{\cal O}_2\,\,=\,\,\sum_{\sigma\in S_{n+2}}Z^{i_1}_{i_{\sigma(1)}}
\cdots Z^{i_{n}}_{i_{\sigma(n)}}Y^{i_{n+1}}_{i_{\sigma(n+1)}}(ZY)^{i_{n+2}}_{i_{\sigma(n+2)}}
\eea
Our previous strategy of performing the $Y$ contractions by hand is not very useful when evaluating the $\langle{\cal O}_2{\cal O}_2^\dagger\rangle$ correlator. A better approach is to start from the identity
\bea
\sum_{\sigma\in S_{n+3}} Z^{i_1}_{i_{\sigma(1)}}\cdots Z^{i_{n+1}}_{i_{\sigma(n+1)}}
Y^{i_{n+2}}_{i_{\sigma(n+2)}}Y^{i_{n+3}}_{i_{\sigma(n+3)}}
&=&\sum_{\sigma\in S_{n+2}} Z^{i_1}_{i_{\sigma(1)}}\cdots Z^{i_{n+1}}_{i_{\sigma(n+1)}}Y^{i_{n+2}}_{i_{\sigma(n+2)}}\Tr (Y)\cr\cr
&&+\sum_{\sigma\in S_{n+2}} Z^{i_1}_{i_{\sigma(1)}}\cdots Z^{i_{n+1}}_{i_{\sigma(n+1)}}(Y^2)^{i_{n+2}}_{i_{\sigma(n+2)}}\cr\cr
&&+(n+1)\sum_{\sigma\in S_{n+2}} Z^{i_1}_{i_{\sigma(1)}}\cdots Y^{i_{n+1}}_{i_{\sigma(n+1)}}(ZY)^{i_{n+2}}_{i_{\sigma(n+2)}}\nonumber
\eea
which follows after summing, on the right hand side, over the subgroup and its cosets. The first two terms on the right hand side can be neglected at large $N$ so that
\bea
{\cal O}_2\,\,=\,\,{1\over n+1}\sum_{\sigma\in S_{n+3}} Z^{i_1}_{i_{\sigma(1)}}\cdots Z^{i_{n+1}}_{i_{\sigma(n+1)}}
Y^{i_{n+2}}_{i_{\sigma(n+2)}}Y^{i_{n+3}}_{i_{\sigma(n+3)}}
\eea
A simple computation now gives
\bea
\langle {\cal O}_2{\cal O}_2^\dagger\rangle
&=&2 (n!)^2 (n+2)(n+3)\prod_{i=0}^{n+2}(N+i)
\eea
and hence we have
\bea
{\langle{\cal O}_2{\cal O}_2^\dagger\rangle\over\langle{\cal O}_1{\cal O}_1^\dagger\rangle}\,\,=\,\,{2(n+3)(N+n+2)\over (n+1)^2 N}\,\,=\,\,O(N^{-1})
\eea

Finally, consider
\bea
{\cal O}'&=&\sum_{\sigma\in S_{n+1}}Z^{i_1}_{i_{\sigma(1)}}\cdots Z^{i_{n}}_{i_{\sigma(n)}}
(YZY)^{i_{n+1}}_{i_{\sigma(n+1)}}
\eea
A very similar argument to the one just given establishes that
\bea
{\langle{\cal O}'{\cal O}^{\prime\dagger}\rangle\over\langle{\cal O}_1{\cal O}_1^\dagger\rangle}&=&O(N^{-2})
\eea

\end{appendix}

\end{document}